\documentclass{emulateapj}

\usepackage{graphicx}
\usepackage{enumerate}
\usepackage{amsmath}
\usepackage{comment}

\def\sec\ond{{\rm s}}

\def\be{\begin{equation}}\def\bea{\begin{eqnarray}}\def\beaa{\begin{eqnarray*}}
\def\ee{\end{equation}}  \def\eea{\end{eqnarray}}  \def\eeaa{\end{eqnarray*}}

\shorttitle{Cosmological Hydrodynamical Simulations: The \textit{Sejong Suite}}
\shortauthors{Graziano Rossi (2020)}
\begin{document}
\title{The \textit{Sejong Suite}: Cosmological Hydrodynamical Simulations with Massive Neutrinos, Dark Radiation, and Warm Dark Matter}
\author{Graziano Rossi}
\affil{Department of Physics and Astronomy, Sejong University, Seoul, 143-747, Republic of Korea; graziano@sejong.ac.kr}
\email{Corresponding Author: Graziano Rossi (graziano@sejong.ac.kr)}



\begin{abstract}

We present the \textit{Sejong Suite}, an extensive collection of 
state-of-the-art high-resolution cosmological hydrodynamical simulations
spanning a variety of cosmological and astrophysical  parameters, primarily developed for modeling the Lyman-$\alpha$ (Ly$\alpha$) forest.
The suite is  organized into three main categories ({\it Grid Suite}, {\it Supporting Suite}, and {\it Systematics Suite}), addressing different science targets.
Adopting a particle-based implementation, we follow the
evolution of gas, dark matter (cold and warm), massive neutrinos, and dark radiation, and consider several combinations of box sizes and number of particles. 
With additional enhancing techniques, we are able to reach an equivalent resolution up to $3 \times  3328^3 = 110$ billion particles in a $(100 h^{-1} {\rm Mpc})^3$ box size, ideal
for current and future surveys (e.g., eBOSS, DESI). 
Noticeably, for the first time, we simulate extended mixed scenarios describing the combined effects of warm dark matter, neutrinos, and dark radiation,
modeled consistently by taking into account the neutrino mass splitting. 
In addition to providing multicomponent snapshots from $z=5.0$ to $z=2.0$ in intervals of $\Delta z=0.2$ for all of the models considered, we
produced over 288 million Ly$\alpha$ skewers 
in the same $z$-range and extended parameter space. 
The skewers are well suited for Ly$\alpha$ forest science studies, 
for mapping the high-$z$ cosmic web and the 
matter-to-flux relation and bias, and for quantifying the critical role of baryons at small scales.  
We also present a first analysis of the suite focused on the matter and flux statistics, and
show that we are able to accurately reproduce the 1D flux power spectrum down to scales $k=0.06~{\rm [km/s]^{-1}}$
as mapped by recent high-resolution quasar data, as well as the thermal history of the intergalactic medium.
The simulations and products described here will be progressively made available.

\end{abstract}






\section{Introduction} \label{sec_introduction}

State-of-the-art experiments such as the Sloan Digital Sky Survey (SDSS; York et al. 2000; Dawson et al. 2016; Blanton et al. 2017; Abolfathi et al. 2018), 
the WiggleZ Dark Energy Survey (Drinkwater et al. 2010, 2018), 
the Dark Energy Survey (DES; The Dark Energy Survey Collaboration 2005; Abbott et al. 2018a, 2018b, 2019), 
and the cosmic microwave background (CMB) Planck mission (Planck Collaboration et al. 2018a, 2018b) have been pivotal 
in constraining with high accuracy the basic parameters of the standard
spatially flat $\Lambda$CDM cosmology, 
dominated by collisionless cold dark matter (CDM) and a dark energy (DE) component in the form of a cosmological constant ($\Lambda$),
with baryons representing only $\bf \sim 5\%$ of the total matter-energy content.
The $\Lambda$CDM model is a relatively simple and successful framework centered on the 
assumption that general relativity (GR) is the correct theory of gravity on cosmological scales, 
and characterized by six minimal parameters,  a power-law spectrum of adiabatic scalar perturbations, and  primordial Gaussian density fluctuations. 
This scenario is consistent with spatial flatness to percent-level precision, in excellent agreement with big bang nucleosynthesis (BBN) 
and the standard value of the number of effective neutrino species $N_{\rm eff}$, with no evidence for dynamical DE and deviations from Gaussianity. 

However, a pure CDM framework stemmed from Gaussian initial conditions is incomplete, 
as some level of primordial non-Gaussianity (PNG) is generically expected in all inflation models, and
conclusions regarding the concordance $\Lambda$CDM model are mainly drawn from large-scale structure (LSS) observations -- 
while small scales remain poorly explored to date. 
In addition, GR may not be holding at cosmological scales,
the nature of  DE and dark matter (DM) still needs to be explained, 
and  it has now been experimentally confirmed that neutrinos are massive particles 
and should be accounted for in the cosmological framework. 
All of these considerations
point to necessary extensions of the $\Lambda$CDM scenario, and call for
a substantial reexamination in our understanding of the {\it dark sector}: precisely the dark side of the universe, and 
in particular, massive neutrinos, dark radiation, and the nature of DM  are the central focus of the present work --
the primary target of which is the Lyman-$\alpha$ (Ly$\alpha$) forest.


\begin{figure*}
\centering
\includegraphics[angle=0,width=0.85\textwidth]{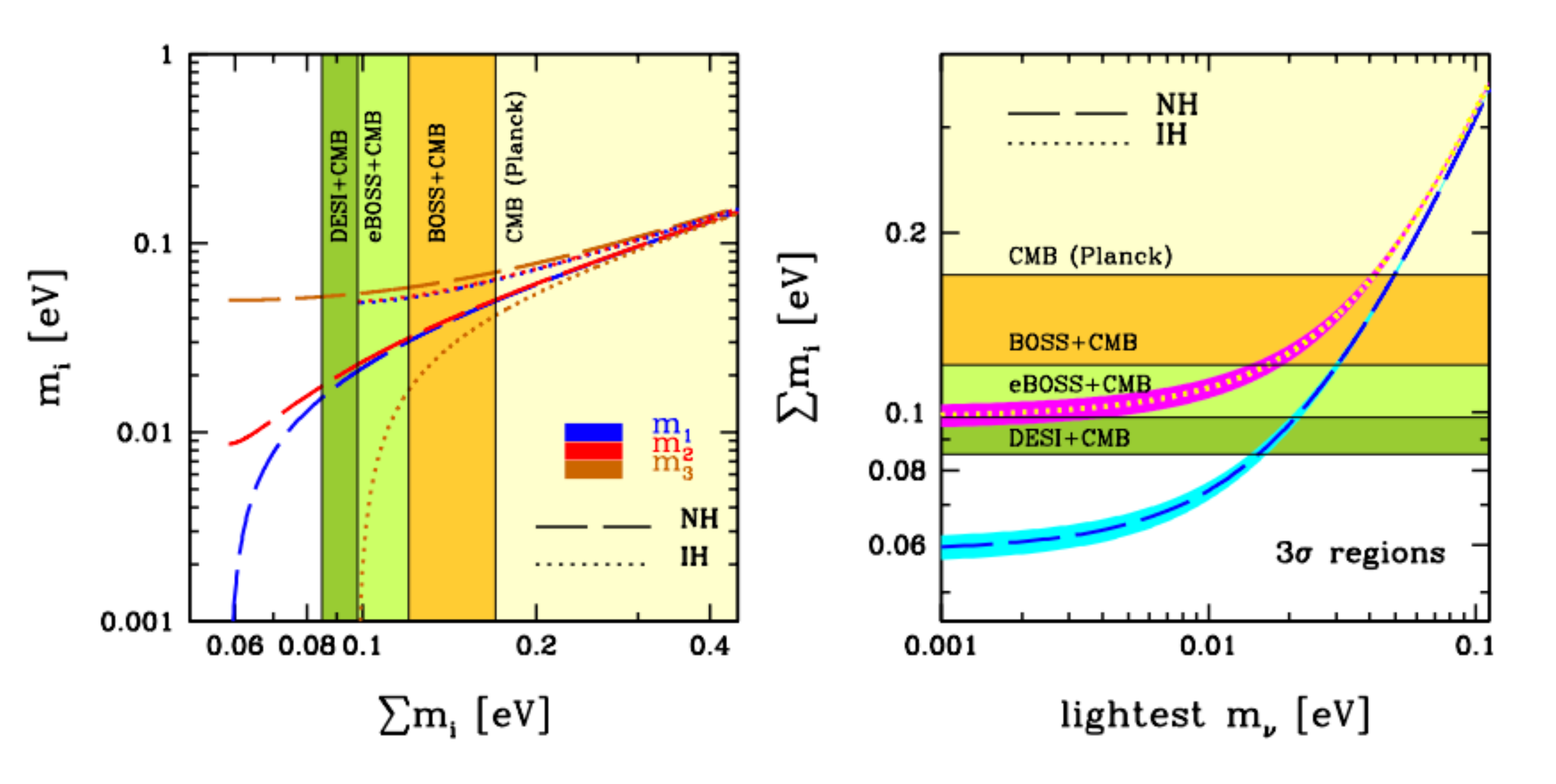}
\caption{
Neutrino mass limits as derived from the SDSS-III BOSS and Planck CMB data (yellow and orange areas in the panels), 
and future forecasts (SDSS-IV eBOSS DR16 and DESI; lighter and darker green areas, respectively). [Left]
Individual neutrino masses in the NH (long-dashed lines) and IH (dotted lines) configurations, as a function of the summed neutrino mass.
[Right] Total vs. the lightest neutrino mass in the two hierarchy scenarios, indicated with similar line styles as in the left panel. 
Likely, mass splitting effects are too small to be detected from cosmological probes.
See the main text for more details.}
\label{fig_neutrino_properties}
\end{figure*}


In this respect, pursuing the physics associated with neutrino mass is currently considered  one of the five major science drivers --
as featured in the report of the 2014 USA Particle Physics Project Prioritization Panel (P5; see Abazajian et al. 2015a, 2015b),
and in the Decadal Survey on Astronomy and Astrophysics (Astro2020; Panel on Cosmology). It is also the main motivation behind our study,  
which aims at producing new high-resolution cosmological hydrodynamical simulations principally for the {\it dark sector} at small scales, as mapped by 
high-redshift quasars. 
Determining the absolute neutrino mass scale, 
hierarchy, and $N_{\rm eff}$ are within reach using cosmological
measurements, rather than laboratory experiments: upper bounds on the total neutrino mass $\sum m_{\nu}$ from cosmology are now
approaching the minimum value allowed by the inverted hierarchy (IH; $\sum m_{\nu} = 0.097~{\rm eV}$),
while in the normal hierarchy (NH) scenario, $\sum m_{\nu} = 0.057~{\rm eV}$ (see, e.g., Capozzi et al. 2017; Planck Collaboration 2018b). 
Future experiments  and next-generation CMB missions (i.e., CMB-S4; Abazajian et al. 2016, 2019; Abitbol et al. 2017) in combination with 21 cm surveys are 
expected to significantly improve the present bounds on $\sum m_{\nu}$ and $N_{\rm eff}$.
For example, the Dark Energy Spectroscopic Instrument (DESI; DESI Collaboration et al. 2016a, 2016b)
will be able to measure $\sum m_{\nu}$ with an uncertainty of $0.020$ eV for $k_{\rm max} < 0.2~h {\rm Mpc^{-1}}$, 
sufficient to make the first direct detection  of $\sum m_{\nu}$ at more than $3 \sigma$ significance, and rule out the IH at the $99\%$ 
confidence level (CL) if the hierarchy is normal and the masses are minimal.

Upper bounds on massive neutrinos from cosmology are usually
obtained from the CMB (the most direct route), and via
 other baryonic tracers of the LSS sensitive to neutrino properties. Here we  
focus on the 
Ly$\alpha$ forest, namely the absorption lines in the spectra of high-redshift quasars due to neutral hydrogen in the intervening photoionized 
intergalactic medium (IGM); for recent studies on the forest, see, for instance, 
du Mas des Bourboux et al. (2019), Garzilli et al. (2019), and Porqueres et al. (2019).
Being a direct tracer of the underlying DM fluctuations and of the baryonic matter distribution over a wide range of scales  and redshifts, and having 
radically different systematics than those of other lower-$z$ probes, the 
Ly$\alpha$ forest is characterized by an excellent  sensitivity to neutrino mass, as thoroughly quantified in Rossi (2017).
The current work is primarily centered around an accurate modeling of the forest, and of the high-$z$ cosmic web. 
Indeed, the most competitive neutrino mass constraints so far have been derived by combining  Ly$\alpha$ data with 
additional tracers,  
and future
forecasts including the forest as mapped by planned large-volume experiments are even more promising.

To this end, Figure \ref{fig_neutrino_properties} summarizes the present status of the most competitive neutrino mass constraints derived from the
SDSS-III Baryon Oscillation Spectroscopic Survey (BOSS; Dawson et al. 2013) combined with Planck CMB data (yellow and orange zones in the panels), and 
highlights predicted achievements by the SDSS-IV Extended Baryon Oscillation Spectroscopic Survey (eBOSS; Dawson et al. 2016) as expected
in its final data release (DR16; see the forecasts by Zhao et al. 2016) and DESI (lighter and darker green zones, respectively) -- always in synergy with CMB data.\footnote{In Figure \ref{fig_neutrino_properties}, 
the colored areas visualize the gain of each 
survey with respect to the previous ones, with the lower left value being the actual (or expected) upper bound on $\sum m_{\nu}$ at the $95\%$ CL.}
Specifically, the left panel shows individual neutrino masses $m_{\rm i}$ (indicated with three different colors) in the two possible configurations of NH (long-dashed lines) and IH (dotted lines),
versus the summed neutrino mass. The right panel displays the total neutrino mass as a function of the lightest one, in the two hierarchy configurations: 
the long-dashed line is again used for the NH, while the dotted line shows the IH. The shaded cyan and purple areas in the figure
represent the $3\sigma$ regions allowed by both configurations. 
The absolute neutrino mass scale (lightest $m_{\nu}$)
has a significant effect on the combined constraints; 
this parameter governs the lower bound on the joint mixing matrix from neutrino oscillation data, and
if it is near the current upper cosmological limit, direct search experiments in the foreseeable future
may have a chance to reach this lower bound. 
Clearly, mass splitting effects are quite small and may be too challenging to be measured. However, as
is evident from the figure, DESI (and partly eBOSS but at lower CL), in combination with CMB data, should be able to
rule out the IH at more than $3\sigma$ under the most standard assumptions previously mentioned,
by placing the tightest $\sum m_{\nu}$ upper bound -- a measurement that would   otherwise be much more challenging from particle physics experiments.

Another interesting aspect related to the \textit{dark sector},
which we examine in this work,
is the possibility of 
extra light relic particles in the universe that may shed light on the physics beyond the standard model,
since recently particle detector experiments have hinted at the
possible existence of additional species of neutrinos (see Gelmini et al. 2019, and references therein).  
Generically, all of these nonstandard models are termed {\it dark radiation}, expressing deviations $\Delta N_{\rm eff}$ from the canonical value of $N_{\rm eff}=3.046$ --
which point to `hidden' neutrino interactions mediated by a scalar or a vector. 
Departures of $N_{\rm eff}$ from $3.046$ would imply 
nonstandard neutrino features or the contribution of other
relativistic relics. 
Our goal is also to
carry out, for the first time,
a full hydrodynamical treatment for dark radiation cosmologies -- as we explain in detail in Sections \ref{sec_sejong_suite_general} and \ref{sec_sejong_suite_components}. 

Moreover, DM (or at least part of it) may be warm
rather than cold: in this respect,  hypothetical sterile neutrinos could account for the baryon asymmetry in the universe and neutrino flavor mixing, and at 
the same time they may constitute convenient candidates for warm dark matter (WDM). In the present work, 
we also explore this possibility by considering early decoupled thermal relics, and assuming that all the DM is warm when massive neutrinos are not present 
(i.e., all the CDM is turned into WDM, known as `pure' $\Lambda$WDM models). Plausible
candidates are keV right-handed neutrinos or sterile neutrinos. 
In addition, we also reproduce mixed scenarios where massive neutrinos and WDM, or dark radiation and WDM, are both present --
another novelty in the literature. Even in this case, the Ly$\alpha$ forest is an excellent probe for detecting dark radiation and WDM imprints,
thanks to significant attenuation effects on the matter and flux power spectra at small scales, jointly with
a characteristic nonlinear and $z$-dependent suppression of power
(Rossi 2017).
Competitive constraints on $N_{\rm eff}$ and on the WDM mass from Ly$\alpha$
have already appeared in numerous works (e.g., Rossi et al. 2015; Ir{\v s}i{\v c} et al. 2017b; Archidiacono et al. 2019). 

Modeling nonlinear evolution and the complex effects of baryons at small scales as required by the Ly$\alpha$ forest, as well as
the presence of additional components such as massive neutrinos, dark radiation, and WDM, 
is only possible via sophisticated high-resolution cosmological hydrodynamical simulations. 
Traditionally, such simulations are developed following two basic flavors: smoothed particle hydrodynamics (SPH; Gingold \& Monaghan 1977; Lucy 1977), 
and grid-based methods; there are also more sophisticated combinations of the two categories. The SPH technique, which 
uses particles to represent the various components, is the choice adopted in this study -- although 
both approaches have been successfully used in the literature to model the Ly$\alpha$ forest. 
Moreover, realistic numerical simulations able to 
reproduce and control the various observational surveys
are indispensable for interpreting high-quality data as those expected from upcoming cosmological experiments, and
for controlling systematics that can spoil parameter constraints: 
a complete hydrodynamical treatment is mandatory to reach the precision that data are now beginning to show.

A vast number of $N$-body gravity-only simulations is available in the literature. Among the plethora of realizations, we recall, e.g., the {\it Hubble Volume}
(Evrard et al. 2002), {\it Millennium} (Springel et al. 2005),  {\it MultiDark} and  {\it Bolshoi} (Riebe et al. 2013; Klypin et al. 2016), 
{\it Dark Sky} (Skillman et al. 2014), {\it UNIT} (Chuang et al. 2019), and {\it HACC} (Heitmann et al. 2019a, 2019b) simulations.
In comparison, less has been already developed in terms of the hydrodynamical counterpart -- but see, e.g.,  
the remarkable {\it Illustris} and {\it IllustrisTNG} (Vogelsberger et al. 2014; Nelson et al. 2019),  {\it Horizon-AGN} (Dubois et al. 2014),  
{\it EAGLE} (Schaye et al. 2015),  {\it Magneticum} (Dolag 2015), {\it MassiveBlack-II} (Khandai et al. 2015), 
{\it Sherwood} (Bolton et al. 2017), and  {\it Borg Cube} (Emberson et al. 2019) simulations; it is even more so 
in relation to the {\it dark sector}. This is mainly due to the high computational demand and considerable costs in producing a large number of hydrodynamical runs,
and because of the complexity in resolving baryonic physics at small scales and the challenges 
associated with the modeling of exotic species -- such as neutrinos and dark radiation. 

Within this context, and motivated by all of these reasons, we have undertaken
a new long-term challenging effort, which aims at pushing further and upgrading 
the significant work carried out in Rossi et al. (2014, 2015) and in Rossi (2017).
To this end, we have produced an extensive number of state-of-the-art and more refined  
cosmological hydrodynamical simulations (over 300 runs) termed the {\it Sejong Suite},  
developed primarily for Ly$\alpha$ forest studies.
The realizations span  
a variety of cosmological and astrophysical parameters including those that represent 
the {\it dark sector}, and 
are particularly well suited for the high-$z$ cosmic web as seen in the Ly$\alpha$ forest, in the redshift interval $5.0 \le  z \le 2.0$. 
The entire suite has been organized into three main categories, addressing different scientific and technical aspects: 
the {\it Grid Suite}, useful for cosmological parameter constraints especially regarding massive and sterile neutrinos; 
the {\it Supporting Suite}, aimed at characterizing the impact of exotic 
particles on the high-$z$ cosmic web; and the {\it Systematics Suite}, 
meant to quantify various systematic effects.  

The {\it Sejong Suite} features a 
number of improvements and novelties with respect to our previously released  simulations, 
related to {\it technical}, {\it modeling}, and {\it innovative} aspects, which we discuss throughout the paper and 
summarize in the final part. Noticeably, 
we have expanded the parameter space for the {\it Grid Suite} and tighten their variation range;
we have included, for the first time, extended mixed scenarios describing the combined effects of WDM, dark radiation, and neutrinos.   
We also address a series of nontrivial systematics, and we have produced more than 288 million Ly$\alpha$ forest skewers mapping an
extended parameter space. 
In particular, reaching a very high sensitivity on small scales and resolving baryonic physics are essential aspects for improving neutrino, $N_{\rm eff}$, and {\it dark sector}
constraints, and for breaking degeneracies -- as  demanded by future high-quality surveys. 

This work is primarily intended as a presentation, technical description, and guide to the usage of the various simulations and of the related post-processing products.
It also represents the foundation for future extensions in the modeling the Ly$\alpha$ forest along novel paths of improvement.
Moreover, we also undertake a first analysis of the {\it Sejong Suite}, mainly focused on the matter and flux statistics. 

The layout of the paper is organized as follows. 
Section \ref{sec_sejong_suite_general} provides a general description and philosophy of the entire suite: in particular,
we present the overall suite structure, several technical specifics about the production of hydrodynamical simulations, the post-processing procedure, 
and some basic aspects regarding the numerical implementation of massive neutrinos, dark radiation, and WDM
-- within the SPH framework.
Section \ref{sec_sejong_suite_components} illustrates in more detail the individual categories comprising the {\it Sejong Suite}, 
namely  the {\it  Grid Suite}, the {\it Supporting Suite}, and the {\it Systematics Suite};  a comparison to previous studies in terms of resolution and box-size requirements 
is also included.
Section \ref{sec_sejong_suite_first_results} describes selected results from the first analysis of the simulations, 
focused on the matter and flux statistics. In particular, we 
show that we are able to accurately reproduce the 1D Ly$\alpha$ flux power spectrum down to scales $k=0.06~{\rm [km/s]^{-1}}$
as mapped by recent high-resolution quasar data, as well as the thermal history of the IGM.
Finally, we conclude in Section \ref{sec_conclusions}, where we
summarize the various products available and highlight the improvements and novelties of the entire suite -- along with ongoing applications and future avenues. 
 


\section{The {\it Sejong Suite}: Structure, Philosophy, Technical Details, and Modeling Aspects}  \label{sec_sejong_suite_general}


In this section we introduce the {\it Sejong Suite}. 
We begin by presenting the general structure and organization of the suite along with technicalities regarding codes and simulation specifics
(shared by all of the realizations), as well as the post-processing procedure. We then highlight our implementation of massive neutrinos, 
dark radiation, and WDM within the hydrodynamical framework.  
Finally, we provide additional details on the supercomputers where the simulations were performed, 
and a brief description of the new pipeline developed for producing all of the runs. 


\subsection{General Structure of the \textit{Sejong Suite}}

The {\it Sejong Suite} is an extensive collection of state-of-the-art high-resolution 
cosmological hydrodynamical simulations,  covering a wide range in cosmological
and astrophysical parameters and spanning a variety of cosmologies including scenarios with massive neutrinos, dark radiation, and
WDM. These  new simulations  are akin in philosophy and strategy to those developed in Rossi et al. (2014), but 
they contain several improvements at all levels and represent an upgrade with respect to our previous runs.  
In particular, for the first time, massive neutrinos, dark radiation, and
WDM are consistently {\it jointly} modeled, allowing one to explore 
their combined effects deep in the nonlinear regime. 

The suite is organized into three main categories, addressing different scientific and technical aspects: (1) the {\it Grid Suite} ($76 \times 3 = 228$ simulations), 
targeted primarily for cosmological parameter constraints
especially regarding massive and sterile neutrinos
and the {\it dark sector}, exploiting the small-scale flux power spectrum -- and they represent our leading effort and major deliverable; (2)
the {\it Supporting Suite} (114 simulations), aimed at studying the detailed physical effects of exotic particles and dark radiation models, as well as their 
impact on the high-redshift cosmic web; and (3) the {\it Systematics Suite} (35 realizations), meant to address several systematic effects, 
ranging from numerical challenges until parameter degeneracies. 
Primarily, the simulations are targeted to explore the high-$z$ cosmic web as seen in the 
Ly$\alpha$ forest (we cover in fact the redshift interval $5.0 \le z \le 2.0$), although they can be of much broader usage. 
Building upon this main framework, we plan
to extend the number of realizations in future works. 


\begin{table}
\centering
\caption{Baseline Parameters of the `Sejong Suite'.}
\doublerulesep2.0pt
\renewcommand\arraystretch{1.5}
\begin{tabular}{cc} 
\hline \hline  
Parameter &  Value \\
\hline
{\centering \it Cosmological} & \\
\hline
$n_{\rm s}$  &                  0.968  \\
$\sigma_8 (z=0)$          &                   0.815  \\
$\Omega_{\rm m}$ &                 0.308 \\
$H_0$ [km s$^{-1}$Mpc$^{-1}$]              &                        67.8\\
$N_{\rm eff}$              &        3.046\\
$\sum m_{\nu}$ [eV]              &                        0.0\\
$m_{\rm WDM}$  [keV]            &       0.0\\
\hline
{\centering \it Astrophysical} & \\
\hline
$T_0 (z=3)$ [K] &             15000   \\
$\gamma(z=3)$   &             1.3    \\
$\tau_{\rm A}$   &             0.0025    \\
 $\tau_{\rm S}$   &             3.7   \\
\hline
{\centering \it Fixed} & \\
\hline
$\Omega_{\rm b}$ &               0.048424   \\
$\Omega_{\rm tot}$  &               1.0  \\
$A_{\rm S}$   &             2.139  $\times$ $10^{-9}$  \\  
$z_{\rm re}$   &             8.8   \\  
\hline
{\centering \it Derived} & \\
\hline
$\Omega_{\rm \Lambda}$   &             0.692    \\
$\Omega_{\rm c}$ &               0.25958   \\
$\Omega_{\rm c} h^2$ &               0.119324  \\
$\Omega_{\rm b} h^2$  &               0.02226   \\
\hline
\hline
\label{table_baseline_params_sims}
\end{tabular}
\end{table}


The box sizes considered range from $25h^{-1}{\rm Mpc}$ to  $100h^{-1}{\rm Mpc}$, and the resolution varies
from $208^3$ up to $1024^3$ particles/type. For the {\it grid} runs, adopting a splicing technique (McDonald 2003) we are able to achieve an effective resolution up to  
$3 \times  3328^3 = 110$ billion particles within a $(100h^{-1} {\rm Mpc})^3$ box size,
which improves on our previous realizations and is ideal for reproducing the main aspects of the Ly$\alpha$ forest as mapped by eBOSS and DESI. 
The various simulations were performed with periodic boundary conditions and for the majority of the runs with an equal number of DM, gas, and neutrino particles. 
Taking into account our largest  $100h^{-1}{\rm Mpc}$ box size, the average spacing between sightlines is of the order of $10h^{-1}{\rm kpc}$ -- much smaller than the scale probed by the Ly$\alpha$ forest. 
Output snapshots are created at regular intervals $\Delta z$ in redshift within the range $5.0 \le z \le 2.0$, where $\Delta z = 0.2$. 
For some visualization runs, we also reach $z = 0$ and/or produce additional snapshots at every redshift interval of $\Delta z = 0.1$. 

The fundamental parameters considered for the baseline cosmology of the entire suite (indicated as {\it `Best Guess'} or BG) along with their central values
are listed in Table \ref{table_baseline_params_sims}. They have been organized into two main categories:
cosmological (first block from the top) and astrophysical (second block).  
For the majority of the runs, they are consistent with Planck 2015 results (Planck Collaboration et al. 2016) -- i.e.,
TT+lowP+lensing $68\%$  limits -- and with the SDSS DR12 flat-$\Lambda$CDM cosmology; although, we also have performed 
some realizations using the latest Planck 2018 reported measurements  for
systematic studies (Planck Collaboration et al. 2018b). 
As cosmological parameters, we adopt the spectral
index of the primordial density fluctuations $n_{\rm s}$, the amplitude of
the matter power spectrum $\sigma_8$, the matter density $\Omega_{\rm m}$, the
Hubble constant $H_0$, the number of effective neutrino species $N_{\rm eff}$, 
the total neutrino mass $\sum m_{\nu}$ if nonzero (expressed in eV), 
and the WDM mass $m_{\rm WDM}$ (expressed in keV). 
As astrophysical parameters, we consider the IGM normalization temperature $T_0(z)$ and the logarithmic slope $\gamma (z)$ at $z = 3$ -- although 
practically $T_0$ and $\gamma$ are set by two related quantities that alter the amplitude and density dependence of the 
photoionization heating rates. Moreover, we take into account two additional
parameters that do not enter directly in the simulations but that are used in a subsequent post-processing  phase
to provide the normalization of the flux power spectrum via the effective optical depth, 
namely, $\tau_{\rm A}$ and $\tau_{\rm S}$. The lower blocks of the same table show some relevant fixed or derived parameters, 
such as the baryon, CDM, cosmological constant, and total densities ($\Omega_{\rm b}, \Omega_{\rm c}, \Omega_{\rm \Lambda}, \Omega_{\rm tot}$), 
the initial amplitude of primordial fluctuations $A_{\rm S}$, and the reionization redshift $z_{\rm re}$.
Flatness is always assumed, so that the total density parameter $\Omega_{\rm tot} \equiv \Omega_{\rm m} + \Omega_{\Lambda} =1$.
For the {\it Grid Suite}, these baseline parameters are varied in turn, as we explain in detail in Section \ref{sec_sejong_suite_components}.   
 

\subsection{Technical Details: Main Codes and Simulation Specifications}
 
All of the simulations are produced with a customized version of GAlaxies with Dark matter and Gas intEracT (Gadget-3; Springel et al. 2001; Springel 2005) 
for evolving Euler hydrodynamical equations and primordial chemistry, along with cooling and an externally specified ultraviolet (UV) background, 
supplemented by the code for Anisotropies in the Microwave Background (CAMB; Lewis et al. 2000)
and a modified parallel version of second-order Lagrangian perturbation theory 
(2LPT; Crocce et al. 2006) for setting the initial conditions. 

Gadget-3 is a widely used massively parallel tree-SPH code for collisionless and gas-dynamical 
cosmological simulations originally developed by Volker Springel, written in ANSI C, and exploiting the standardized 
message passing interface (MPI) along with several open-source libraries. The code is efficiently parallelized via a space decomposition achieved 
with a space-filling curve (i.e. the Peano-Hilbert decomposition),
and  reaches a high-level optimization along with being  efficient in memory consumption and  
communication bandwidth -- with a nearly optimal scalability and  work-load balance.  
Gravitational interactions are computed with a hierarchical multipole expansion 
using the standard $N$-body method, and gas-dynamics is followed with 
SPH techniques 
and fully adaptive smoothing lengths, so that energy and entropy are conserved. 

Besides gas and DM, we 
model similarly massive neutrinos, dark radiation, and WDM within the SPH framework (see more details in Section \ref{subsec_nu_dark_rad_wdm_modeling}). 
Short-range forces are treated with the tree method, and long-range forces with Fourier techniques. For our realizations, we set the number of mesh 
cells of the particle-mesh (PM) grid equal to the number of particles. When massive neutrinos are included, 
we neglect the small-scale neutrino clustering, and their short-range gravitational tree force in the TreePM scheme is not computed: hence,
the spatial resolution for the neutrino component is of the order of the grid resolution used for the PM force calculation. 
Generally, in all of the various 
realizations, the gravitational softening is set to 1/30 of the mean interparticle spacing for all of the species considered. 

Using CAMB, we compute transfer functions per each individual component and corresponding power spectra, and for the majority of the runs,
initial conditions are fixed at $z = 33$ with 2LPT. 
We adopt the entropy formulation of SPH proposed by Springel \& Hernquist (2002), with the gas assumed of primordial composition photoionized and heated by a spatially uniform ionizing background,
having an helium mass fraction of $Y = 0.24$ and no metals or evolution of elementary abundances.
Regarding star formation, we utilize the same criterion assumed in Rossi et al. (2014, 2015) and include the most relevant effects of baryonic physics that impact the IGM, 
so that the thermal history in the simulations is consistent with recent temperature measurements (see Section \ref{section_igm_properties}).
This criterion speeds up the calculations considerably, and it has been shown to have negligible effects on the Ly$\alpha$ flux statistics. 
Moreover, we disable feedback options and neglect galactic winds, as suggested by the results of Bolton et al. (2008), who found that winds have a negligible effect on the Ly$\alpha$ forest.


\subsection{Post-processing} \label{subsec_postprocessing}
 
From simulated Gadget-3 snapshots within the redshift range $5.0 \le z \le 2.0$, we 
construct Ly$\alpha$ skewers fully spanning all of the various simulated cosmologies.
This is achieved through an elaborate pipeline, which eventually allows us to obtain averaged Ly$\alpha$  flux power 
spectra and temperature-density relations. 
The primary step in this process requires the extraction of simulated quasar sightlines: 
for the majority of all of our runs and for all the redshift intervals considered, we
extract 100,000 randomly placed sightlines through the simulation box per redshift,
using  2048 bins along the line of sight (LOS) -- unless specified otherwise. 
We then estimate 
the absorption due to each SPH particle near the sightline from the positions, velocities, densities, and temperatures 
of all of the SPH particles at a given redshift (i.e., Theuns et al. 1998, 2002). With this procedure, we obtain 
a number of simulated quasar 
spectra that we subsequently smooth with a 3D cubic spline kernel. 
The photoionization rate is then fixed,  
by requiring the effective optical depth at each redshift to follow the 
empirical power law 
$\tau_{\rm eff}(z) = \tau_{\rm A}(1 + z)^{\tau_{\rm S}}$, with $\tau_{\rm A} = 0.0025$ and $\tau_{\rm S} = 3.7$, as done
in Rossi et al. (2014, 2015) and in Rossi (2017).
This procedure is routinely adopted in a series of other studies, as the instantaneous IGM temperature only depends on the spectral shape of the UV background and the gas density.  
Finally, all of the spectra are rescaled by a constant so that the mean flux across all spectra and absorption 
bins matches the observed mean flux at redshift $z$. 
After performing the normalization procedure, the mean over
all of the rescaled spectra is used as the extracted flux power spectrum for a given box. 

In addition to fluxes, we also extract particle samples for studying the IGM temperature-density relation ($T_0-\gamma$) in the presence of massive neutrinos, dark radiation, and WDM.
This is performed for every simulation and for each considered redshift interval, and we generally extract 100,000 particles per type per redshift. 


\subsection{Implementation of Massive Neutrinos, Dark Radiation, and Warm Dark Matter} \label{subsec_nu_dark_rad_wdm_modeling}

 
\begin{figure*}
\begin{center}
\includegraphics[angle=0,width=0.76\textwidth]{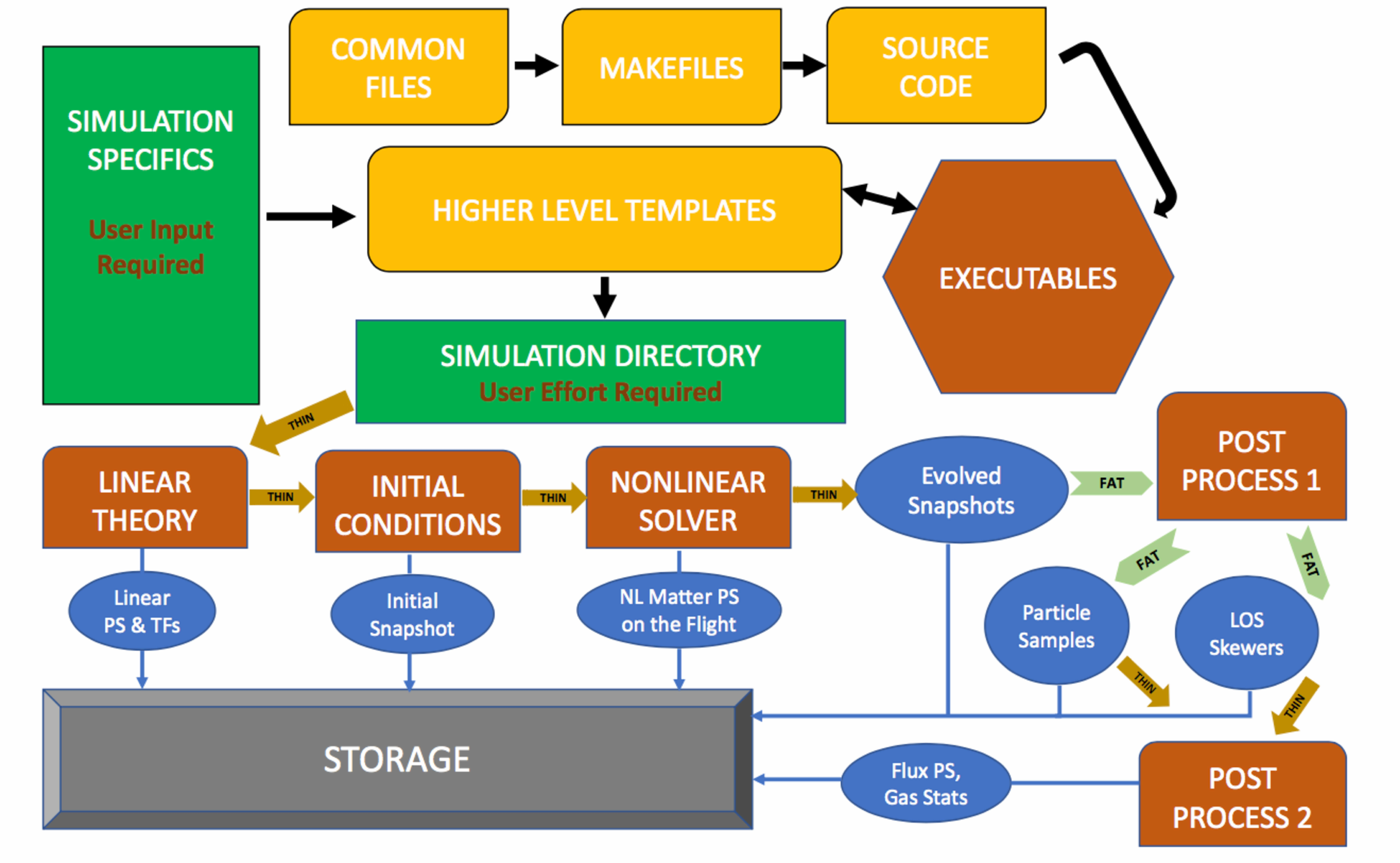}
\caption{Workflow diagram of the new pipeline developed for producing the ``Sejong Suite". 
A single integrated script, requiring minimal effort from the user, allows one to perform the complete simulation pipeline (including post-processing),
in a fully automated way.  In particular, the various products are stored at different stages in the simulation-making process, as visualized by the blue circles
in the figure. See the main text for more details.}
\label{fig_pipeline_workflow_details}
\end{center}
\end{figure*}  


As in our previous releases, we adopt a particle-based implementation of massive neutrinos
following the SPH formalism: in essence, neutrinos are treated as an additional particle component, on top of gas and DM. 
Our methodology is now common to other recent related studies (e.g., Viel et al. 2010; Castorina et al. 2014, 2015; 
Villaescusa et al. 2014, 2018, 2019; Carbone et al. 2016; Ruggeri et al. 2018), and 
represents a neat way of incorporating  massive neutrinos within the SPH framework. In fact,
in the Ly$\alpha$ regime and for the range of neutrino masses considered in our simulations,  the neutrino free-streaming condition 
 is always satisfied (Lesgourgues \& Pastor 2006; Rossi 2017) so that
neutrinos can simply be treated as collisionless particles similarly to DM, with a 
phase space distribution isotropic  following a Fermi-Dirac distribution and a nonzero anisotropic stress. 
Despite more demanding computational requirements due to an extra particle component,
this implementation is ideal for our main goals
as we need to resolve small-scale physics deep in the nonlinear regime, and in particular reproduce with high accuracy 
all of the main aspects of the Ly$\alpha$ forest -- at the level of eBOSS and DESI, in a $k$-range where nonlinear evolution of cosmological neutrinos cannot be neglected --as 
well as  the scale dependence of the total matter power spectrum due to nonlinear evolution.
We do not rely on any linear approximation  nor interchange between grid- and particle-based methods, since nonlinear 
evolution at small scales is not properly reproduced by the grid method.
With this implementation, we are able to precisely 
quantify the response of the power spectrum to isolated variations in individual parameters and varying neutrino masses, 
and also disentangle the  $\sum m_{\nu}-\sigma_8$
degeneracy, as well as the effects of baryons at small scales. This latter aspect is particularly relevant 
for the usage of the 1D and 3D flux power spectra  for cosmological studies.  
We have already proven this implementation to be robust, and capable of providing competitive constraints on massive neutrinos and dark 
radiation (i.e., Rossi et al. 2015). 
Recent alternative implementations can be found in 
Ali-Haimoud \& Bird (2013), Banerjee \& Dalal (2016), Upadhye et al. 
(2016), Emberson et al. (2017), and Mummery et al. (2017). 
The variety of techniques adopted in the literature reflects the fact that 
neutrinos can be treated either as a fluid or as 
an ensemble of particles, and one may describe their evolution with linear theory or perform a full nonlinear treatment. 
While all of these complementary methods may be faster in terms of central processing unit (CPU) hours and production runs
than ours, we need  to accurately reproduce the nonlinear evolution at small scales
in order to characterize the Ly$\alpha$ forest with high-fidelity -- hence our implementation choice. 
 
Regarding dark radiation  (i.e., any deviation from the canonical value of $N_{\rm eff} =3.046$ --which 
remains still  poorly studied in the literature), in this work  we
consider models with four neutrinos, where three are massive and active while the fourth one is 
massless, sterile, and thermalized with the active ones.
Unlike in Rossi et al. (2015) where we adopted an analytic remapping technique to simulate such models,
here, for the first time, we consistently carry out a full hydrodynamical treatment -- combining  
three massive neutrinos of degenerate mass with a massless sterile neutrino so that $N_{\rm eff}=4.046$.  
This represents a novelty in the literature, and
such new simulations were already used in Rossi (2017) to study
the combined effect of baryons, DM, neutrinos, and dark radiation at high-$z$ and small scales, to
quantify the impact of extra dark components on Ly$\alpha$ forest observables as a function of scale and redshift. 
In particular, in Rossi (2017), we have shown that 
departures of $N_{\rm eff}$ from the canonical value of $3.046$, due to nonstandard neutrino features or to the contribution of other relativistic relics, 
have a non-negligible repercussion for the subsequent cosmological structure formation: the
CMB power spectrum shows a suppressed first peak and enhancements in the the others, 
making the situation clearly 
distinguishable from the case of three massive neutrinos only. 
 
As far as WDM, we focus here on two implementations,
following  different methodologies. 
In both cases, we only consider early decoupled thermal relics, and assume that  
all the  DM is warm when massive neutrinos are not present (i.e., all the CDM is turned into WDM, also known as `pure' $\Lambda$WDM models):
suitable candidates are keV right-handed neutrinos or sterile neutrinos. 
Thermal relics are the most basic model of WDM particles, analogous to
neutrinos but with a larger particle mass (indicated as $m_{\rm WDM}$ in our notation), 
and characterized by WDM with a simple thermal history -- so that it is possible to
calculate their free streaming. 
Noticeably, for early decoupled thermal relics, there is a direct correspondence between 
their mass $m_{\rm WDM}$ (always expressed in keV),
and the mass  $m_{\rm s}$ of a non-thermalized nonresonant massive sterile neutrino
in the standard Dodelson-Widrow mechanism  (DW; Dodelson \& Widrow 1994):
hence, our WDM runs are useful for studying and eventually constraining
both pure WDM candidates and/or  nonresonant massive sterile neutrinos.
Specifically, it is possible to introduce a re-parameterization via a non-thermalized massive sterile neutrino as follows:

\begin{equation}
m_{\rm s} = 4.43 \Big ( {m_{\rm WDM} \over {\rm keV}} \Big )^{4/3} \Big ( {\omega_{\rm WDM} \over 0.1225} \Big )^{-1/3}~{\rm keV}
\end{equation}
\noindent where $\omega_{\rm WDM} = \Omega_{\rm WDM} h^2$, and the relation holds for the nonresonant DW production 
mechanism. And, in terms of $\Delta N_{\rm eff}$, one can write:
\begin{equation}
m^{\rm eff} = \chi m_{\rm s} \equiv \Delta N_{\rm eff} m_{\rm s},  
\label{eq_sterile_cea_2} 
\end{equation}
where $m_{\rm s} = m_{\rm WDM} / (\Delta N_{\rm eff})^{1/4}$.

In the context of the previous formalism, in our first prescription, WDM
is implemented by modifying the linear matter power spectrum to
mimic the presence of thermal relics. 
This is achieved by introducing a small-scale cut-off at the level of 2LPT (when we generate our initial conditions)
to the BG CAMB linear power spectrum $P_{\rm \Lambda CDM}$. Namely:

\begin{equation}
T^2_{\rm lin} (k) = {P_{\rm WDM} (k) \over P_{\rm \Lambda CDM} (k) } = \Big [1 + (\alpha k)^{2 \nu}    \Big ]^{-10/\nu} 
\label{eq_viel_1}
\end{equation}

\noindent where $\Omega_{\rm WDM} = \Omega_{\rm m} - \Omega_{\rm b} \equiv \Omega_{\rm c}$  and

\begin{equation}
\alpha \equiv \alpha (m_{\rm WDM})   = a_1 \Big(  {1~ {\rm keV} \over m_{\rm WDM}} \Big )^{a_2}  \Big (  {\Omega_{\rm WDM} \over 0.25 } \Big )^{a_3}   \Big( {h \over 0.7}  \Big )^{a_4}   
\label{eq_viel_2}
\end{equation}

\noindent with $\alpha$ in units of $h^{-1}{\rm Mpc}$, and ($\nu, a_1, a_2, a_3, a_4$) = ($1.12, 0.049, 1.11, 0.11, 1.22$) following Viel et al. (2005, 2012, 2013) or 
in a different prescription assuming ($\nu, a_1, a_2, a_3, a_4$) = ($1.2, 0.048, 1.15, 0.15, 1.3$) -- the latter according to Bode, Ostriker \& Turok (2001). 
This is a good approximation at $k< 5\div 10 ~h {\rm Mpc}^{-1}$, and below one needs a more complex function plus the addition of acoustic oscillations.
For $m_{\rm WDM} \sim 1$ keV, the characteristic cutoff in the power spectrum is at $ k \sim 1.5 ~h {\rm Mpc}^{-1}$.
Initial velocities for WDM are drawn from a Fermi-Dirac distribution and added to the proper velocity assigned by linear theory.

In our second prescription, WDM 
is implemented starting directly from a properly designed CAMB power spectrum,
by setting $\Delta N_{\rm eff}$ to express departures from the
canonical $N_{\rm eff}$ value according to the relic mass -- where $N_{\rm eff} = 3.046 + \Delta N_{\rm eff}$.
These runs are  simply indicated as WDM in Table \ref{table_supporting_sims_wdm}, and we  
adopt this implementation for our {\it Grid Suite}, since
it is more physically motivated and it is also straightforward (although delicate) to include massive neutrinos within this framework.
The inclusion is simply achieved by accounting for
the neutrino mass degeneracy,  
where: 
\begin{equation}
\Delta N_{\rm eff} = \Big ( {93.14 \cdot \Omega_{\rm WDM} h^2 \over m_{\rm WDM} } \Big )^{4/3}.
\end{equation}  
We eventually use the one-to-one mapping between thermal relics and 
keV sterile neutrinos to simulate massive sterile neutrino cosmologies in the DW mechanism -- as previously explained.

Within the second prescription, we also 
simulate for the first time cosmologies with both massive active and sterile neutrinos, in a consistent fashion.
This is achieved by introducing two neutrino eigenstates (one for the active ones, and one for the sterile counterpart) in a three-component simulation, and then
by considering a neutrino mass splitting -- so that $\Omega_{\nu} = \Omega_{\rm m} - \Omega_{\rm b}$ now contains two distinct contributions. In essence, 
three active massive-with-degenerate-mass neutrinos contribute as $N_{\rm eff} =3.046$, while a non-thermalized massive sterile neutrino
contributes with an additional $\Delta N_{\rm eff}$ to the number of effective neutrino species, directly proportional to its mass.  
Such runs represent a unique novelty in the literature, and allow us to study the combined effects of massive and sterile neutrinos, as well as to
put constraints on pure WDM scenarios directly from Ly$\alpha$ forest data. 
The latter simulations are indicated as $\sum m_{\nu}^{+} m_{\rm WDM}^{+}$  in Table \ref{table_grid_sims_cross_terms} ({\it Grid} runs), 
and as {\it NU\_WDM} in Table \ref{table_supporting_sims_wdm}. 
Finally, note that we always include massless neutrinos in WDM-only simulations. 


\subsection{Additional Technical Details: Machines and Automated Pipeline}
 
All of the simulations presented here have been produced at the Korea Institute of Science and Technology Information (KISTI) via the Tachyon 2 supercomputer, 
under allocations KSC-2016-G2-0004, KSC-2017-G2-0008, and KSC-2018-G3-0008 (for a total of 11.25 million CPU hours), which 
allowed us to use up to 2176 dedicated cores for 9 months on 
an exclusive queue, and also up to 8192 cores on a regular queue -- exploiting the MPI parallel architecture. 
Tachyon 2 is an SUN Blade 6275 machine with Intel Xeon x5570 Nehalem 2.93 GHz CPUs on an Infiniband 4x 
QDR network, having a total of 25,408 cores grouped into eight-core nodes, with 24 GB of memory per node and 300 TFlops of peak performance.
Our standard runs required all 2176 cores, with an average clock time of about 3--4 days per
simulation (excluding post-processing). Our highest-resolution runs used 8192 cores. 
The post-processing of the simulations (i.e., extraction of LOS skewers and particle samples) demanded dedicated fat nodes on different architectures, and it was 
performed  with the KISTI KAT System under allocations KSC-2018-T1-0017, KSC-2018-T1-0033, and KSC-2018-T1-0061. 
Specifically, we used  the KISTI/TESLA `Skylake' architecture, an Intel Xeon Skylake (Gold 6140) at 2.30GHz (18-core and two socket), with 192GB DDR4 memory,
and the KISTI/TESLA `Bigmem' architecture, an Intel Xeon Westmere (E7-4870) at 2.40GHz (10-core, four socket) with  512GB DDR3 memory. 
Finally, some further post-processing was carried out with our new cluster system at Sejong University, composed of a 
Xeon Silver 4114 master node  
and a Xeon Gold 6126 computing node architecture.  

For performing all our novel simulations and post-processing, we devised a new portable and efficient pipeline, able to produce end-to-end simulations 
with a single integrated script, in order to avoid human error in the process. Figure \ref{fig_pipeline_workflow_details} provides a self-explanatory workflow diagram of the pipeline:
in particular, blue circles indicate the various products stored at different stages, with the most computationally demanding
post-processing step requiring fat node architecture (green arrows in the diagram).
All of the details of the simulation are preselected by the user (including directory settings, choices of initial conditions, as well as simulation,  
astrophysical,   cosmological,   neutrino, WDM, and 
post-processing parameters), and all of the specifics of a simulation are organized by categories with 
a unique identifier associated per simulation. The elegance of this pipeline relies in its portability and generality (not tied to a specific machine),
and on the overall simplicity in performing in a fully automated way a rather complex sequence of tasks -- which represents a considerable improvement
from the pipeline developed in Rossi et al. (2014). In this way, we are able to run a generic simulation (including post-processing) 
with just a few preparatory steps and minimal effort from the user, who is only required to set initially the specifics of the simulation to run, and subsequently simply 
launch the desired production run.   
This improvement was necessary, given the large amount of simulations and heavy post-processing involved here 
(exceeding 200 TB of data). Our pipeline is also straightforward to modify or extend for future use. 



\section{The {\it Sejong Suite}: Components and Products}  \label{sec_sejong_suite_components}


In this section we describe 
the three main categories of simulations  that 
constitute the \textit{Sejong Suite} -- denoted as the {\it Grid}, {\it Supporting}, and {\it Systematics Suites}, respectively.
Moreover, we present the entire list of simulations performed in Tables \ref{table_grid_sims_base}-\ref{table_systematics_sims_all},
provide some visualizations from selected snapshots, and also include a comparison 
to previous studies in terms of resolution and box-size requirements.
We conclude this part with a schematic list of all of the products available. 


\subsection{The \textit{Grid Suite}} \label{subsec_grid_suite}
 

The  hydrodynamical {\it Grid Suite} has been devised with the intent of providing
a high-resolution mapping -- including all of the
effects of baryonic physics at small scales -- of a wide landscape defined by a
variety of cosmological and astrophysical parameters, centered on what we indicate as BG or `reference' cosmology.
Here the BG model is set by Planck 2015 results, i.e.,
TT+lowP+lensing $68\%$  limits (see Table \ref{table_baseline_params_sims} for the central parameters adopted). 
The primary motivation in developing such a grid of simulations is the construction of a 
Taylor expansion model for the flux power spectrum,  a key theoretical quantity useful for building the Ly$\alpha$ forest 
likelihood (see, e.g., Rossi et al. 2015;  Rossi 2017). With the Ly$\alpha$ forest likelihood in hands, it is possible to
provide competitive constraints on cosmological parameters, neutrino masses, WDM, and more generally on the {\it dark sector}, 
exploiting the synergy between the Ly$\alpha$ forest as a unique high-$z$ probe, lower-$z$ tracers, and the CMB. 


\begin{figure}
\centering
\includegraphics[angle=0,width=0.43\textwidth]{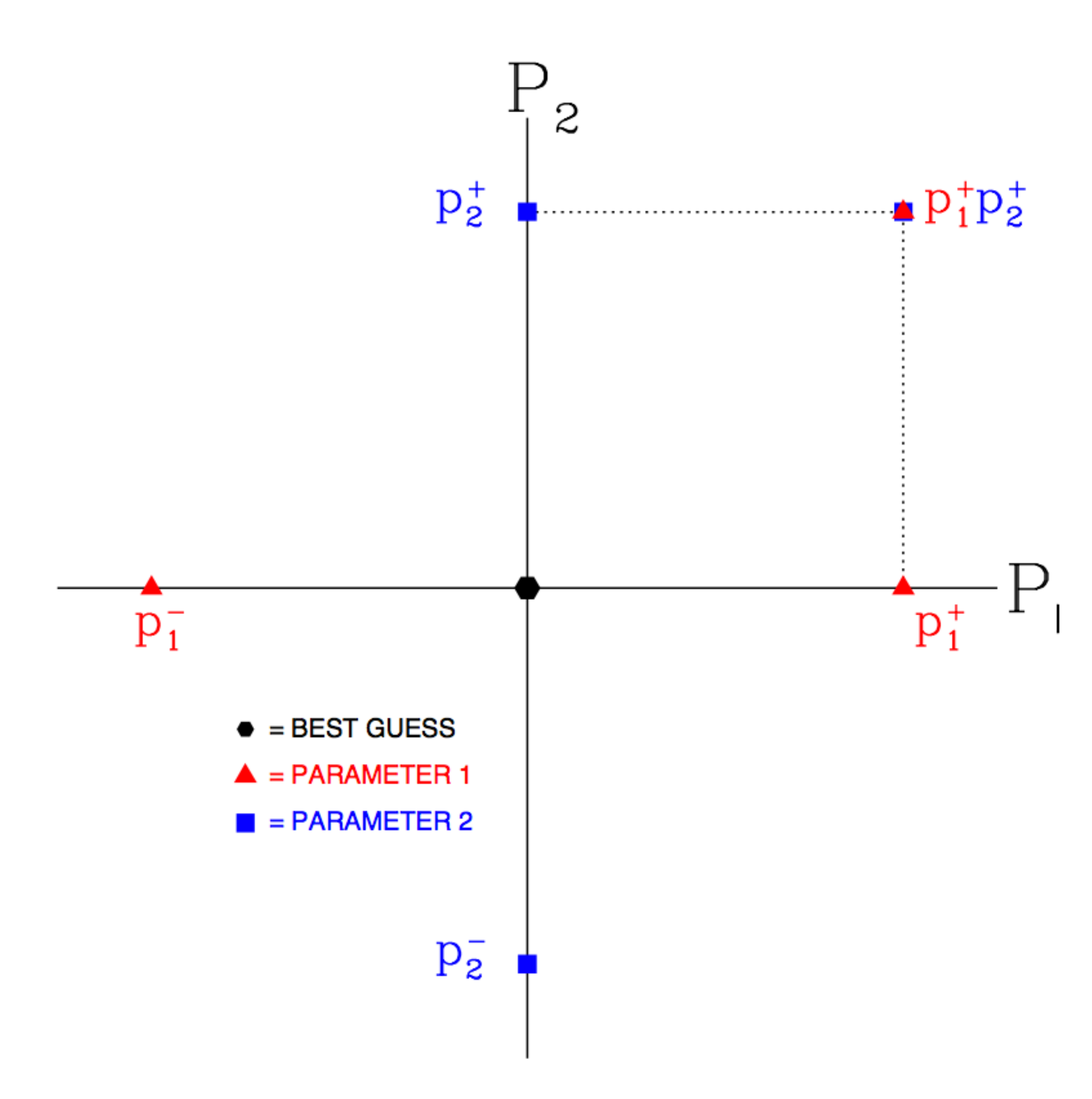}
\caption{Typical structure of the {\it Grid Suite}, a subset of the {\it Sejong Suite},
for a generic combination ($p_1, p_2$) of cosmological and/or astrophysical parameters.}
\label{fig_grid_structure}
\end{figure}


\begin{table}
\centering
\caption{Allowed variations of the key grid parameters.}
\doublerulesep2.0pt
\renewcommand\arraystretch{1.5}
\begin{tabular}{ccc} 
\hline \hline  
Parameter &  Central Value & Variation \\
\hline
$n_{\rm s}$ &                  0.968  & $\pm$ 0.018 \\
$\sigma_8 (z=0)$          &                   0.815  & $\pm$ 0.027 \\
$\Omega_{\rm m}$ &                 0.308 & $\pm$ 0.036\\
$H_0$ [km s$^{-1}$Mpc$^{-1}$]              &                        67.8 & $\pm$ 2.7 \\
$N_{\rm eff}$             &        3.046 & $\pm$ 1.000 \\
$\sum m_{\nu}$ [eV]              &                        0.0  &       +0.3  \\
$\sum m_{\rm WDM}$ [keV]              &                        0.0  &       +3.0  \\
\hline
$T_0 (z=3)$ [K]  &             15000  & $\pm$ 5000  \\
$\gamma(z=3)$   &             1.3    & $\pm$ 0.3 \\
\hline
$\tau_{\rm A}$   &             0.0025    & $\pm$ 0.0015 \\
$\tau_{\rm S}$   &             3.7   & $\pm$ 0.4 \\
\hline
\hline
\label{table_grid_params_variations}
\end{tabular}
\end{table}


\begin{table*}
\begin{center}
\tiny
\centering
\caption{Grid Simulations (1): Individual Parameter Variations.}
\doublerulesep 2.0pt
\renewcommand\arraystretch{1.5}
\begin{tabular}{ccccccccc} 
\hline \hline
 Simulation Name &   $M_{\rm \nu}$ [eV]  &   $m_{\rm WDM}$ [keV] & $N_{\rm eff}$   & $\sigma_8(z=0)$  & Boxes [Mpc/h] & $N_{\rm p}^{1/3}$ & Mean Part. Sep.  [Mpc/h] & Softening [kpc/h]  \\ 
\hline
BEST GUESS  a/b/c &  0.0          &  0.0 & 3.0460  &    0.815             &     25/25/100  & 208/832/832   & 0.12019/0.03005/0.12019 & 4.01/1.00/4.01\\
$n_{\rm s}^+$                         a/b/c &  0.0  &   0.0       &  3.0460  &    0.815             &     25/25/100  & 208/832/832   & 0.12019/0.03005/0.12019 & 4.01/1.00/4.01\\
$n_{\rm s}^-$                            a/b/c &  0.0   &   0.0      &  3.0460  &    0.815             &     25/25/100  & 208/832/832   & 0.12019/0.03005/0.12019 & 4.01/1.00/4.01\\
$\sigma_8^+$                            a/b/c &  0.0    &   0.0    &  3.0460  &    0.842             &     25/25/100  & 208/832/832   & 0.12019/0.03005/0.12019 & 4.01/1.00/4.01\\
$\sigma_8^-$                         a/b/c &  0.0        & 0.0  &  3.0460  &    0.788            &     25/25/100  & 208/832/832   & 0.12019/0.03005/0.12019 & 4.01/1.00/4.01\\
$\Omega_{\rm m}^+$                         a/b/c &  0.0 &  0.0        &  3.0460  &    0.815             &     25/25/100  & 208/832/832   & 0.12019/0.03005/0.12019 & 4.01/1.00/4.01\\
$\Omega_{\rm m}^-$   a/b/c &  0.0    &   0.0    &  3.0460  &    0.815             &     25/25/100  & 208/832/832   & 0.12019/0.03005/0.12019 & 4.01/1.00/4.01\\
$H_{\rm 0}^+$  a/b/c &  0.0    &  0.0     &  3.0460  &    0.815             &     25/25/100  & 208/832/832   & 0.12019/0.03005/0.12019 & 4.01/1.00/4.01\\
$H_{\rm 0}^-$  a/b/c &  0.0    &   0.0    &  3.0460  &    0.815             &     25/25/100  & 208/832/832   & 0.12019/0.03005/0.12019 & 4.01/1.00/4.01\\
$\sum m_{\nu}^+$  a/b/c &  0.3 &  0.0         &  3.0460  &    0.815             &     25/25/100  & 208/832/832   & 0.12019/0.03005/0.12019 & 4.01/1.00/4.01\\
$N_{\rm eff}^+$  a/b/c &  0.0      &  0.0   &  4.0460  &    0.815             &     25/25/100  & 208/832/832   & 0.12019/0.03005/0.12019 & 4.01/1.00/4.01\\
$N_{\rm eff}^-$  a/b/c &  0.0 & 0.0         &  2.0460  &    0.815             &     25/25/100  & 208/832/832   & 0.12019/0.03005/0.12019 & 4.01/1.00/4.01\\
$m_{\rm WDM}^+$  a/b/c &  0.0 &  3.0         &  3.0466  &    0.815             &     25/25/100  & 208/832/832   & 0.12019/0.03005/0.12019 & 4.01/1.00/4.01\\
$T_0^+$ a/b/c &  0.0     &    0.0  &  3.0460  &    0.815             &     25/25/100  & 208/832/832   & 0.12019/0.03005/0.12019 & 4.01/1.00/4.01\\
$T_0^-$ a/b/c &  0.0      &   0.0  &  3.0460  &    0.815             &     25/25/100  & 208/832/832   & 0.12019/0.03005/0.12019 & 4.01/1.00/4.01\\
$\gamma^+$ a/b/c &  0.0  &  0.0       &  3.0460  &    0.815             &     25/25/100  & 208/832/832   & 0.12019/0.03005/0.12019 & 4.01/1.00/4.01\\                                                 
$\gamma^-$ a/b/c &  0.0    &    0.0   &  3.0460  &    0.815             &     25/25/100  & 208/832/832   & 0.12019/0.03005/0.12019 & 4.01/1.00/4.01\\  
$\tau_{\rm A}^+$ a/b/c &  0.0     &  0.0   &  3.0460  &    0.815             &     25/25/100  & 208/832/832   & 0.12019/0.03005/0.12019 & 4.01/1.00/4.01\\
$\tau_{\rm A}^-$ a/b/c &  0.0     &  0.0   &  3.0460  &    0.815             &     25/25/100  & 208/832/832   & 0.12019/0.03005/0.12019 & 4.01/1.00/4.01\\
$\tau_{\rm S}^+$ a/b/c &  0.0     &  0.0   &  3.0460  &    0.815             &     25/25/100  & 208/832/832   & 0.12019/0.03005/0.12019 & 4.01/1.00/4.01\\
$\tau_{\rm S}^-$ a/b/c &  0.0     &  0.0   &  3.0460  &    0.815             &     25/25/100  & 208/832/832   & 0.12019/0.03005/0.12019 & 4.01/1.00/4.01\\                          
\hline         
\hline
\label{table_grid_sims_base}
\end{tabular}
\end{center}
\end{table*}


As reported in Table \ref{table_baseline_params_sims}, our grid is constructed around seven cosmological parameters 
($n_{\rm s}$, $\sigma_8$, $\Omega_{\rm m}$, $H_0$, $N_{\rm eff}$, $\sum m_{\nu}$, and $m_{\rm WDM}$) 
plus four astrophysical parameters ($T_0[z=3]$, $\gamma[z=3]$, $\tau_{\rm A}$, and $\tau_{\rm S}$).
The baseline BG model  does not contain any massive neutrinos, WDM, or dark radiation 
(i.e., $\Delta N_{\rm eff} =0$), but only a massless neutrino component in the form of three massless active neutrinos  
compatible with the standard model of particle physics, such that $N_{\rm eff} = 3.046$. 

The overall structure of the grid is exemplified in Figure \ref{fig_grid_structure}.   
Moving from the BG fiducial simulation 
and considering in turn distinct pairs of parameters  ($p_1, p_2$) selected from those of the grid, 
we produce two runs per parameter altering their reference central values via symmetric positive and negative shifts, indicated as $p_1^{+}, p_2^{+}, p_1^{-}, p_2^{-}$ in the figure -- and marked with red
triangles or blue squares, respectively. These realizations are useful for evaluating second-order derivatives in the Taylor expansion of the flux. 
We then carry out an additional joint run where both $p_1$ and $p_2$ are (simultaneously) positively altered, namely $p_1^+ p_2^+$  (upper right corner in the figure), in order to
evaluate their cross-derivative terms, which also enter in the Taylor expansion of the flux. 
This procedure is repeated for all the possible sets of parameter pairs describing the grid; hence, 
using $n$ to indicate the number of parameters, the comprehensive simulations required to compute the Taylor expansion coefficients are [$1+2n+n(n-1)/2]$.
Note also that all of the derivatives are approximated to second order, except for the 
cross-terms, which are of the first order; else, we would need $[n(n - 1)/2]$ supplementary simulations. 
Essentially, the structure of the grid is determined by positive and negative variations of the key parameters around their reference values, plus all the possible 
combinations of pairs having both positive shifts.
This means that a mapping defined by 11 parameters requires $76$ simulations to cover the entire space -- excluding two unphysical runs characterized by negative mass variations (i.e., $\sum m_{\nu}^{-}$ and $m^{-}_{\rm WDM}$).
In practice, the astrophysical  parameters $\tau_{\rm A}$ and $\tau_{\rm S}$ are only varied during post-processing, 
thus reducing the effective total number of required simulations. 

The chosen variations of the cosmological and astrophysical grid parameters are reported in Table \ref{table_grid_params_variations}. 
These fluctuations are tighter than those allowed in our previous simulation grid, and more stringent in terms of cosmology -- in order to avoid  
wider excursions in their values, which may lead to  interpolation errors and eventually impact parameter constraints. 
Specifically, we consider changes of $\pm 0.018$ in $n_{\rm s}$,  variations of $\pm 0.027$ in $\sigma_8$,
variations of $\pm 0.036$ in $\Omega_{\rm m}$, while we alter the Hubble constant $H_0$ by $\pm 2.7$.
The number of effective neutrino species $N_{\rm eff}$ is varied by $\pm 1$ with respect to its canonical value of $3.046$.
For massive neutrinos, we assume a central total upper mass limit of $\sum m_{\nu} =0.3~{\rm eV}$, while for WDM
interpreted as massive sterile neutrinos, we take $m_{\rm WDM} =3.0~{\rm keV}$ as the reference value. 
Regarding  astrophysical parameters, 
we alter both $T_0$ and $\gamma$, the former by $\pm 5000$ K and the latter by $\pm 0.3$. 
The parameters $\tau_{\rm A}$ and $\tau_{\rm S}$ are instead changed only in the post-processing phase by $\pm 0.0015$ and $\pm 0.4$, respectively.
Regarding interpolation between models, in earlier works (see, e.g., Rossi et al. 2014, 2015; Palanque-Delabrouille et al. 2015)
we have investigated and assessed the 
precision of the classical Taylor expansion of the flux (i.e., Viel et al. 2010; Wang et al. 2013) 
resulting from an analogous simulation grid. 
In particular, by analyzing simulations for several sets of parameters with values different from those used in the grid 
and comparing the power spectrum derived to the one predicted by the Taylor expansion, we
found good agreement -- as long as the tested parameters remained within the range of values used to compute the expansion. 
For these novel simulations, the variations of the grid parameters are less extreme compared to our previous runs, 
particularly in terms of $n_{\rm s}$, $\sigma_8$ and neutrino mass (which
are expected to affect $P^{\rm 1D}$ more significantly), in order to 
improve the accuracy of the interpolation between models. 

For every cosmological scenario defined by an individual grid parameter or a grid parameter pair, our standard simulation set consists of three realizations (rather than just one), 
characterized by different box sizes and numbers of particles; their combinations determine the lowest and highest $k$-modes that can be resolved in terms of power spectra, our major observables. 
This setting is motivated by the 
application of the {\it splicing} technique proposed by McDonald (2003), which allows us to considerably reduce computational time by
avoiding the production of single high-resolution and computationally prohibitive numerical simulations (see in particular 
Section \ref{subsec_resolution_requirements} for more details on this aspect).
Hence, in effect we always generate one run having the
largest box size and the lowest resolution considered, a second run characterized by the 
smallest box size and the highest resolution adopted, and a
third run with the resolution  of the first simulation and the box size of the second one. 
In this way,  we are able to correct the larger-box simulation for the lack of resolution and the small box for the lack of nonlinear coupling between the highest and lowest $k$-modes,
virtually mimicking an equivalent single larger-box highest-resolution (and more expensive) simulation.    
We chose to adopt similar box-size settings  as in Rossi et al. (2014), while increasing the number of particles in the various realizations -- since  
extensive convergence and resolution tests in support of our setup have already been carried out (see Rossi et al. 2014, and references therein).
Specifically, we assume a box size of $100~h^{-1} {\rm Mpc}$ for large-scale power with a number of particles per component $N_{\rm p} = 832^3$,
and a box size of $25~h^{-1}{\rm Mpc}$ for small-scale power, in this case with $N_{\rm p}  = 832^3$ or $208^3$, respectively. 
In a series of works, we have already proven  
analogous choices to be considerably successful in
constraining the summed neutrino mass and the amount of dark radiation, and in accurately reproducing the measured Ly$\alpha$ flux power spectrum. 
Thanks to these new simulations and via \textit{splicing}, we are thus able to achieve an effective resolution up to  
$3 \times  3328^3 = 110$ billion particles in a $(100 h^{-1} {\rm Mpc})^3$ box size, while abating  computational expenses: 
this effective resolution is ideal for reproducing all of the Ly$\alpha$ forest key features at the quality of eBOSS, 
and also for upcoming DESI high-$z$ quasar observations --see also Section \ref{subsec_resolution_requirements} for more details on resolution requirements. 
 
By construction, all of the grid simulations have been tuned to have $\sigma_8$ at $z=0$ matching the corresponding BG central value, 
given by the Planck 2015 baseline cosmology as indicated in Table \ref{table_baseline_params_sims}. This normalization choice, namely  $\sigma_8$ 
at the present epoch consistent with the value derived in the reference model,
differs from the convention adopted for most of the {\it Supporting Suite} realizations; we will return to this aspect in the next subsection. 
Moreover, the grid simulations have been performed with the same random seed, starting
at $z = 33$ with initial conditions fixed by 2LPT, and carried out until $z=2.0$; output snapshots are created at regular $\Delta z = 0.2$ intervals, within 
the range $5.0 \le z \le 2.0$. 

The full list of {\it Grid Simulations} produced is shown in Tables \ref{table_grid_sims_base} and \ref{table_grid_sims_cross_terms}. 
Specifically, Table \ref{table_grid_sims_base}  reports the central reference model (BG) and all the 
runs involving variations in individual parameters that define the grid -- as indicated in Table \ref{table_grid_params_variations}
and exemplified in Figure \ref{fig_grid_structure} using the symbols $p^{\pm}$. 
Table \ref{table_grid_sims_cross_terms} contains the cross-terms comprising all of the allowed 
combinations, for the cosmological and astrophysical 
parameters considered in the grid.
The tables also display some technical details regarding the various simulations, such as box sizes, resolutions (number of particles $N_{\rm p}$ per type),
mean particle separation, and gravitational softening, along with essential information on neutrino and WDM masses ($\sum m_{\nu} \equiv M_{\nu}$; $m_{\rm WDM}$), 
the number of effective neutrino species $N_{\rm eff}$, and the value of $\sigma_8$.


\begin{table*}
\begin{center}
\tiny
\centering
\caption{Grid Simulations (2): Cross-Terms.}
\doublerulesep 2.0pt
\renewcommand\arraystretch{1.5}
\begin{tabular}{ccccccccc} 
\hline \hline
 Simulation Name &   $M_{\rm \nu}$ [eV]  &   $m_{\rm WDM}$ [keV] & $N_{\rm eff}$   & $\sigma_8(z=0)$  & Boxes [Mpc/h] & $N_{\rm p}^{1/3}$ & Mean Part. Sep.  [Mpc/h]  & Softening [kpc/h] \\ 
\hline                                                                                                                                                             
$n_{\rm s}^+\sigma_8^+$ a/b/c &  0.0 &       0.0  &  3.0460  &    0.842             &     25/25/100  & 208/832/832   & 0.12019/0.03005/0.12019 & 4.01/1.00/4.01\\
$n_{\rm s}^+\Omega_{\rm m}^+$ a/b/c &  0.0&     0.0     &  3.0460  &    0.815             &     25/25/100  & 208/832/832   & 0.12019/0.03005/0.12019 & 4.01/1.00/4.01\\
$n_{\rm s}^+H_{\rm 0}^+$ a/b/c &  0.0    &   0.0   &  3.0460  &    0.815             &     25/25/100  & 208/832/832   & 0.12019/0.03005/0.12019 & 4.01/1.00/4.01\\
$n_{\rm s}^+\sum m_{\nu}^+$ a/b/c &  0.3 &      0.0   &  3.0460  &    0.815             &     25/25/100  & 208/832/832   & 0.12019/0.03005/0.12019 & 4.01/1.00/4.01\\
$n_{\rm s}^+N_{\rm eff}^+$ a/b/c &  0.0   &    0.0   &  4.0460  &    0.815             &     25/25/100  & 208/832/832   & 0.12019/0.03005/0.12019 & 4.01/1.00/4.01\\
$n_{\rm s}^+m_{\rm WDM}^+$ a/b/c &  0.0   &     3.0  &  3.0466  &    0.815             &     25/25/100  & 208/832/832   & 0.12019/0.03005/0.12019 & 4.01/1.00/4.01\\
$n_{\rm s}^+T_0^+$ a/b/c &  0.0      &  0.0  &  3.0460  &    0.815             &     25/25/100  & 208/832/832   & 0.12019/0.03005/0.12019 & 4.01/1.00/4.01\\
$n_{\rm s}^+\gamma^+$ a/b/c &  0.0  &    0.0    &  3.0460  &    0.815             &     25/25/100  & 208/832/832   & 0.12019/0.03005/0.12019 & 4.01/1.00/4.01\\
$n_{\rm S}^+\tau_{\rm A}^+$ a/b/c &  0.0     &  0.0   &  3.0460  &    0.815             &     25/25/100  & 208/832/832   & 0.12019/0.03005/0.12019 & 4.01/1.00/4.01\\
$n_{\rm S}^+\tau_{\rm S}^+$ a/b/c &  0.0     &  0.0   &  3.0460  &    0.815             &     25/25/100  & 208/832/832   & 0.12019/0.03005/0.12019 & 4.01/1.00/4.01\\

$\sigma_8^+\Omega_{\rm m}^+$ a/b/c &  0.0 &   0.0     &  3.0460  &    0.842             &     25/25/100  & 208/832/832   & 0.12019/0.03005/0.12019 & 4.01/1.00/4.01\\
$\sigma_8^+H_{\rm 0}^+$ a/b/c &  0.0     &  0.0   &  3.0460  &    0.842             &     25/25/100  & 208/832/832   & 0.12019/0.03005/0.12019 & 4.01/1.00/4.01\\
$\sigma_8^+\sum m_{\nu}^+$ a/b/c &  0.3 &   0.0      &  3.0460  &    0.842             &     25/25/100  & 208/832/832   & 0.12019/0.03005/0.12019 & 4.01/1.00/4.01\\
$\sigma_8^+N_{\rm eff}^+$ a/b/c &  0.0    &   0.0   &  4.0460  &    0.842             &     25/25/100  & 208/832/832   & 0.12019/0.03005/0.12019 & 4.01/1.00/4.01\\
$\sigma_8^+m_{\rm WDM}^+$ a/b/c &  0.0    &   3.0   &  3.0466  &    0.842             &     25/25/100  & 208/832/832   & 0.12019/0.03005/0.12019 & 4.01/1.00/4.01\\
$\sigma_8^+T_0^+$ a/b/c &  0.0     &  0.0   &  3.0460  &    0.842             &     25/25/100  & 208/832/832   & 0.12019/0.03005/0.12019 & 4.01/1.00/4.01\\
$\sigma_8^+\gamma^+$ a/b/c &  0.0  &      0.0  &  3.0460  &    0.842             &     25/25/100  & 208/832/832   & 0.12019/0.03005/0.12019 & 4.01/1.00/4.01\\
$\sigma_{\rm 8}^+\tau_{\rm A}^+$ a/b/c &  0.0     &  0.0   &  3.0460  &    0.842             &     25/25/100  & 208/832/832   & 0.12019/0.03005/0.12019 & 4.01/1.00/4.01\\
$\sigma_{\rm 8}^+\tau_{\rm S}^+$ a/b/c &  0.0     &  0.0   &  3.0460  &    0.842             &     25/25/100  & 208/832/832   & 0.12019/0.03005/0.12019 & 4.01/1.00/4.01\\

$\Omega_{\rm m}^+H_{\rm 0}^+$ a/b/c &  0.0  &    0.0    &  3.0460  &    0.815             &     25/25/100  & 208/832/832   & 0.12019/0.03005/0.12019 & 4.01/1.00/4.01\\
$\Omega_{\rm m}^+\sum m_{\nu}^+$ a/b/c &  0.3 &     0.0    &  3.0460  &    0.815             &     25/25/100  & 208/832/832   & 0.12019/0.03005/0.12019 & 4.01/1.00/4.01\\
$\Omega_{\rm m}^+N_{\rm eff}^+$ a/b/c &  0.0     &    0.0 &  4.0460  &    0.815             &     25/25/100  & 208/832/832   & 0.12019/0.03005/0.12019 & 4.01/1.00/4.01\\
$\Omega_{\rm m}^+m_{\rm WDM}^+$ a/b/c &  0.0     &    3.0 &  3.0466  &    0.815             &     25/25/100  & 208/832/832   & 0.12019/0.03005/0.12019 & 4.01/1.00/4.01\\
$\Omega_{\rm m}^+T_0^+$ a/b/c &  0.0      &   0.0 &  3.0460  &    0.815             &     25/25/100  & 208/832/832   & 0.12019/0.03005/0.12019 & 4.01/1.00/4.01\\
$\Omega_{\rm m}^+\gamma^+$ a/b/c &  0.0  &    0.0    &  3.0460  &    0.815             &     25/25/100  & 208/832/832   & 0.12019/0.03005/0.12019 & 4.01/1.00/4.01\\
$\Omega_{\rm m}^+\tau_{\rm A}^+$ a/b/c &  0.0     &  0.0   &  3.0460  &    0.815             &     25/25/100  & 208/832/832   & 0.12019/0.03005/0.12019 & 4.01/1.00/4.01\\
$\Omega_{\rm m}^+\tau_{\rm S}^+$ a/b/c &  0.0     &  0.0   &  3.0460  &    0.815             &     25/25/100  & 208/832/832   & 0.12019/0.03005/0.12019 & 4.01/1.00/4.01\\

$H_{\rm 0}^+\sum m_{\nu}^+$ a/b/c &  0.3    &   0.0   &  3.0460  &    0.815             &     25/25/100  & 208/832/832   & 0.12019/0.03005/0.12019 & 4.01/1.00/4.01\\
$H_{\rm 0}^+N_{\rm eff}^+$ a/b/c &  0.0    &    0.0  &  4.0460  &    0.815             &     25/25/100  & 208/832/832   & 0.12019/0.03005/0.12019 & 4.01/1.00/4.01\\
$H_{\rm 0}^+m_{\rm WDM}^+$ a/b/c &  0.0    &   3.0   &  3.0466  &    0.815             &     25/25/100  & 208/832/832   & 0.12019/0.03005/0.12019 & 4.01/1.00/4.01\\
$H_{\rm 0}^+T_0^+$ a/b/c &  0.0      &   0.0 &  3.0460  &    0.815             &     25/25/100  & 208/832/832   & 0.12019/0.03005/0.12019 & 4.01/1.00/4.01\\
$H_{\rm 0}^+\gamma^+$ a/b/c &  0.0  &   0.0     &  3.0460  &    0.815             &     25/25/100  & 208/832/832   & 0.12019/0.03005/0.12019 & 4.01/1.00/4.01\\
$H_{\rm 0}^+\tau_{\rm A}^+$ a/b/c &  0.0     &  0.0   &  3.0460  &    0.815             &     25/25/100  & 208/832/832   & 0.12019/0.03005/0.12019 & 4.01/1.00/4.01\\
$H_{\rm 0}^+\tau_{\rm S}^+$ a/b/c &  0.0     &  0.0  &  3.0460  &    0.815             &     25/25/100  & 208/832/832   & 0.12019/0.03005/0.12019 & 4.01/1.00/4.01\\

$\sum m_{\nu}^+N_{\rm eff}^+$ a/b/c &  0.3 &      0.0   &  4.0460  &    0.815             &     25/25/100  & 208/832/832   & 0.12019/0.03005/0.12019 & 4.01/1.00/4.01\\
$\sum m_{\nu}^+m_{\rm WDM}^+$ a/b/c &  0.3 &  3.0       &  3.0466  &    0.815             &     25/25/100  & 208/832/832   & 0.12019/0.03005/0.12019 & 4.01/1.00/4.01\\
$\sum m_{\nu}^+T_0^+$ a/b/c &  0.3      &   0.0 &  3.0460  &    0.815             &     25/25/100  & 208/832/832   & 0.12019/0.03005/0.12019 & 4.01/1.00/4.01\\
$\sum m_{\nu}^+\gamma^+$ a/b/c &  0.3  &     0.0   &  3.0460  &    0.815             &     25/25/100  & 208/832/832   & 0.12019/0.03005/0.12019 & 4.01/1.00/4.01\\
$\sum m_{\nu}^+\tau_{\rm A}^+$ a/b/c &  0.3     &  0.0   &  3.0460  &    0.815             &     25/25/100  & 208/832/832   & 0.12019/0.03005/0.12019 & 4.01/1.00/4.01\\
$\sum m_{\nu}^+\tau_{\rm S}^+$ a/b/c &  0.3     &  0.0   &  3.0460  &    0.815             &     25/25/100  & 208/832/832   & 0.12019/0.03005/0.12019 & 4.01/1.00/4.01\\

$N_{\rm eff}^+m_{\rm WDM}^+$ a/b/c &  0.0     &  3.0   &  4.0466  &    0.815             &     25/25/100  & 208/832/832   & 0.12019/0.03005/0.12019 & 4.01/1.00/4.01\\
$N_{\rm eff}^+T_0^+$ a/b/c &  0.0     &   0.0  &  4.0460  &    0.815             &     25/25/100  & 208/832/832   & 0.12019/0.03005/0.12019 & 4.01/1.00/4.01\\
$N_{\rm eff}^+\gamma^+$ a/b/c &  0.0  &     0.0   &  4.0460  &    0.815             &     25/25/100  & 208/832/832   & 0.12019/0.03005/0.12019 & 4.01/1.00/4.01\\
$N_{\rm eff}^+\tau_{\rm A}^+$ a/b/c &  0.0     &  0.0   &  4.0460  &    0.815             &     25/25/100  & 208/832/832   & 0.12019/0.03005/0.12019 & 4.01/1.00/4.01\\
$N_{\rm eff}^+\tau_{\rm S}^+$ a/b/c &  0.0     &  0.0   &  4.0460  &    0.815             &     25/25/100  & 208/832/832   & 0.12019/0.03005/0.12019 & 4.01/1.00/4.01\\

$m_{\rm WDM}^+T_0^+$ a/b/c &  0.0     &   3.0  &  3.0466  &    0.815             &     25/25/100  & 208/832/832   & 0.12019/0.03005/0.12019 & 4.01/1.00/4.01\\
$m_{\rm WDM}^+\gamma^+$ a/b/c &  0.0  &     3.0   &  3.0466  &    0.815             &     25/25/100  & 208/832/832   & 0.12019/0.03005/0.12019 & 4.01/1.00/4.01\\
$m_{\rm WDM}^+\tau_{\rm A}^+$ a/b/c &  0.0     &  3.0   &  3.0466  &    0.815             &     25/25/100  & 208/832/832   & 0.12019/0.03005/0.12019 & 4.01/1.00/4.01\\
$m_{\rm WDM}^+\tau_{\rm S}^+$ a/b/c &  0.0     &  3.0   &  3.0466  &    0.815             &     25/25/100  & 208/832/832   & 0.12019/0.03005/0.12019 & 4.01/1.00/4.01\\

$T_0^+\gamma^+$ a/b/c &  0.0     &  0.0   &  3.0460  &    0.815             &     25/25/100  & 208/832/832   & 0.12019/0.03005/0.12019 & 4.01/1.00/4.01\\
$T_{\rm 0}^+\tau_{\rm A}^+$ a/b/c &  0.0     &  0.0   &  3.0460  &    0.815             &     25/25/100  & 208/832/832   & 0.12019/0.03005/0.12019 & 4.01/1.00/4.01\\
$T_{\rm 0}^+\tau_{\rm S}^+$ a/b/c &  0.0     &  0.0   &  3.0460  &    0.815             &     25/25/100  & 208/832/832   & 0.12019/0.03005/0.12019 & 4.01/1.00/4.01\\

$\gamma^+\tau_{\rm A}^+$ a/b/c &  0.0     &  0.0   &  3.0460  &    0.815             &     25/25/100  & 208/832/832   & 0.12019/0.03005/0.12019 & 4.01/1.00/4.01\\
$\gamma^+\tau_{\rm S}^+$ a/b/c &  0.0     &  0.0   &  3.0460  &    0.815             &     25/25/100  & 208/832/832   & 0.12019/0.03005/0.12019 & 4.01/1.00/4.01\\

$\tau_{\rm A}^+\tau_{\rm S}^+$ a/b/c &  0.0     &  0.0   &  3.0460  &    0.815             &     25/25/100  & 208/832/832   & 0.12019/0.03005/0.12019 & 4.01/1.00/4.01\\

\hline
\hline
\label{table_grid_sims_cross_terms}
\end{tabular}
\end{center}
\end{table*}


\begin{figure*}
\centering
\includegraphics[angle=0,width=0.90\textwidth]{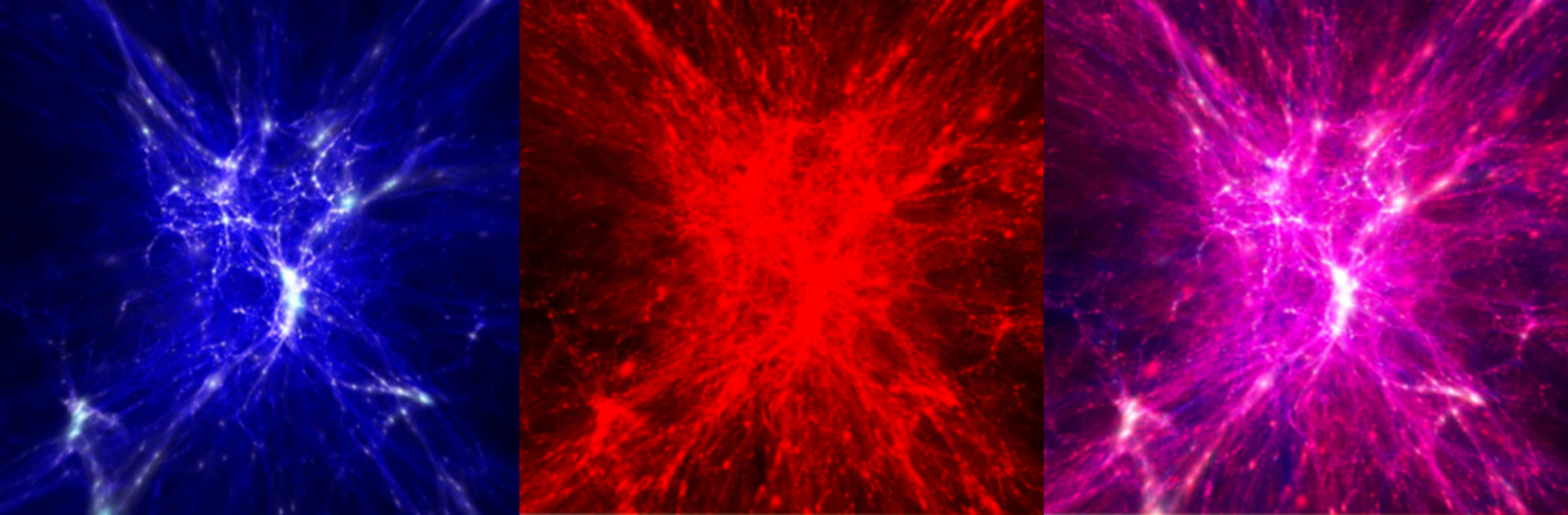}
\caption{Full-size projections of the density field at $z=2.0$ from one of the `\textit{Best Guess}' reference simulations, characterized by $N_{\rm p} = 832^3$ particles/type over 
a $25h^{-1}{\rm Mpc}$ box size. The gas distribution (left panel), dark matter distribution (central panel), as well as both components (right panel) are shown over the entire $25h^{-1}{\rm Mpc}$ simulation length. 
The density field  is projected along the $x$- and $y$-directions, and across $z$.}
\label{fig_visualization_1}
\end{figure*}

\begin{figure*}
\centering
\includegraphics[angle=0,width=0.90\textwidth]{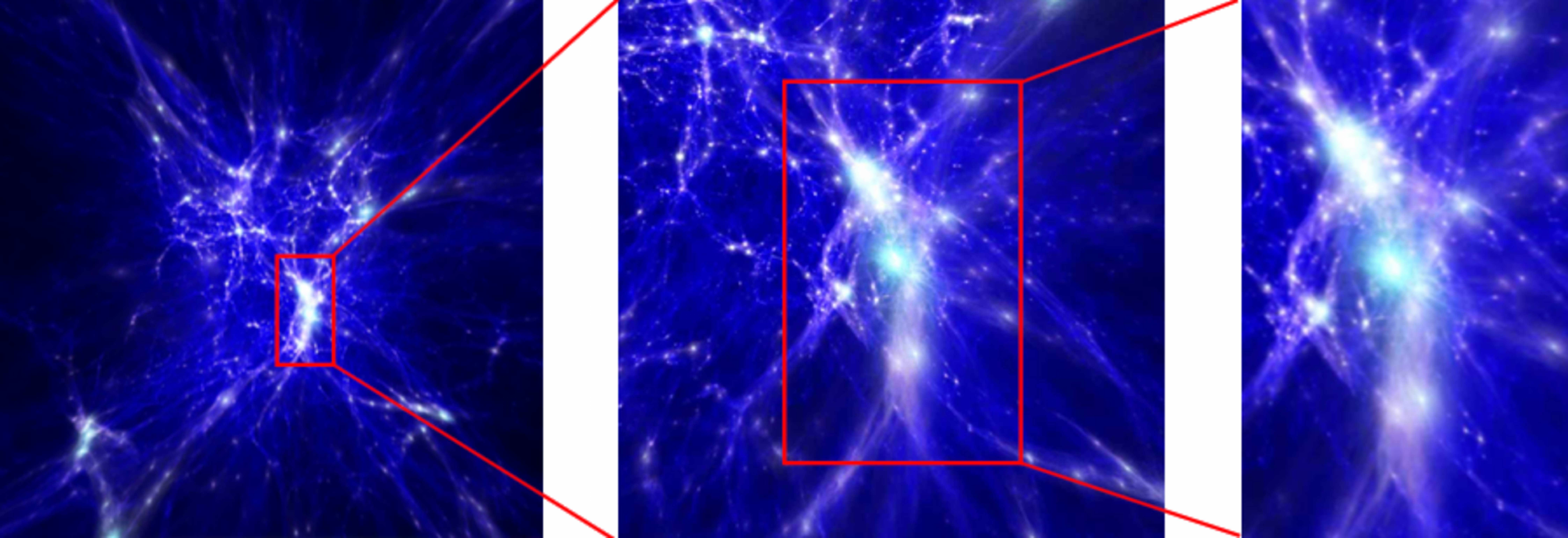}
\caption{Progressive zoom into the `\textit{Best Guess}' gas distribution (as seen in the left panel of Figure \ref{fig_visualization_1}) when the
box size is  $25 h^{-1} {\rm Mpc}$,   
and the resolution is given by $N_{\rm p} = 832^3$ particles per type, highlighting a large and complex structure. 
Out of the full $25 \times 25$ (Mpc/h)$^2$ projected surface, the first $\sim 2.7 \times 5.4$ (Mpc/h)$^2$ and second $\sim 1.2 \times 3.9$ (Mpc/h)$^2$ insets 
are  gradual enlargements of a simulated patch containing richer structure, showing that our simulations are able to
resolve quite accurately small-scale structures -- even below megaparsec scales.}
\label{fig_visualization_1_bis}
\end{figure*}


Figure \ref{fig_visualization_1} shows visual examples 
of full-size projections at $z =2.0$ of the density field  along the $x$- and $y$-directions (and across $z$) from one of the BG reference simulations, 
having a box size of $25h^{-1}{\rm Mpc}$ and the resolution defined by $N_{\rm p} = 832^3$ particles per type -- as reported in Table \ref{table_grid_sims_base}.
Specifically, the left panel displays the gas distribution, the central panel is for the DM distribution,  and the right panel shows both components over the entire extension of the simulation box.  
Figure \ref{fig_visualization_1_bis} is a progressive zoom into the BG gas distribution as seen in the left panel of Figure \ref{fig_visualization_1}, when the original box size is  $25 h^{-1} {\rm Mpc}$   
and $N_{\rm p} = 832^3$ particles per type. The first inset is a $\sim 2.7 \times 5.4$ (Mpc/h)$^2$ area out of the full   $25 \times 25$ (Mpc/h)$^2$ projected surface, while the second inset considered in the central panel -- and fully visualized in the
right one -- is an enlargement of a 
$\sim 1.2 \times 3.9$ (Mpc/h)$^2$ patch containing a rich and complex structure.  
The main point of this visualization is to show that the resolution quality of the \textit{Grid Suite} allows one to resolve quite accurately small-scale structures, even below megaparsec (Mpc) scales. 

Noticeably, the grid contains models with massive neutrinos (indicated as $\sum m_{\nu}^{+}$), WDM (termed with $m_{\rm WDM}^{+}$), and 
dark radiation (labeled as $N_{\rm eff}^{+}$). See Section \ref{subsec_nu_dark_rad_wdm_modeling}  for technical implementation details. We just 
recall  here that when we include massive neutrinos, we always keep $\Omega_{\Lambda}+ \Omega_{\rm m}$ fixed to give a flat geometry (i.e., $\Omega_{\rm tot} = 1$ with 
$\Omega_{\rm m} = \Omega_{\rm b} + \Omega_{\rm c} + \Omega_{\nu}$), 
and vary the additional massive neutrino component $\Omega_{\nu}$ to the detriment of the DM component $\Omega_{\rm c}$.
Moreover, for the first time, we also modeled consistently combined scenarios with massive neutrinos and WDM (indicated as $\sum m_{\nu}^{+}  m_{\rm WDM}^{+}$ in Table \ref{table_grid_params_variations}),
scenarios with massive neutrinos and dark radiation ($ \sum m_{\nu}^{+} N_{\rm eff}^{+}$), and cosmologies with dark radiation and WDM 
($ N_{\rm eff}^{+} m_{\rm WDM}^{+}$): their implementation in the $N$-body setting is nontrivial, 
and it is also briefly addressed in Section \ref{subsec_nu_dark_rad_wdm_modeling}.


\begin{figure*}
\centering
\includegraphics[angle=0,width=0.90\textwidth]{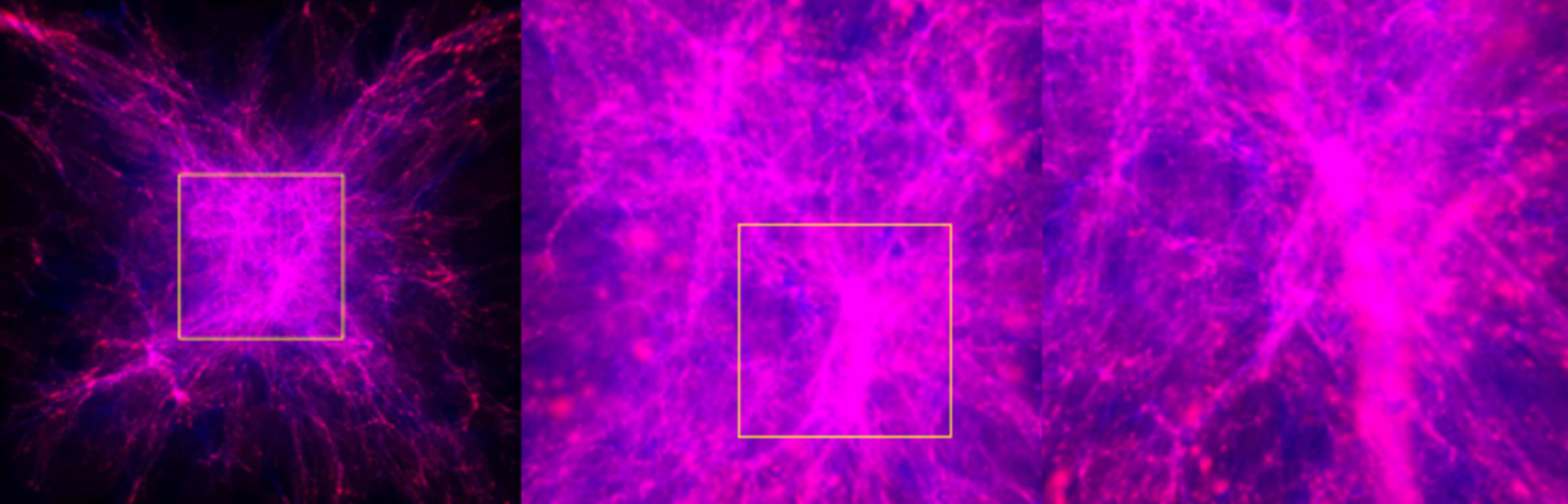}
\includegraphics[angle=0,width=0.90\textwidth]{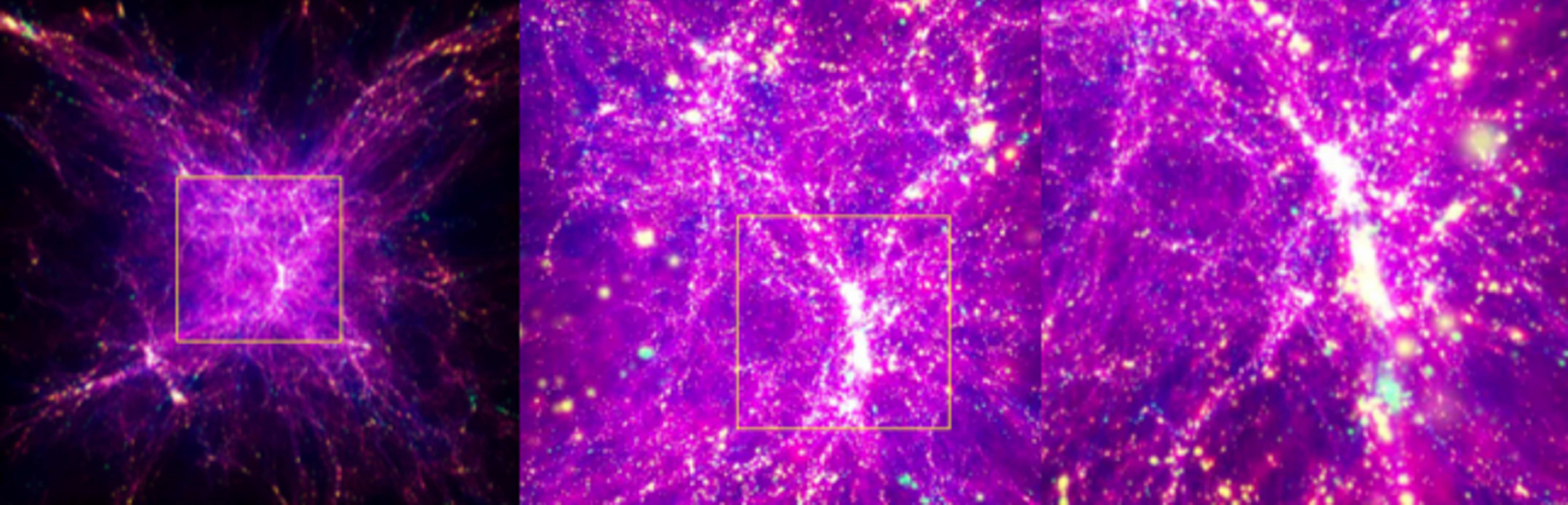}
\caption{Comparison between a \textit{``Best Guess"} run (top panels) and a massive neutrino scenario characterized by 
$\sum m_{\nu} =0.3$ eV (bottom panels; realization indicated as $\sum m_{\nu}^{+}$ in Table \ref{table_grid_sims_base}).
From left to right, progressive enlargements of an $\sim 8.0 \times 8.0$ 
(Mpc/h)$^2$ area (middle panels), and of a $3.3 \times 3.3$ (Mpc/h)$^2$ patch (right panels) are shown.
The box size of the two simulations is $25h^{-1}{\rm Mpc}$, for an overall resolution determined by $N_{\rm p} = 832^3$ particles/type.
Gas, dark matter, and massive neutrinos (whenever present) are shown simultaneously within the same frameworks. In particular, 
the neutrino component -- present only in the bottom panels -- is explicitly displayed using a white/brighter color scale, to facilitate the visual impact of their 
clustering properties over the gas and dark matter distributions. In actuality, the overall  effect of a nonzero neutrino mass 
on the global structure of the cosmic web is small and, hence, hard to appreciate visually.}
\label{fig_visualization_3C}
\end{figure*}


Finally, Figure \ref{fig_visualization_3C} shows a comparison between the
BG run (top panels) and a massive neutrino cosmology realization characterized by a total neutrino mass of 
$\sum m_{\nu} =0.3$ eV (bottom panels -- run denoted as $\sum m_{\nu}^{+}$ in Table \ref{table_grid_sims_base}).
Similarly as in Figures \ref{fig_visualization_1} and \ref{fig_visualization_1_bis}, the plot displays full-size projections of the density field at $z =2.0$   along the $x$- and $y$-directions (and across $z$).
All of the components are now shown simultaneously, namely, gas, DM, and massive neutrinos -- whenever present.
The resolution of the simulations is $N_{\rm p} = 832^3$ particles/type, over a $25h^{-1}{\rm Mpc}$ box size. 
From left to right, there is a progressive enlargement focused on two square patches: the first is an $\sim 8.0 \times 8.0$ 
(Mpc/h)$^2$ zoom (middle panels), and the second represents a  $3.3 \times 3.3$ (Mpc/h)$^2$  amplification (right panels). 
In the bottom panels, the neutrino component is explicitly rendered using a white/brighter color scale, in order to make the overall details of the neutrino clustering 
visually appreciable. In actuality, because of significant free streaming, the impact of massive neutrinos on the gas and DM distribution is 
a relatively small effect that is hard to detect and, hence, not easily distinguishable from the BG scenario -- at least visually.  
However, as shown in Rossi (2017), the growth of structures is less evolved in the simulation with neutrinos (i.e., the voids are less empty) since their suppressed clustering slows down the growth of perturbations
in the overall matter density, and this in turn does affect the properties of the gas and DM.
Hence, the presence of massive neutrinos does induce detectable changes in the thermal state of the gas and in the LSS clustering of DM, 
impacting both the properties of the gas as well as the overall DM clustering features. 
In particular, variations in the gas thermal state are relevant for the well-known $T_0-\gamma$  power-law relation, which arises from the competition 
between photo-heating and cooling due to the adiabatic expansion of the universe, following reionization. 
Typical neutrino fluctuations at the largest scales are about $10\%$ 
around the mean, while for gas and DM, the fluctuations are usually much stronger.


\subsection{The \textit{Supporting Suite}: Massive Neutrinos, Dark Radiation, Warm Dark Matter} \label{subsec_supporting_suite}


\begin{table*}
\begin{center}
\tiny
\centering
\caption{Supporting Suite (1): Massive Neutrinos and Dark Radiation.}
\doublerulesep 2.0pt
\renewcommand\arraystretch{1.5}
\begin{tabular}{cccccccc} 
\hline \hline
 Simulation Name &   $M_{\rm \nu}$ [eV]  &   $N_{\rm eff}$ & $\sigma_8(z=0)$  & Boxes [Mpc/h] & $N_{\rm p}^{1/3}$ & Mean Par. Sep.  [Mpc/h] & Softening [kpc/h]  \\ 
\hline
Best Guess Grid a/b/c & 0.0          &3.046 & 0.8150             &     25/25/100  & 208/832/832 & 0.1202/0.0300/0.1202 & 4.01/1.00/4.01 \\
BG\_NORM  a/b/c & 0.0          &3.046 & 0.8150             &     25/100/100  & 256/512/832 & 0.0976/0.1953/0.1202 & 3.25/6.51/4.01 \\
BG\_UN  a/b & 0.0           &3.046  & 0.8150             &     25/100 & 256/512 & 0.0976/0.1953 &   3.25/6.51  \\
\hline
\hline
NU\_NORM 01  a & 0.1       &3.046 & 0.8150 & 25 & 256 & 0.0976 &   3.25  \\
NU\_NORM 02  a & 0.2       &3.046 & 0.8150 & 25 & 256 & 0.0976 &   3.25  \\
NU\_NORM 03  Grid a/b/c & 0.3       &3.046 & 0.8150 & 25/25/100  & 208/832/832 & 0.1202/0.0300/0.1202 & 4.01/1.00/4.01 \\
NU\_NORM 03  a & 0.3       &3.046 & 0.8150 & 25 & 256 &   0.0976 & 3.25 \\
NU\_NORM 04  a & 0.4       &3.046 & 0.8150 & 25 & 256 & 0.0976 &  3.25 \\
\hline
NU\_UN 01 Grid-Like  a/b/c & 0.1       &3.046 &  0.7926 & 25/25/100 & 208/832/832 & 0.1202/0.0300/0.1202 & 4.01/1.00/4.01 \\
NU\_UN 01  a/b/c & 0.1       &3.046 &  0.7926 & 25/100/100 & 256/512/832 & 0.0976/0.1953/0.1202 &   3.25/6.51/4.01 \\
NU\_UN 02  a       & 0.2       &3.046 & 0.7674 & 25 & 256  & 0.0976  &  3.25 \\
NU\_UN 03  Grid-Like a/b/c & 0.3    &3.046 & 0.7423 & 25/25/100 & 208/832/832 & 0.1202/0.0300/0.1202 & 4.01/1.00/4.01 \\
NU\_UN 03  a/b/c & 0.3       &3.046 & 0.7423 & 25/100/100 & 256/512/832 & 0.0976/0.1953/0.1202 &  3.25/6.51/4.01 \\
NU\_UN 04  a       & 0.4       &3.046 &  0.7179 & 25 & 256  & 0.0976  &  3.25\\
\hline
\hline
DR\_NORM BG  a & 0.0+s       &4.046 & 0.8150 & 25 & 256 & 0.0976 & 3.25 \\
DR\_NORM 01  a & 0.1+s       &4.046& 0.8150 & 25 & 256 & 0.0976 &   3.25\\
DR\_NORM 02  a & 0.2+s       &4.046& 0.8150 & 25 & 256 & 0.0976 &  3.25 \\
DR\_NORM 03  Grid a/b/c & 0.3+s       &4.046& 0.8150 &         25/25/100  & 208/832/832 & 0.1202/0.0300/0.1202 & 4.01/1.00/4.01 \\
DR\_NORM 03  a & 0.3+s       &4.046& 0.8150 & 25 & 256 & 0.0976  &  3.25 \\
DR\_NORM 04  a & 0.4+s       &4.046& 0.8150 & 25 & 256 & 0.0976 &  3.25 \\
\hline
DR\_UN BG  a & 0.0+s       &4.046& 0.7583 & 25 & 256 & 0.0976 &  3.25 \\
DR\_UN 01  Grid-Like a/b/c & 0.1+s       &4.046& 0.7375 &  25/25/100  & 208/832/832 & 0.1202/0.0300/0.1202 & 4.01/1.00/4.01 \\
DR\_UN 01  a/b/c & 0.1+s       &4.046& 0.7375 & 25/100/100 & 256/512/832 & 0.0976/0.1953/0.1202 &  3.25/6.51/4.01 \\
DR\_UN 02  a       & 0.2+s       &4.046& 0.7140 & 25 & 256  & 0.0976  &  3.25 \\
DR\_UN 03  Grid-Like a/b/c & 0.3+s       &4.046& 0.6908 &  25/25/100  & 208/832/832 & 0.1202/0.0300/0.1202 & 4.00/1.00/4.00 \\
DR\_UN 03  a/b/c & 0.3+s       &4.046& 0.6908 & 25/100/100 & 256/512/832 & 0.0976/0.1953/0.1202 &  3.25/6.51/4.01 \\
DR\_UN 04  a       & 0.4+s       &4.046& 0.6682 & 25 & 256  & 0.0976  &  3.25 \\
\hline
\hline
BG\_VIS\_NORM  a & 0.0           &3.046 & 0.8150             &     25 & 208 & 0.1202 &  4.01\\
NU\_VIS\_NORM 03  a  & 0.3 &3.046          & 0.8150             &     25 & 208 & 0.1202 &  4.01 \\
NU\_VIS\_UN 03  a  & 0.3 &3.046          &  0.7423            &     25 & 208 & 0.1202 &  4.01 \\
DR\_VIS\_UN 03   a  & 0.3+s  &4.046         & 0.6908            &     25 & 208 & 0.1202 &  4.01 \\
\hline
\hline
\label{table_supporting_sims_dark_rad}
\end{tabular}
\end{center}
\end{table*}


The hydrodynamical {\it Supporting Suite} has been developed  to  
characterize the physics of the small-scale high-$z$ cosmic web 
in several nonstandard cosmological scenarios, in order to
 identify unique signatures and preferred scales where such models may differ significantly from the baseline $\Lambda$CDM framework. 
In particular, we focus here on cosmologies with massive neutrinos, dark radiation, and 
WDM -- along with their combinations, which represent a novelty in the literature.
When  dark radiation is included, it is implemented in the form of a massless sterile neutrino thermalized with active neutrinos.
Regarding WDM, we only consider thermal relics -- recalling that there is a direct correspondence between thermal relics and massive sterile neutrinos.  
For the various implementation details within the SPH formalism, see Section \ref{subsec_nu_dark_rad_wdm_modeling}. 
The primary aim in developing these runs is to 
carefully study and quantify the effects of  
neutrinos, dark radiation, and WDM on structure formation at small scales,
with a closer eye on Ly$\alpha$ forest observables and on the high-$z$ cosmic web. 
This is an essential step for obtaining 
robust parameter constraints free from systematic biases. 
In what follows, we briefly highlight the main aspects of the {\it Supporting Suite}, while
we will present detailed analyses of these simulations in forthcoming studies. 
 
Table \ref{table_supporting_sims_dark_rad} lists all of the {\it supporting} simulations related to massive neutrinos and dark radiation, 
along with corresponding details such as resolution, box size, mean particle separation, gravitational softening length,
values of $\sigma_8$ at $z=0$, neutrino mass, and $N_{\rm eff}$.
As specified in the table, we consider different box sizes and resolutions, varying from $25h^{-1}$Mpc to $100h^{-1}$Mpc, and
a maximum number of particles characterized by $3 \times 832^3$.  We run a total of 56 simulations for this subset, including BG runs. 
Aside from the standard reference cosmology, all these realizations contain different degrees of summed
 neutrino mass (runs abbreviated with `NU') and additional dark radiation contributions (runs indicated as `DR'), the latter in the form of a massless sterile neutrino thermalized with active neutrinos.
The presence of a sterile neutrino is denoted by `$s$' in Table \ref{table_supporting_sims_dark_rad}.
Specifically, 
 while our central model has only a 
massless neutrino component (i.e., $\sum m_{\nu}=0.0~{\rm eV}, N_{\rm eff}=3.046$), the other scenarios  
incorporate three degenerate  massive neutrinos with $\sum m_{\nu} =0.1, 0.2, 0.3, 0.4 $ eV, respectively, 
and whenever indicated, they also contain an additional massless sterile neutrino -- so that
$N_{\rm eff}=4.046$. 
The runs with the extra label `VIS' are small-box simulations only made for visualization purposes, and carried out until $z=0$ with increased redshift outputs.
In terms of normalization, all the realizations labeled as `NORM' share the convention highlighted in Section \ref{sec_sejong_suite_general}; namely, $\sigma_8$ is rescaled so that
its value at $z=0$ is consistent with that of the reference cosmology (BG) at the present epoch. Instead, all the runs termed `UN' 
have the same primordial amplitude value $A_{\rm s}$ of the baseline model, but they differ in terms of $\sigma_8$ at $z=0$.

This subset of {\it supporting} simulations has already been used in Rossi (2017) for studies related
to Ly$\alpha$ forest observables, and in particular, to accurately measure the tomographic evolution of the shape and amplitude of the
small-scale matter and flux power spectra in massive neutrino and dark radiation cosmologies, and to characterize the corresponding thermal state of the IGM
through the temperature-density relation 
-- seeking for unique signatures at small scales (see their work for extensive details).  


\begin{figure*}
\centering
\includegraphics[angle=0,width=0.95\textwidth]{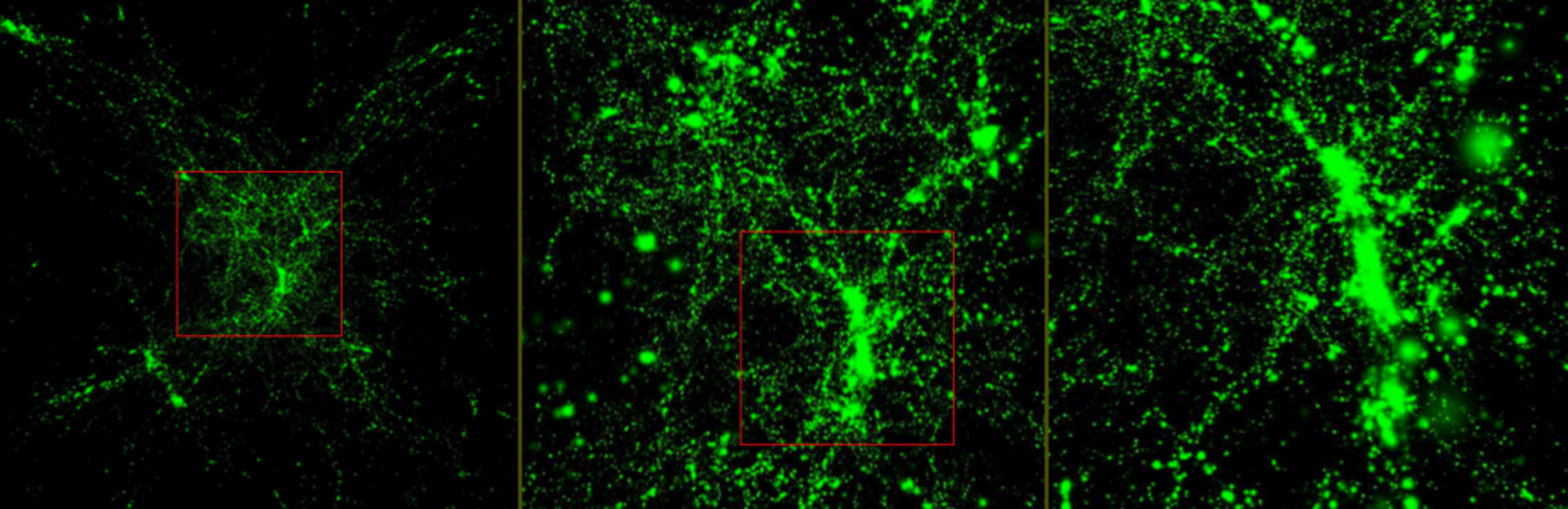}
\caption{Visualization of the neutrino component from a full-size density projection at $z = 2.0$, in a simulation having a $25h^{-1}$ Mpc box size and resolution $N_{\rm p} = 832^3$ particles per type, and
a total summed neutrino mass  $\sum m_{\nu} = 0.3$ eV (see Table \ref{table_supporting_sims_dark_rad}).     
From left to right,    the progressive enlargements 
show an $\sim 8.0 \times 8.0$ 
(Mpc/h)$^2$  square patch (middle panel), and a  $3.3 \times 3.3$ (Mpc/h)$^2$ square
patch (right panel). The presence of massive neutrinos does induce noticeable changes in the thermal state of the gas and in the
LSS clustering of dark matter, with a measurable impact on the high-$z$ cosmic web.}
\label{fig_supporting_suite_visualizations_neutrinos_A}
\end{figure*}


Figure \ref{fig_supporting_suite_visualizations_neutrinos_A} is a visual example of the LSS cosmic web as seen in the neutrino component alone from a {\it supporting} simulation with 
a $25h^{-1}$ Mpc box size and resolution $N_{\rm p} = 832^3$ particles per type. 
The plot displays the density projection at $z = 2.0$, when 
the total summed neutrino mass  is $\sum m_{\nu} = 0.3$ eV. 
From left to right, the progressive enlargements show an $\sim 8.0 \times 8.0$ 
(Mpc/h)$^2$  square patch (middle panel), and a  $3.3 \times 3.3$ (Mpc/h)$^2$ square
patch (right panel). While neutrinos do not cluster at small scales because 
of their high free-streaming velocities, their presence does induce noticeable changes in the global thermal state of the gas, and in the
LSS clustering of DM -- with a measurable impact on the high-$z$ cosmic web. It is also quite interesting to
notice -- even just visually -- how neutrinos trace the overall structure of the cosmic web. 

 
\begin{table*}
\begin{center}
\tiny
\centering
\caption{Supporting Suite (2): Warm Dark Matter and Massive Neutrinos.}
\doublerulesep 2.0pt
\renewcommand\arraystretch{1.5}
\begin{tabular}{ccccccccc} 
\hline \hline
 Simulation Name &   $M_{\rm \nu}$ [eV]  &    $m_{\rm WDM}$ [keV] &  $N_{\rm eff}$ & $\sigma_8(z=0)$  & Boxes [Mpc/h] & $N_{\rm p}^{1/3}$ & Mean Par. Sep.  [Mpc/h] & Softening [kpc/h]  \\ 
\hline
Best Guess Grid a/b/c & 0.0 & 0.00          &3.0460 & 0.8150             &     25/25/100  & 208/832/832 & 0.1202/0.0300/0.1202 & 4.01/1.00/4.01 \\
Best Guess  a/b/c & 0.0      & 0.00      &3.0460 & 0.8150             &     25/25/100  & 128/256/512 &           0.1953/0.0976/0.1953 &     6.51/3.25/6.51 \\
BG\_UN  a/b/c & 0.0     & 0.00       &3.0460 & 0.8150             &     25/25/100  & 128/256/512 &            0.1953/0.0976/0.1953 &     6.51/3.25/6.51 \\
\hline
\hline
WDM$^{*}$\_UN\_0.25\_keV  a & 0.0    & 0.25        & 3.0460    & 0.8305             &     25  & 128 &           0.1953 &     6.51 \\
WDM$^{*}$\_0.25\_keV  a & 0.0    & 0.25        & 3.0460    & 0.8150             &     25  & 128 &           0.1953 &     6.51 \\
WDM$^{\dagger}$\_UN\_0.25\_keV  a & 0.0    & 0.25        & 3.0460    & 0.8305              &     25  & 128 &           0.1953 &     6.51 \\
WDM$^{\dagger}$\_0.25\_keV  a & 0.0    & 0.25        & 3.0460    & 0.8150             &     25  & 128 &           0.1953 &     6.51 \\
WDM\_UN\_0.25\_keV  a & 0.0    & 0.25        & 3.0617    & 0.8269             &     25  & 128 &           0.1953 &     6.51 \\
WDM\_0.25\_keV  a & 0.0    & 0.25        & 3.0617    & 0.8150             &     25  & 128 &           0.1953 &     6.51 \\
\hline
\hline
WDM$^{\dagger}$\_1.00\_keV  a/b/c  & 0.0    & 1.00        & 3.0460      & 0.8150             &     25/100/100  & 256/512/832 &       0.0976/0.1953/0.1202  &     3.25/6.51/4.01  \\
WDM$^{\dagger}$\_2.00\_keV  a/b  & 0.0    & 2.00        & 3.0460      & 0.8150             &     25/100  & 256/512 &       0.0976/0.1953  &     3.25/6.51  \\
WDM$^{\dagger}$\_3.00\_keV  a/b  & 0.0    & 3.00        & 3.0460      & 0.8150             &     25/100  & 256/512 &       0.0976/0.1953  &     3.25/6.51  \\
WDM$^{\dagger}$\_3.00\_keV Grid-Like a/b/c  & 0.0    & 3.00        & 3.0460      & 0.8150             &     25/25/100  & 208/832/832 & 0.1202/0.0300/0.1202 & 4.01/1.00/4.01 \\
WDM$^{\dagger}$\_4.00\_keV  a/b  & 0.0    & 4.00        & 3.0460      & 0.8150             &     25/100  & 256/512 &       0.0976/0.1953  &     3.25/6.51  \\
\hline
WDM\_1.00\_keV  a/b/c  & 0.0    & 1.00        & 3.0485    & 0.8150             &     25/100/100  & 256/512/832 &       0.0976/0.1953/0.1202  &     3.25/6.51/4.01  \\
WDM\_2.00\_keV  a/b  & 0.0    & 2.00        & 3.0469    & 0.8150             &     25/100  & 256/512 &       0.0976/0.1953  &     3.25/6.51  \\
WDM\_3.00\_keV  a/b  & 0.0    & 3.00        & 3.0466    & 0.8150             &     25/100  & 256/512 &       0.0976/0.1953  &     3.25/6.51  \\
WDM\_3.00\_keV Grid a/b/c  & 0.0    & 3.00        & 3.0466     & 0.8150             &     25/25/100  & 208/832/832 & 0.1202/0.0300/0.1202 & 4.01/1.00/4.01 \\
WDM\_4.00\_keV  a/b  & 0.0    & 4.00        & 3.0464    & 0.8150             &     25/100  & 256/512 &       0.0976/0.1953  &     3.25/6.51  \\
\hline
\hline
NU\_01\_WDM\_3.00\_keV  a/b/c  & 0.1    & 3.00        &  3.0466      & 0.8150             &     25/100/100  & 256/512/832 &       0.0976/0.1953/0.1202  &     3.25/6.51/4.01  \\
NU\_02\_WDM\_3.00\_keV  a/b  & 0.2    & 3.00        & 3.0466    & 0.8150             &     25/100  & 256/512 &       0.0976/0.1953  &     3.25/6.51  \\
NU\_03\_WDM\_3.00\_keV  a/b  & 0.3    & 3.00        & 3.0466    & 0.8150             &     25/100  & 256/512 &       0.0976/0.1953  &     3.25/6.51  \\
NU\_03\_WDM\_3.00\_keV Grid a/b/c  & 0.3    & 3.00        & 3.0466     & 0.8150             &     25/25/100  & 208/832/832 & 0.1202/0.0300/0.1202 & 4.01/1.00/4.01 \\
NU\_04\_WDM\_3.00\_keV  a/b  & 0.4    & 3.00        & 3.0466    & 0.8150             &     25/100  & 256/512 &       0.0976/0.1953  &     3.25/6.51  \\
\hline
NU\_03\_WDM\_1.00\_keV  a/b/c  & 0.3    & 1.00        &  3.0484      & 0.8150             &     25/100/100  & 256/512/832 &       0.0976/0.1953/0.1202  &     3.25/6.51/4.01  \\
NU\_03\_WDM\_2.00\_keV  a/b  & 0.3    & 2.00        & 3.0469    & 0.8150             &     25/100  & 256/512 &       0.0976/0.1953  &     3.25/6.51  \\
NU\_03\_WDM\_4.00\_keV  a/b  & 0.3    & 4.00        & 3.0464    & 0.8150             &     25/100  & 256/512 &       0.0976/0.1953  &     3.25/6.51  \\
\hline
\hline
\label{table_supporting_sims_wdm}
\end{tabular}
\end{center}
\end{table*}


Analogously to Table \ref{table_supporting_sims_dark_rad},
Table \ref{table_supporting_sims_wdm} lists all of the {\it supporting} simulations related to WDM and massive neutrinos, 
along with corresponding details such as resolution, box size, mean particle separation, and gravitational softening length.  
As explained in Section \ref{subsec_nu_dark_rad_wdm_modeling}, WDM 
is implemented using two different procedures, as indicated by different names/symbols in the table. 
Specifically,  in one implementation, WDM is included by  modifying the linear matter power spectrum to mimic the presence of early decoupled thermal relics;
the BG power spectrum is altered at the 2LPT level, by introducing a small-scale cutoff.
The cutoff can have different functional forms: we indicate with {\it ``WDM$^{*}$''} the form originally proposed by Bode, Ostriker \& Turok (2001), and with {\it ``WDM$^{\dagger}$''}
the updated version by Viel et al. (2005, 2012, 2013). The considered masses of thermal relics are specified explicitly using corresponding names, and expressed in keV. 
In this implementation, all the CDM is turned into WDM, which is valid only for a nonresonant DW production mechanism. 
Note again the direct correspondence between the mass of WDM early decoupled relics and the mass of a non-thermalized non-resonant
DW sterile neutrino  --while the resonant production is more complex.
An alternative implementation of   early decoupled thermal relics is indicated simply with {\it ``WDM''} in Table \ref{table_supporting_sims_wdm};
also in this case all the DM is warm, since we consider pure WDM particles with suitable candidates such as keV right-handed neutrinos or sterile neutrinos.
For this latter modeling, we introduce  a shift $\Delta N_{\rm eff}$ from the canonical value $N_{\rm eff}=3.046$ at the level of CAMB, and enforce a neutrino mass splitting if massive neutrinos are included. 
There is a direct correspondence with  a thermal relic $m_{\rm X} \equiv m_{\rm WDM}$ and the DW sterile neutrino of mass $m_{\rm s}$; hence,
we only need simulations with thermal relics $m_{\rm X}$, and then $m_{\rm s}$ of a massive sterile neutrino will be readily determined
via a direct mapping procedure using $\Delta N_{\rm eff}$, $h$, and $\Omega_{\rm c}$. 
This latter WDM implementation is also used in our {\it Grid Suite} to simulate WDM scenarios, since it is more physically motivated. To this end,   
note in particular that with this  implementation, the value of $N_{\rm eff}$ actually changes according to the addition of $\Delta N_{\rm eff}$, depending on the thermal relic mass. 
In all of these WDM runs, we consider five values of the relic mass, namely, $m_{\rm WDM} =0.25, 1.00, 2.00, 3.00, 4.00~{\rm keV}$, as well as a range of box sizes and resolutions --
as specified in Table \ref{table_supporting_sims_wdm}. 

The runs where massive neutrinos are also present in addition to WDM, indicated as {\it NU\_0\#\_WDM} with the label $0\#$ used to specify the total 
neutrino mass in eV (for example, `NU\_03' corresponds to
$\sum m_{\nu} =0.3~{\rm eV}$), are quite interesting and represent a novelty in the literature. Their implementation is nontrivial,  as it
requires two separate eigenstates and a mass splitting: namely, we model
three  degenerate massive neutrinos of total mass $\sum m_{\nu}$ in one eigenstate, and 
a non-thermalized massive sterile neutrino (interpreted as nonresonant WDM relic) in a different eigenstate. 
For these joint runs, we consider two scenarios: in the first one, we keep the WDM relic mass fixed to be $m_{\rm WDM} =3.00~{\rm keV}$
and vary the total neutrino mass as $\sum m_{\nu} =0.1, 0.2, 0.3, 0.4~{\rm eV}$, respectively. In the second one, we fix the total neutrino mass
to be $\sum m_{\nu} =0.3~{\rm eV}$ and allow $m_{\rm WDM}$ to vary as $1.00$, $2.00$, and $4.00~{\rm keV}$, respectively.
Note also that we adopt similar normalization conventions as in the massive neutrino/dark radiation \textit{Supporting Suite}. 
In total, we run  58 {\it supporting} simulations related to WDM and massive neutrinos (including BG runs), and their detailed analysis is
the subject of a companion publication.


\begin{figure*}
\centering
\includegraphics[angle=0,width=0.49\textwidth]{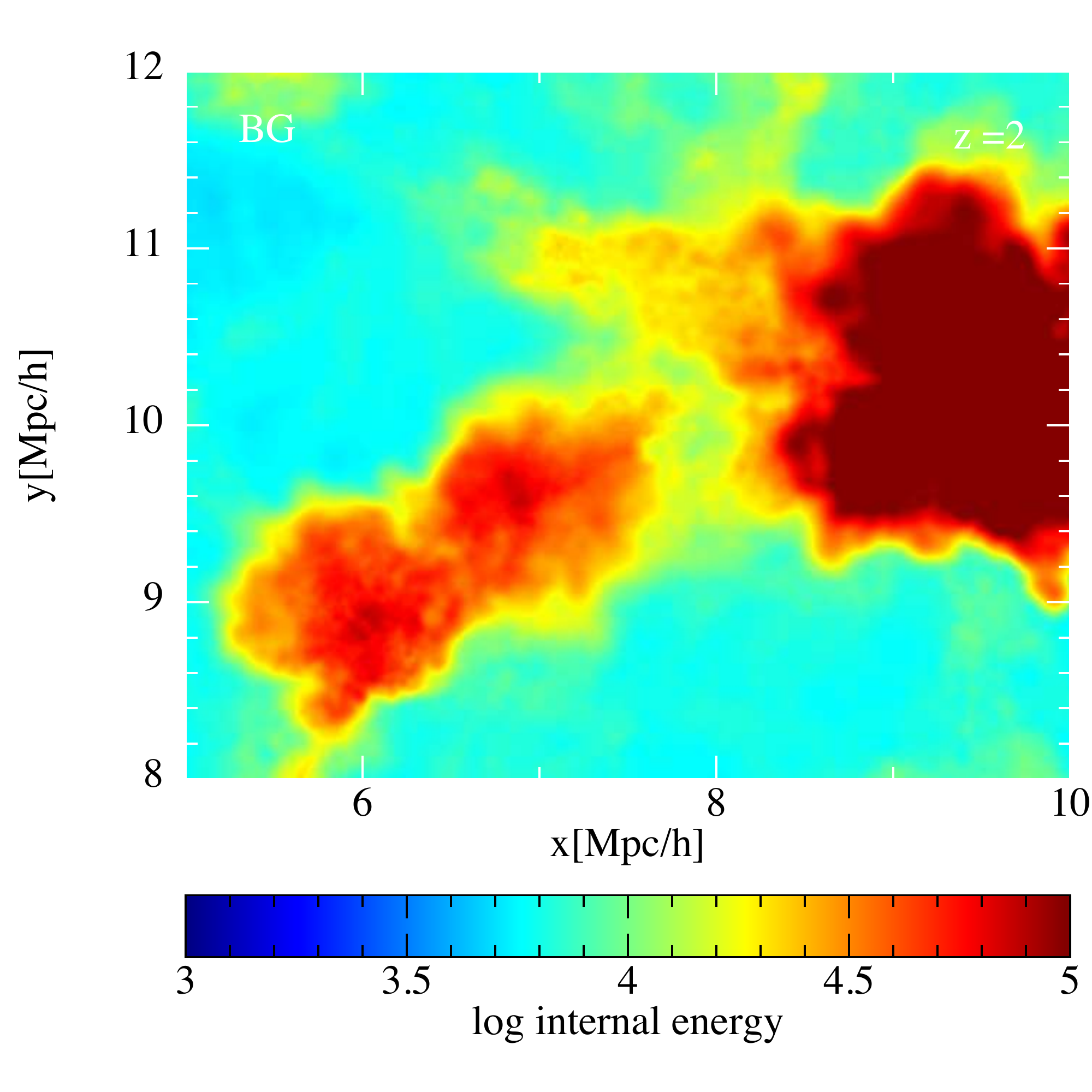}
\includegraphics[angle=0,width=0.49\textwidth]{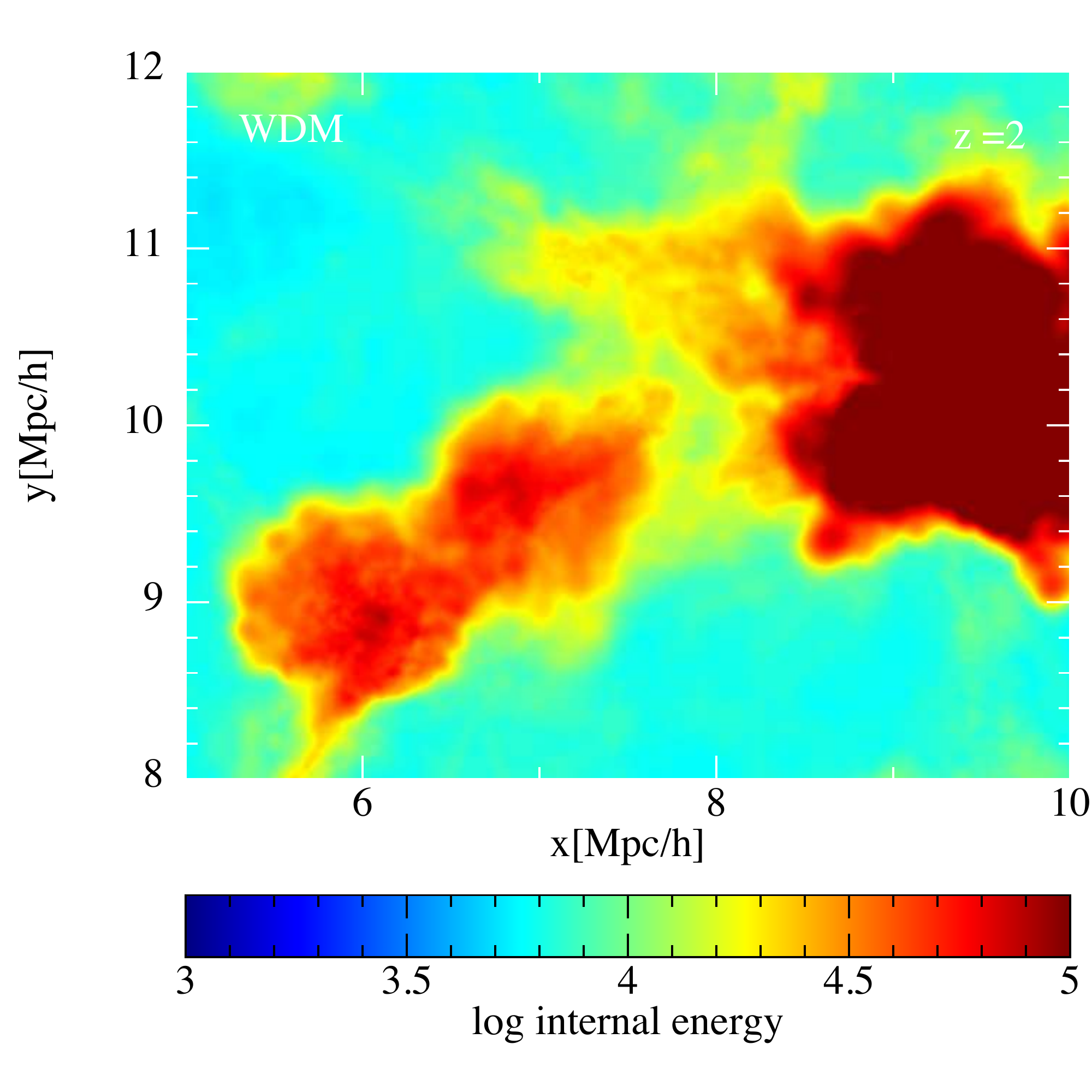}
\caption{Projected $(5.0 \times 6.0)$ $[h^{-1} {\rm Mpc}]^2$ patches at $z=2.0$ from $25h^{-1}$ Mpc simulations (Table \ref{table_supporting_sims_wdm}), when $N_{\rm p} = 256^3$  particles/type.
A complex structure is displayed, as seen in the gas internal energy of the \textit{``Best Guess"} reference model (left panel), and in a WDM cosmology when $m_{\rm WDM} = 2.00~{\rm keV}$ (right panel). 
Although differences are hardly perceptible, the   impact of an $m_{\rm WDM} = 2.00~{\rm keV}$ relic on the high-$z$ cosmic web is significant.}
\label{fig_wdm_visualization}
\end{figure*}


As a visual example drawn from Table \ref{table_supporting_sims_wdm}, we show in Figure \ref{fig_wdm_visualization} a projected $(5.0 \times 6.0)$ $[h^{-1} {\rm Mpc}]^2$  
patch from simulations having a box size of $25h^{-1}$ Mpc and a relatively low resolution of $256^3$ particles/type, describing the gas internal energy at $z=2.0$.
The left panel displays a complex structure as seen in the BG reference model, while the right panel highlights the same structure 
as seen in the WDM cosmology when $m_{\rm WDM} = 2.00~{\rm keV}$.  Once again, differences are tiny and therefore hard to be detected visually,
although the impact of an $m_{\rm WDM} = 2.00~{\rm keV}$ relic on the high-redshift LSS is significant. 
 
With  56 simulations describing  massive neutrino/dark radiation cosmologies and 
58 WDM/massive neutrino realizations, our {\it Supporting Suite} consists in total of 114 simulations (including BG reference runs)
and represents a valuable resource to study and accurately characterize the high-$z$ cosmic web including the {\it dark sector},
as well as baryonic effects at small scales.


\subsection{The \textit{Systematics Suite}} \label{subsec_systematics_suite}


\begin{table*}
\begin{center}
\tiny
\centering
\caption{Systematics Suite: List of Simulations.}
\doublerulesep 2.0pt
\renewcommand\arraystretch{1.5}
\begin{tabular}{cccccccc} 
\hline \hline
 Simulation Name &  $\sum m_{\nu}$ [eV]  & $N_{\rm eff}$ & $\sigma_8(z=0)$  & Boxes [Mpc/h] & $N_{\rm p}^{1/3}$ & Mean Par. Sep.  [Mpc/h] & Softening [kpc/h]  \\ 
\hline
Best Guess Grid a/b/c  & 0.00 & 3.0460 & 0.8150     &     25/25/100  & 192/768/768 & 0.1302/0.0325/0.1302 & 4.34/1.08/4.34 \\
Best Guess Grid a/b/c  & 0.00 & 3.0460 & 0.8150     &     25/25/100  & 224/896/896 & 0.1116/0.0279/0.1116 & 3.72/0.93/3.72  \\
\hline
BG\_CONV a  & 0.00 & 3.0460 & 0.8150     & 10 & 128 & 0.0781 & 2.60 \\
BG\_CONV b & 0.00 & 3.0460 & 0.8150     & 10 & 256 & 0.0391 & 1.30 \\
BG\_CONV c  & 0.00 & 3.0460 & 0.8150     & 10 & 336 & 0.0298 & 0.99 \\
BG\_CONV d  & 0.00 & 3.0460 & 0.8150     & 10 & 512 & 0.0195 & 0.65 \\
\hline
BG\_A$_{\rm s}^{+}$ a/b/c  & 0.00  & 3.0460 & 0.8684     &    25/25/100  & 208/832/832   & 0.1202/0.0300/0.1202 & 4.01/1.00/4.01\\
BG\_A$_{\rm s}^{-}$ a/b/c  & 0.00  & 3.0460 & 0.7907     &     25/25/100  & 208/832/832   & 0.1202/0.0300/0.1202 & 4.01/1.00/4.01\\
\hline 
BG\_FINER  & 0.00 &3.0460 & 0.8150     & 100/100  &  512/832  &  0.1953/0.1202 & 6.51/4.01 \\
BGM & 0.06 & 3.0460 & 0.8150     &   25 & 256 & 0.0976 &  3.25 \\
BGT\_UN & 0.01 & 3.0460 & 0.8150     &   25 & 256 & 0.0976 &  3.25 \\
NU\_SN\_01 a & 0.10 & 3.0460 & 0.8150     &   25 & 256/512 & 0.0976/0.0488 &  3.25/1.63 \\
NU\_SN\_01 b & 0.10 & 3.0460 & 0.8150     &   100 & 512/1024 & 0.1953/0.0976  &  6.51/3.25\\
DR\_SN\_01 a & 0.10 & 4.0460 & 0.8150     &   25 & 256/512 & 0.0976/0.0488 &  3.25/1.63 \\
BG\_SEED\_1  & 0.00 &3.0460 & 0.8150     & 100  &  832  &  0.1202 & 4.01 \\
BG\_SEED\_2  & 0.00 &3.0460 & 0.8150     & 100  &  832  &  0.1202 & 4.01 \\
BG\_SEED\_3  & 0.00 &3.0460 & 0.8150     & 100  &  832  &  0.1202 & 4.01 \\
BG\_SEED\_4  & 0.00 &3.0460 & 0.8150     & 100  &  832  &  0.1202 & 4.01 \\
BG\_SEED\_5  & 0.00 &3.0460 & 0.8150     & 100  &  832  &  0.1202 & 4.01 \\
BG\_PLANCK18  a/b/c & 0.00 &3.0460 & 0.8150     & 25/25/100  & 208/832/832 & 0.1202/0.0300/0.1202 & 4.01/1.00/4.01 \\
NU03\_z33 & 0.30 & 3.0460 & 0.8150    & 100  &  832  &  0.1202 & 4.01 \\ 
NU03\_z50 & 0.30 & 3.0460 & 0.8150    & 100  &  832  &  0.1202 & 4.01 \\
\hline
BG High Res & 0.00 &  3.0460 & 0.8150     &     100 & 1024 & 0.0976 & 3.25\\
NU\_01  High Res  & 0.10 & 3.0460 & 0.8150     &     100 & 1024 & 0.0976 & 3.25 \\
DR\_01 High Res  &  0.10 & 4.0460 & 0.8150     &     100 & 1024 & 0.0976 & 3.25 \\
\hline
\hline 
\label{table_systematics_sims_all}
\end{tabular}
\end{center}
\end{table*}


The third component of the \textit{Sejong Suite} is termed  {\it Systematics Suite}.
This collection of simulations has been performed with the main purpose of
studying and characterizing several systematic effects that
can impact parameter constraints, which are related to the modeling of additional species (i.e., massive neutrinos, dark radiation, WDM), 
or simply that are intrinsic to the numerical nature of our simulations.  
What we denote as `systematics' thus varies from convergence and resolution tests, to numerical artifacts, theoretical degeneracies,
and much more. We report here a first set of systematic runs performed, and 
we will expand  along these lines in dedicated forthcoming publications -- including full analyses. 
This first list of realizations serves in fact as a general reference and guideline, since the various systematic runs share similar
aspects that are common to all of the other simulations presented here. 
It will be then convenient for future work to expand around this framework and add more targeted simulations, while  
using the current setup as the standard `backbone' of the {\it Sejong Suite}.

Table \ref{table_systematics_sims_all} lists our first set of systematic runs,  addressing different modeling aspects as we briefly explain next.
Specifically, the `{\it Best Guess Grid}' simulations are constructed using the same parameters of the BG (which is part of the 
{\it Grid Suite}), but their resolution is different. In the first case, it is lower than that of the grid (with a maximum of $2 \times 768^3$ particles), while in the second case, it is higher
(for a maximum of $2 \times 896^3$ particles).  These runs allow one to explore the effect of  resolution on the various Ly$\alpha$ observables, while maintaining analogous 
box-size selections. 
The simulations termed `BG\_CONV' are small-box realizations used for convergence and resolution studies, as explained in the next Section \ref{subsec_resolution_requirements}.
The simulations denoted as  ``BG\_A$_{\rm s}^{+}$" and  ``BG\_A$_{\rm s}^{-}$" explore the effects of altering the overall normalization given by the initial amplitude of the primordial fluctuations $A_{\rm s}$,
which is well known to be degenerate with a variation in $\sigma_8$, which in turn can mimic a small neutrino mass.  
The realizations termed ``BG\_FINER"  are runs where the time-integration steps are forced to be much smaller than what is assumed for the grid simulations, and their overall performance requires almost twice 
the execution time requested for an equivalent simulation but with grid-like integration steps. 
These simulations allow one to test the impact of finer time-integration steps on quantities such as the nonlinear matter and flux power spectra, and ultimately the impact on parameter constraints. 
BGM is a run having a BG-like cosmology setup, but also containing a minimal neutrino mass of 
$0.06$ eV, which is essentially the baseline model adopted in the Planck 2015 and 2018 analyses. 
BGT\_UN is a control simulation containing a tiny neutrino mass of $\sum m_{\nu} =0.01$ eV, used to test the  consistency and convergence
of our neutrino implementation -- as done in Rossi et al. (2014). 
The simulations ``NU\_SN\_01"  and ``DR\_SN\_01"
have been developed to study shot-noise (``SN" in simulation names) effects in the presence of massive neutrinos 
and dark radiation (with an enhanced number of particles for neutrino modeling).
In detail,  ``NU\_SN\_01" contains a total neutrino mass of $\sum m_{\nu}=0.10$ eV and has twice as many particles 
for modeling the neutrino component as the corresponding gas and DM components ($512^3$ or $1024^3$ 
when the related DM and gas particles are $256^3$ and $512^3$, respectively).  
``DR\_SN\_01" is similar to the previous run, but with the addition of a thermalized massless sterile neutrino, so that $N_{\rm eff}=4.046$. 
The series of ``BG\_SEED\_\#" explores the effect of cosmic variance, as the initial random seeds of the simulations are varied with respect to the one used for grid simulations.
``BG\_PLANCK18" is a BG-like run, but with Planck 2018 best-fit parameters. 
``NU03\_z33" and ``NU03\_z50" are two realizations sharing the same massive neutrino cosmology with
$\sum m_{\nu}=0.3{~\rm eV}$, but differing in the initial redshift. The first realization starts
 at $z=33$, while the second one begins at $z=50$ --both with 2LPT.
Finally, simulations with {\it ``BG High Res"}, {\it ``NU\_01 High Res"}, and {\it ``DR\_01 High Res"} 
involve high-resolution runs, where the number of particles per species is increased to $1024^3$.
Specifically, BG has a BG-like setting, {\it NU\_01} has a $\sum m_{\nu} =0.10~{\rm eV}$ mass, and {\it DR\_01} has in addition a massless sterile neutrino, 
so that $N_{\rm eff} =4.046$.  These realizations are carried out to  evaluate the global computational cost, and to estimate the advantages/disadvantages of such settings.

Overall, we performed  35 runs for the first set of  {\it systematics simulations}. A detailed analysis of this interesting suite -- as well as additional  
systematic runs within this framework, such as effects of active galactic nucleus (AGN) feedback on massive neutrinos, and much more --will be presented in forthcoming dedicated studies. 
Characterizing systematic effects is in fact an essential 
component for obtaining unbiased parameter estimates, particularly when constraints are derived from small-scale measurements.     


\subsection{Resolution and Box-Size Requirements: Comparisons to Previous Studies} \label{subsec_resolution_requirements}


\begin{figure*}
\centering
\includegraphics[angle=0,width=0.85\textwidth]{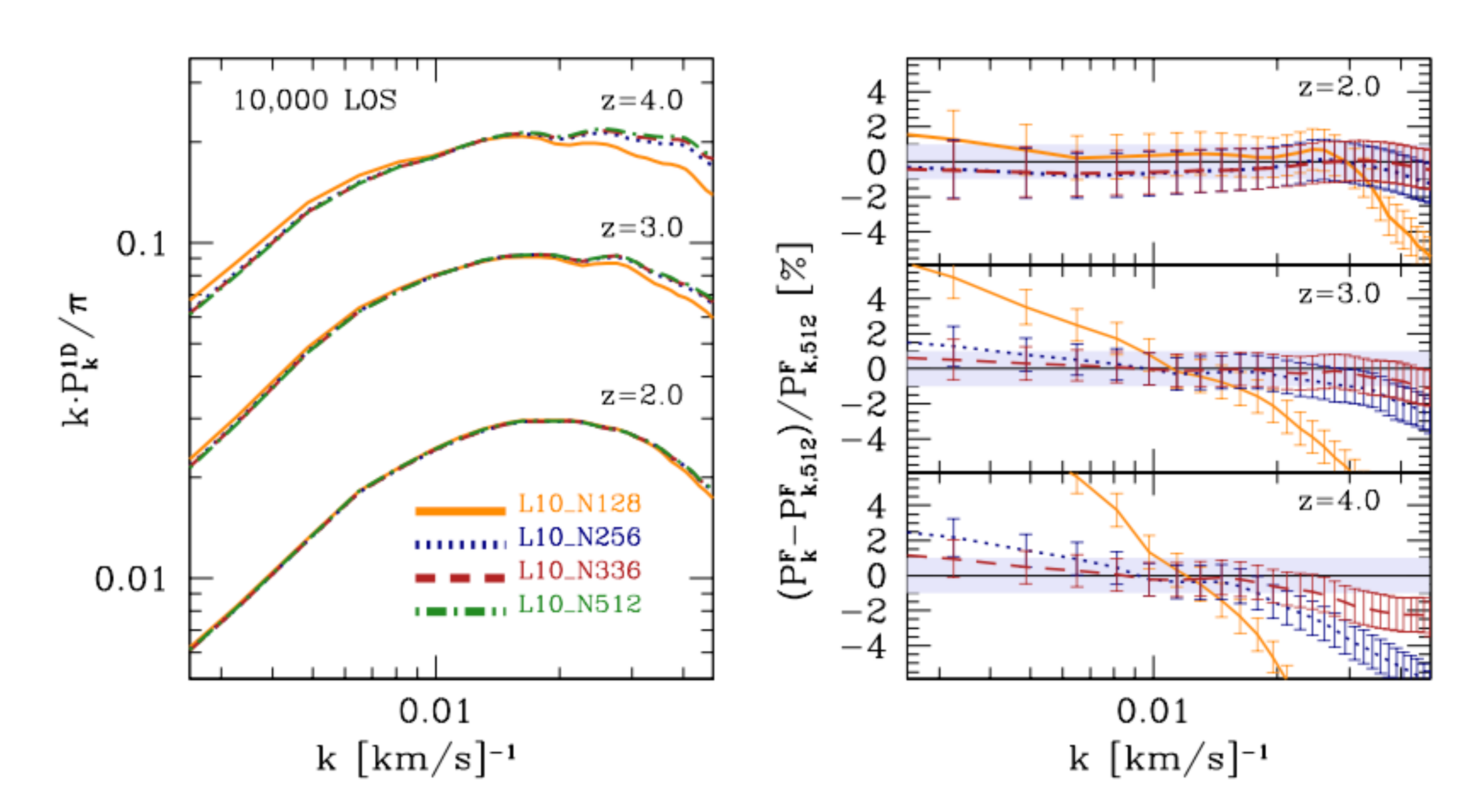}
\caption{Convergence study driving the design of the  {\it Sejong Grid Suite}, derived from 
 the $10h^{-1}{\rm Mpc}$ box simulations of the {\it Systematics Suite} (see Table \ref{table_systematics_sims_all}). 
[Left] Flux power spectra at $z=2, 3, 4$
computed from $10,000$ LOS skewers extracted at each targeted redshift from the realizations 
characterized by   $N^3=128^3, 256^3, 336^3, 512^3$, respectively, 
indicated with different line styles and colors in the panel. 
[Right] 1D flux power spectrum ratios, expressed in percentage, 
 with respect to the highest-resolution run (``L10\_N512", $20h^{-1}{\rm kpc}$ equivalent grid resolution). 
Error bars are 1$\sigma$ deviations computed from   $10,000$ skewers at each $z$,
and the gray areas highlight the $1\%$ level.
A  $30h^{-1}{\rm kpc}$ equivalent grid resolution is sufficient for achieving $\sim 1\%$ convergence on flux statistics 
in the $k$-range covered by eBOSS, and it also guarantees a comparable 
degree of accuracy for the expected extension of DESI Ly$\alpha$ forest data. }
\label{fig_convergence}
\end{figure*}


Achieving numerical convergence in the modeling of the Ly$\alpha$ forest is 
challenging, especially because most of the signal comes from poorly resolved under-dense regions --not necessarily 
 in local hydrostatic equilibrium.
Two primary factors are driving the resolution level that simulations need to reach, for an accurate reproduction of the forest:
the experimentally attainable Ly$\alpha$ forest range,  and the measurement errors on the targeted  Ly$\alpha$ forest  observables. 
Clearly,  convergence requirements will always depend on the physical process under consideration, as well as on the precision of the 
observational data with which the simulations are confronted. Moreover, 
one needs to find an optimal compromise between the simulation  box size, the total number of particles used in the runs, 
and the overall CPU consumption.
In particular, assuming the power spectrum as the main targeted observable, 
the largest $k$-mode achievable ($k_{\rm max}$) is bounded by the Nyquist-Shannon limit --i.e., 
$k_{\rm Nyquist} = \pi/ \Delta v$, where $\Delta v$ is the pixel width of the measured spectra by a given survey.
The smallest  $k$-mode ($k_{\rm min}$) is instead driven, at least in principle, by the overall extension of the Ly$\alpha$ forest.
In reality, instrumental constraints make it difficult to reach $k_{\rm Nyquist}$, 
and in addition, one needs  a sufficiently large box size $L$ to prevent missing modes. 
Numerically, $L$ determines 
the smallest $k$-mode (i.e. $k_{\rm min} = 2 \pi / L$), while
the largest $k$-mode depends on the 
specific computational algorithm adopted. 
In fact, within the SPH formalism,  
over-densities (in local hydrostatic equilibrium, thus bounded by the Jeans scale) 
are sampled with higher spatial resolution than average --
with a particle spacing significantly smaller 
than $L/N$, where  $N^3$ is the total number of particles per species. 
This is not always the case for under-dense regions, often out of local hydrostatic equilibrium, 
for which convergence tests ensure that simulations are able to resolve the smallest structures
in the transverse direction -- useful for an accurate estimate of the 1D flux power spectrum. 

To this end,  we have extensively addressed convergence and resolution  requirements within our SPH formalism,
in order to accurately reproduce the main Ly$\alpha$ forest observables
as demanded by state-of-the-art redshift surveys.  
Based on convergence studies, 
our conclusions in terms of resolution requirements
are challenging, pointing toward computationally expensive runs. 

For instance, in previous releases (see, e.g., Rossi et al. 2014), 
we have developed realizations with an 
equivalent resolution of $N^3=3072^3$ particles per species over   a $100 h^{-1}{\rm Mpc}$ box 
(corresponding to a mean grid resolution of $\sim 33 h^{-1}{\rm kpc}$), 
a choice that guarantees a power spectrum convergence within  $\sim 2.0\%$  
 for  every redshift in the range $2.2 \le  z \le 4.4$ and in the $k$-range 
$1 \times 10^{-3} \le k [{\rm km/s}]^{-1}  \le 2 \times 10^{-2} $ 
(or approximately $0.1 \le k {\rm [{\it h}Mpc^{-1}]}  \le 2.0$ at $z \sim 3$).
The previous $k$-range  represents the optimal
coverage provided by the  SDSS-III/BOSS and SDSS-IV/eBOSS 
surveys. This is obtained by assuming 
a constant pixel width of $\Delta v = 69 {\rm [km/s]^{-1}}$ common for SDSS quasar coadded spectra, 
and  by considering instrumental constraints and 
additional observational limitations -- so that, within a given Ly$\alpha$ forest, the redshift span 
is $\Delta z =0.2$.  
Regarding box size, $100h^{-1}{\rm Mpc}$ represents an optimal configuration choice
to safely reach  $k_{\rm min} = 1.0 \times 10^{-3} {\rm [km/s]^{-1}}$. 

Having eBOSS and DESI Ly$\alpha$ forest data as our primary target, as well as 
high-resolution spectra such as those provided by the 
VLT/XSHOOTER
legacy survey (XQ-100; L{\'o}pez et al. 2016),
with the  {\it Sejong Grid Suite} we have improved 
the effective resolution of the simulations. 
In fact, although the Ly$\alpha$ forest range covered by eBOSS is 
similar to that of BOSS 
(with an identical
pixel width of $\Delta v = 69 {\rm [km/s]^{-1}}$) but spans over 35 modes from 
$k = 10^{-3}{\rm [km/s]^{-1}}$ to $k = 0.02{\rm [km/s]^{-1}}$ 
beyond which the SDSS spectrograph resolution cuts the power by a factor of 10,
better estimations of the SDSS pipeline noise and a more statistically significant sample of high-$z$  quasars
have reduced the  measurement errors on Ly$\alpha$ forest observables. 
Therefore, an increased mass resolution in simulations is necessary to match the
quality of these new data. Moreover, DESI is pushing further the observable Ly$\alpha$
forest extension, and it is expected to provide data 
with almost twice the spectral resolution and a higher signal-to-noise ratio for quasars at $z > 2.1$ 
compared to eBOSS.

In this respect, a minimal requirement to 
meet current and upcoming high-quality Ly$\alpha$ forest data  is being
able to cover at least $100h^{-1}{\rm Mpc}$ with an 
equivalent mean grid resolution of $30 h^{-1}{\rm kpc}$. 
This conclusion is 
supported by the  convergence study 
for Ly$\alpha$ flux statistics 
shown in Figure \ref{fig_convergence}, 
based on the  simulations
indicated as `BG\_CONV' in Table \ref{table_systematics_sims_all} from the {\it Systematics Suite}.
These realizations are characterized by the same cosmology of the reference BG, 
but they cover  only  a small periodic box of $10h^{-1}{\rm Mpc}$.
Their mass resolution varies as $N^3=128^3$, $256^3$, $336^3$, and $512^3$ -- comparable to an
equivalent grid resolution of $78, 39, 30, 20$ $h^{-1}{\rm kpc}$,  respectively.
The corresponding DM particles at those resolutions are 
 $3.435 \times 10^7$, $4.294 \times 10^6$, $1.899 \times 10^6$, and $5.367 \times 10^5$ $h^{-1}M_{\odot}$;
 and, in terms of gas mass particles, we have 
 $6.409 \times 10^6$, $8.011 \times 10^5$, $3.543 \times 10^5$, and $1.001 \times 10^5$ $h^{-1}M_{\odot}$. 
The left panel of Figure \ref{fig_convergence}  shows  flux power spectra at $z=2, 3, 4$
computed from $10,000$ LOS skewers extracted  from all of the 10$h^{-1}{\rm Mpc}$ runs at each targeted redshift.
Different line styles and colors refer to the varying mass resolutions, as specified in the plot. 
The right panels  display 
1D flux power spectrum ratios for realizations having $N^3=128^3, 256^3, 336^3$, respectively, estimated
with respect to the highest-resolution run (i.e. ``L10\_N512", having an equivalent grid resolution
of $20h^{-1}{\rm kpc}$), expressed in percentage, and with identical colors and line styles as in the left panel. 
Error bars are 1-$\sigma$ deviations computed from $10,000$ simulated skewers randomly extracted at each  redshift from the simulation boxes,
and the gray areas highlight the $1\%$ level.
As can be clearly seen, 1D flux power spectra obtained from 
the $N^3=336^3$ realization   (i.e. $30h^{-1}{\rm kpc}$ equivalent grid resolution) are within $\sim 1\%$  
in the $k$-range covered by eBOSS (up to $k=2 \times 10^{-2} [{\rm km/s}]^{-1}$), and closer to the 
same degree of accuracy for the expected extension of DESI Ly$\alpha$ forest data.
Clearly, achieving sub-percentage convergence at
higher-$z$ is progressively challenging, but a grid of $30h^{-1}{\rm kpc}$ is sufficient for a 
satisfying convergence even at those redshifts. 
 
The previous rationale is precisely what has driven the overall  architecture of the {\it Sejong Grid Suite}, 
designed to achieve an equivalent resolution 
up to $3 \times 3328^3$ = 110 billion particles in a $(100h^{-1}{\rm Mpc})$ box, corresponding to
 a $30 h^{-1}{\rm kpc}$ mean grid resolution that ensures a convergence on Ly$\alpha$ flux statistics
 closer to the desired $\sim 1.0\%$ level that the final DESI data will provide. 

Clearly, running a very large number of 
simulations with  $2 \times 3328^3$ or $3 \times 3328^3$ elements over 
a $(100h^{-1}{\rm Mpc})$ box is still computationally challenging, particularly when neutrinos 
are included as particles. 
The global computational cost in performing an entire suite at such resolution
would easily require $\sim 100$ million CPU hours. 
Therefore, as in our previous release, we adopt a
splicing technique.
In this respect, the three simulations per fixed parameter set of the {\it Sejong Grid Suite} should be regarded
as one single realization having the equivalent mean grid resolution of $30 h^{-1}{\rm kpc}$ over 
a $100h^{-1}{\rm Mpc}$ box size. 
To this end, we note that using simulations from Rossi et al. (2014) with lower resolution than the {\it Sejong Grid Suite}, 
along with splicing  (and despite the limitations intrinsic to the splicing method), we 
were able to provide among 
the most competitive neutrino mass and dark radiation bounds
to date, by combining 
Ly$\alpha$ forest data with additional tracers
and adopting a Taylor expansion approach for the Ly$\alpha$ forest flux.  
With this new release, other than several  
technical improvements, we 
have increased the resolution to achieve a better modeling of the  
1D flux power spectrum at high-$z$ (see Figure \ref{fig_1d_flux_ps_data_16D_paper}),
and therefore, we expect tighter upper bounds on the summed neutrino mass along with more stringent limits 
on dark radiation. 

In this context, a very interesting and meticulous study on resolution requirements
for the modeling of the Ly$\alpha$ forest has been carried out by Lukic et al. (2015). 
Their conclusions are also quite stringent, and in particular their findings point to a grid resolution
of $20h^{-1}{\rm kpc}$ to reproduce $1\%$ convergence on the Ly$\alpha$ flux statistics up to $k=10 {\rm {\it h}Mpc^{-1}} $,
and a box size greater than  $40h^{-1}{\rm Mpc}$ to suppress numerical errors to a sub-percent level and 
overcome missing modes. 
 
A direct comparison with Lukic et al. (2015)  is not straightforward, since 
their results are obtained with an Eulerian hydrodynamical code,
while our convergence findings are based on SPH-Lagrangian techniques.
In fact, within the SPH formalism, the gravitational resolution is much higher for the same grid ($L/N$),
providing $\sim 10$ times higher gravitational resolution than the grid codes for an identical grid configuration,
while  hydrodynamical quantities are smoothed on scales of $\sim 2$ times the mean interparticle spacing for gas around the mean density.
Nevertheless, our requirements are consistent with their prescriptions. 
The simulations of the {\it Grid Suite} -- useful for parameter constraints  via a Taylor expansion model for the flux  -- 
should in fact 
be seen as realizations with  $N^3=3328^3$ particles/type over a 100 $h^{-1}{\rm Mpc}$ box, thus spanning a
sufficiently large size to
guarantee an accurate reproduction of the Ly$\alpha$ flux statistics -- in line with their suggested safe choice of $80 h^{-1}{\rm Mpc}$.  
In terms of mass resolution, the  {\it Grid Suite} 
is designed to reach an 
equivalent mean grid resolution of $30h^{-1}{\rm kpc}$.
While this effective grid resolution is larger than the one proposed by Lukic et al. (2015), 
eBOSS  can only map $k$-modes up to  $2 {\rm {\it h}Mpc^{-1}} $ and DESI will expand the range (up to $ k \sim 3 {\rm {\it h}Mpc^{-1}} $)
but certainly will not reach  $k=10{\rm {\it h}Mpc^{-1}} $. 
Therefore, our effective resolution 
is able to guarantee convergence on Ly$\alpha$ flux statistics approaching the desired $\sim 1\%$ level within the
targeted $k$-range for those surveys. 
Note also that our  $10 h^{-1}{\rm Mpc}$ run with $N^3=512^3$ used for convergence studies (see Table \ref{table_systematics_sims_all} and Figure \ref{fig_convergence}) is
comparable to their realization termed ``L10\_N512", which, according to their results, looks virtually identical to the ``L10\_N1024" simulation
in terms of flux values, and agrees to better than $1\%$ even beyond $k=0.1{\rm [km/s]^{-1}}$ in terms of flux statistics.  

Clearly, the constraining power of the Taylor-based approach for modeling the 
Ly$\alpha$ flux is ultimately limited by the accuracy of splicing.
To this end, Lukic et al. (2015) have reported a $5-10\%$ precision
level related to this technique, in terms of 1D flux power spectrum. 
Even in this case, a comparison with their results 
is not direct. Nevertheless, in our studies we have 
found the method to be more accurate, 
and  in addition at the level of parameter constraints, the 
residual biases introduced by splicing are corrected via nuisance parameters,
and eventually marginalized over.   

Finally, we note that the stringent resolution requirements discussed in this section are primarily relevant for the targeted science 
associated with the {\it Grid Suite}, while the {\it Supporting} and {\it Systematics Suites} are
mostly developed for relative comparisons and for the study of systematics, where having high resolution is 
not essential. 
With less memory-intensive and more optimized 
codes becoming available, running larger-volume
high-resolution hydrodynamical simulations
will be  progressively  less 
computationally prohibitive. 
Hence, either performing a large number of runs at the challenging resolution previously discussed, or
abandoning splicing and interpolation
techniques and adopting emulator-based strategies
will soon be feasible. We aim at extending the {\it Sejong Suite} in this direction, and leave  
the task to future releases. 


\subsection{Summary of Available Products} \label{subsec_list_available_products}

Schematically, all of the products available from the entire {\it Sejong Suite} and illustrated in this section are listed below, for   
ease of accessibility.

\begin{itemize}
\item Full simulation snapshots (boxes) in Gadget format (type 2), containing all of the species considered (gas, DM, neutrinos whenever present), from $z=5.0$ to $z=2.0$ in intervals of $\Delta z = 0.2$, for all of the models 
reported in Tables \ref{table_grid_params_variations}-\ref{table_systematics_sims_all}. 
\item Initial conditions (2LPT) for all the runs. 
\item Initial CAMB total matter power spectra and transfer functions per components for all the runs.
\item Nonlinear matter power spectra (total and individual components)  from $z=5.0$ to $z=2.0$, in intervals of $\Delta z = 0.2$, for all of the models 
described in Tables \ref{table_grid_params_variations}-\ref{table_systematics_sims_all}. 
\item Ly$\alpha$ forest skewers/mocks from $z=5.0$ to to $z=2.0$ in intervals of $\Delta z = 0.2$ for all of the models considered,  
containing information about flux density, column density, gas temperature, velocity, optical depth, and pixel position. At each $z$-interval, we
provide 100,000 skewers randomly selected from the simulated boxes for any given cosmology (see Section \ref{subsec_postprocessing} for details). 
\item Particle samples from $z=5.0$ to $z=2.0$ in intervals of $\Delta z = 0.2$ for all of the models considered. For each $z$-interval and selected model, we
provide subsets of 100,000 particles. 
\end{itemize}



\section{The {\it Sejong Suite}: First Results}  \label{sec_sejong_suite_first_results}



\begin{figure*}
\centering
\includegraphics[angle=0,width=0.95\textwidth]{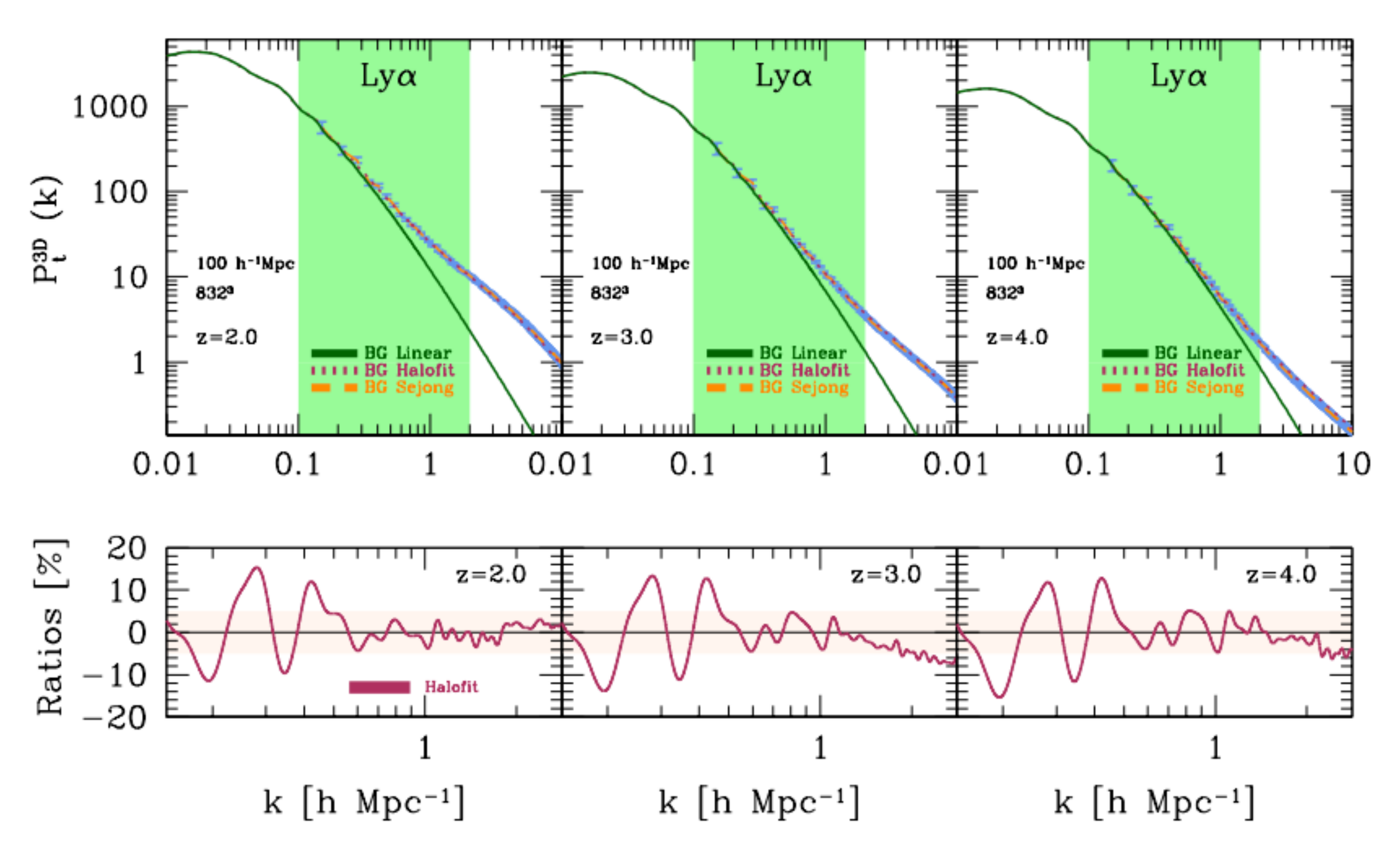}
\caption{Total matter power spectra $P^{\rm 3D}_{\rm t}$ from the \textit{`Best Guess'} model -- the baseline cosmology of the \textit{Sejong Suite}. 
[Top panels] Linear theory expectations for $P^{\rm 3D}_{\rm t}$ (solid lines), nonlinear halo model predictions from  \textit{Halofit} (dotted lines), and 
measurements from simulated \textit{grid}  snapshots with $832^3$ particles/type over a 
$100h^{-1}{\rm Mpc}$ box size (dashed lines). 
From left to right, the tomographic evolution of the $P^{\rm 3D}_{\rm t}$ shape and amplitude is shown at $z=2.0,3.0,4.0$, respectively, as indicated in the panels.
The  green areas refer to the $k$-regime relevant for the
Ly$\alpha$ forest, where clearly nonlinear effects are important. [Bottom panels]  
Ratios between measurements derived from our high-resolution hydrodynamical simulations and corresponding
\textit{Halofit} predicted total matter power spectra, expressed in percentages. 
Redshift intervals are the same as in the top panels. Not surprisingly, standard halo model predictions 
neglecting baryonic effects depart from  hydrodynamical  measurements significantly, up to $\sim 10-15\%$.}
\label{fig_3d_ps_A}
\end{figure*}

\begin{figure}
\centering
\includegraphics[angle=0,width=0.46\textwidth]{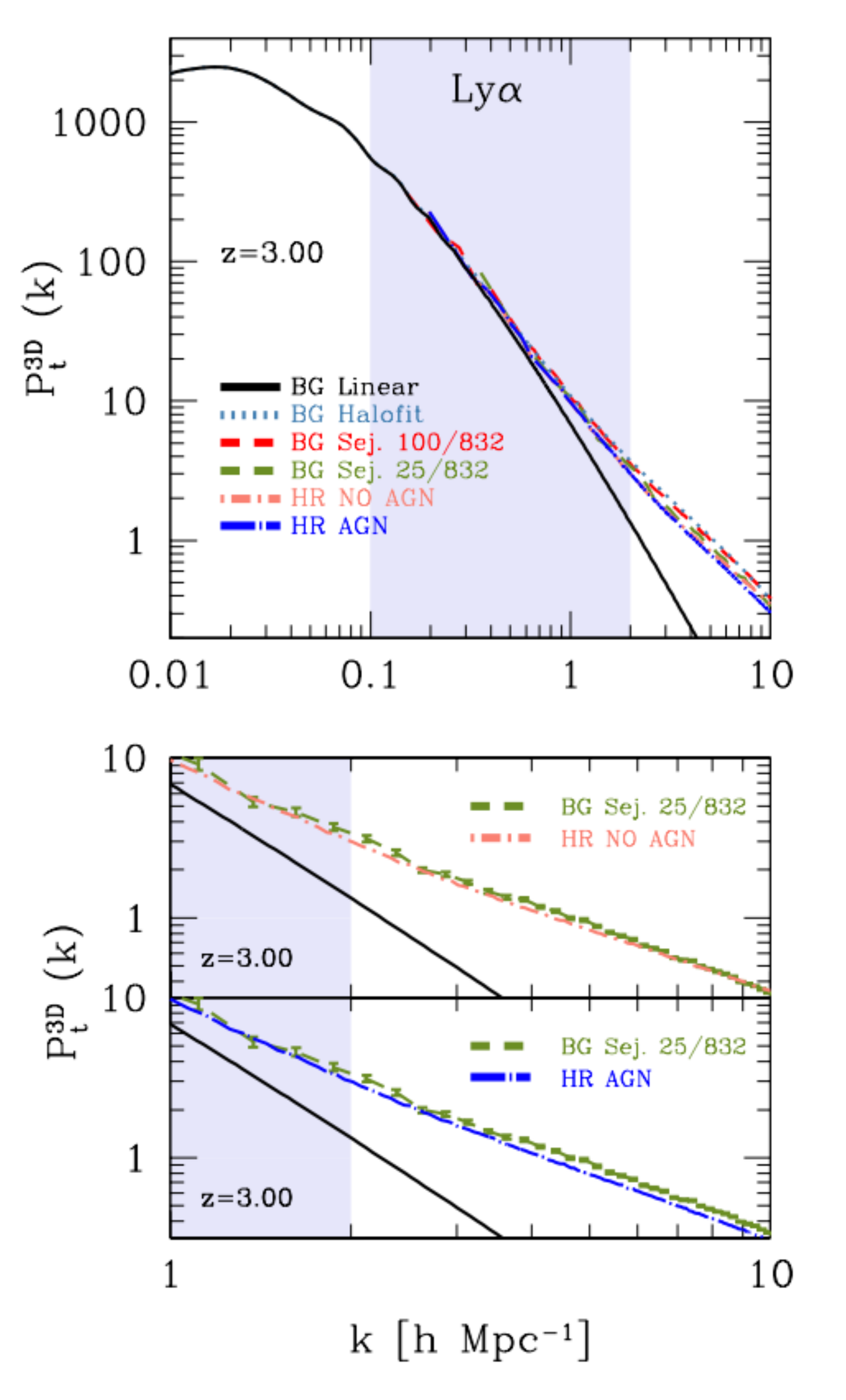}
\caption{Comparison between the {\it `Best Guess'} Sejong Suite and the Horizon simulations (Dubois et al. 2014, 2016), 
in terms of total matter power spectrum at $z=3$. [Top panel]
Total matter power spectrum from the Sejong BG 100/832 run (red dashed line),  
the Sejong BG 25/832 run (green dashed line), the 
 Horizon run containing full baryonic physics (``HR AGN"; blue long-dashed-dotted line), and
 the Horizon run lacking AGN physics (``HR NO AGN"; pink short-dashed-dotted line).
The corresponding linear and Halofit-based predictions are also shown, with solid and dotted lines, 
respectively. 
[Bottom panels] Zoom into the 
$k$-range $1 \le k {\rm {[} {\it h} \rm Mpc^{-1}]} \le 10$,
showing that the ``HR NO AGN" realization appears to be in reasonably good agreement (within error bars) with the
Sejong 25/832 simulation,  
while the  ``HR AGN" total matter power spectrum already shows deviations 
and signs of suppression at 
$z=3$ due to AGN feedback.}
\label{fig_3d_ps_A_bis}
\end{figure}


In this section we present a first analysis of the \textit{Sejong Suite}, 
mainly focused on selected topics related to the matter and flux statistics. 
In particular,  we 
show that we are able to accurately reproduce the 1D flux power spectrum down to scales $k=0.06~{\rm [km/s]^{-1}}$
as mapped by recent high-resolution quasar data, as well as the thermal history of the IGM.
The small-scale 1D flux power spectrum is  a key quantity for constructing the Ly$\alpha$ forest likelihood, and an excellent probe for 
detecting neutrino and dark radiation imprints. Moreover, 
as shown in  Rossi (2017), the IGM at $z \sim 3$ provides the best sensitivity to active and sterile neutrinos.
While we discuss here only a few scientific applications, 
we anticipate detailed studies based on the \textit{Sejong Suite} in forthcoming publications -- addressing a variety of aspects, 
particularly in relation to the {\it dark sector}. 


\subsection{Matter Power Spectrum: Shape and Tomography} \label{subsec_3d_matter_ps}


\begin{figure*}
\centering
\includegraphics[angle=0,width=0.85\textwidth]{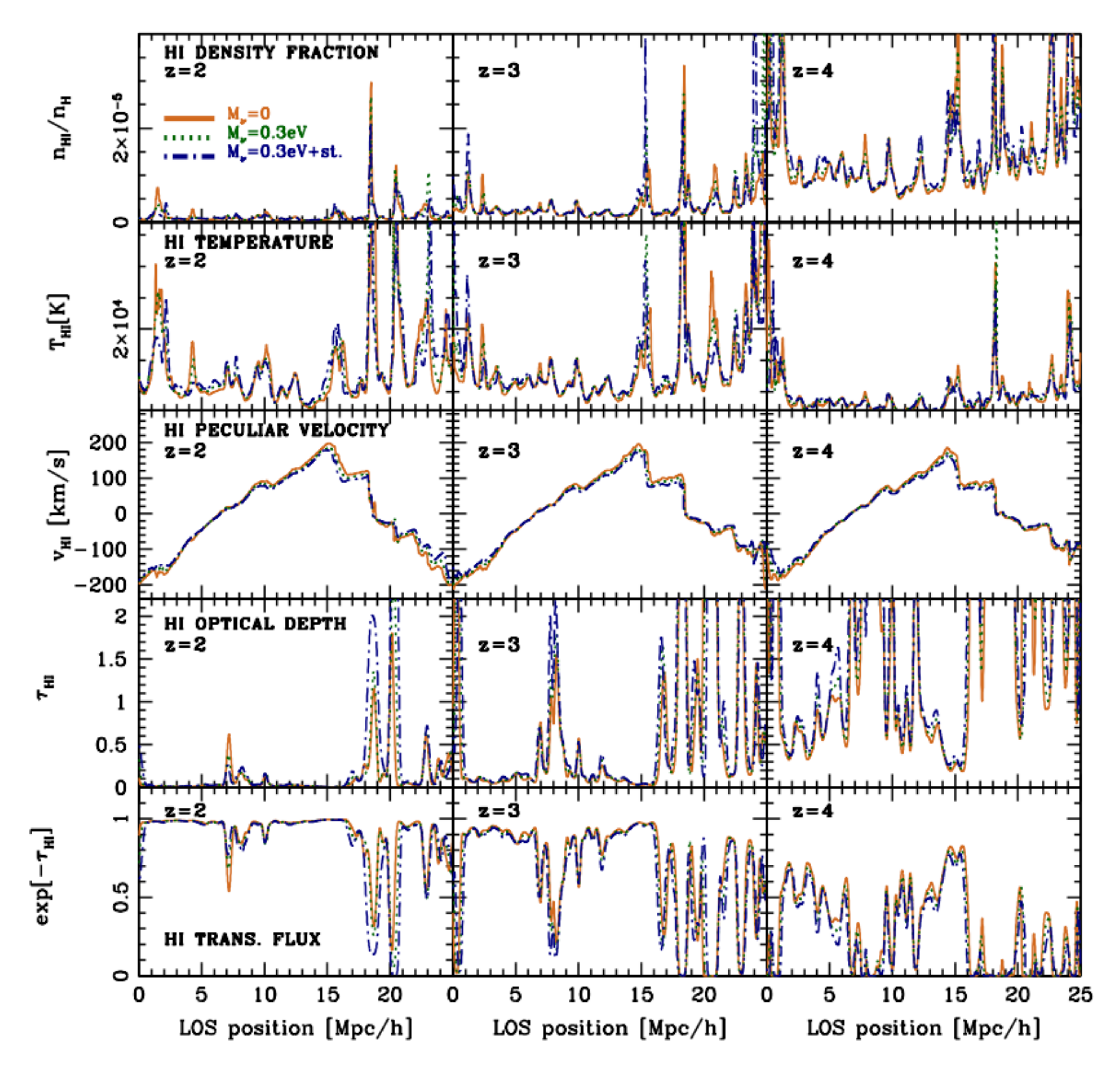}
\caption{HI main properties, as inferred from Ly$\alpha$ skewers extracted from $25~h^{-1}{\rm Mpc}$ box simulations belonging 
to the \textit{Supporting Grid} (Table \ref{table_supporting_sims_dark_rad}). Different redshift slices are considered (i.e., $z=2.0,~3.0,~4.0$, respectively), as indicated in the various panels. 
From top to bottom, the physical quantities displayed
are the HI density fraction, the HI temperature (in kelvins),  the HI peculiar velocity (in km/s), the HI optical depth, and the HI transmitted flux. 
Skewers having the same random seeds are drawn for three different cosmologies: the {\it `Best Guess'} (solid orange  lines), 
a massive neutrino scenario with $\sum m_{\nu} = 0.3~{\rm eV}$ (dotted green lines), and a dark radiation model including three massive neutrinos, 
so that $N_{\rm eff} = 4.046$ (dashed-dotted blue lines). The normalization convention is such that the values of $\sigma_8$ in the various models are all different at $z=0$;
i.e., runs termed  ``UN" in Table \ref{table_supporting_sims_dark_rad}. Variations in the HI physical properties among the three  distinctive
cosmologies are easily perceptible.}
\label{fig_LyA_observables_A}
\end{figure*}

\begin{figure*}
\epsscale{1.0}
\centering
\includegraphics[angle=0,width=0.70\textwidth]{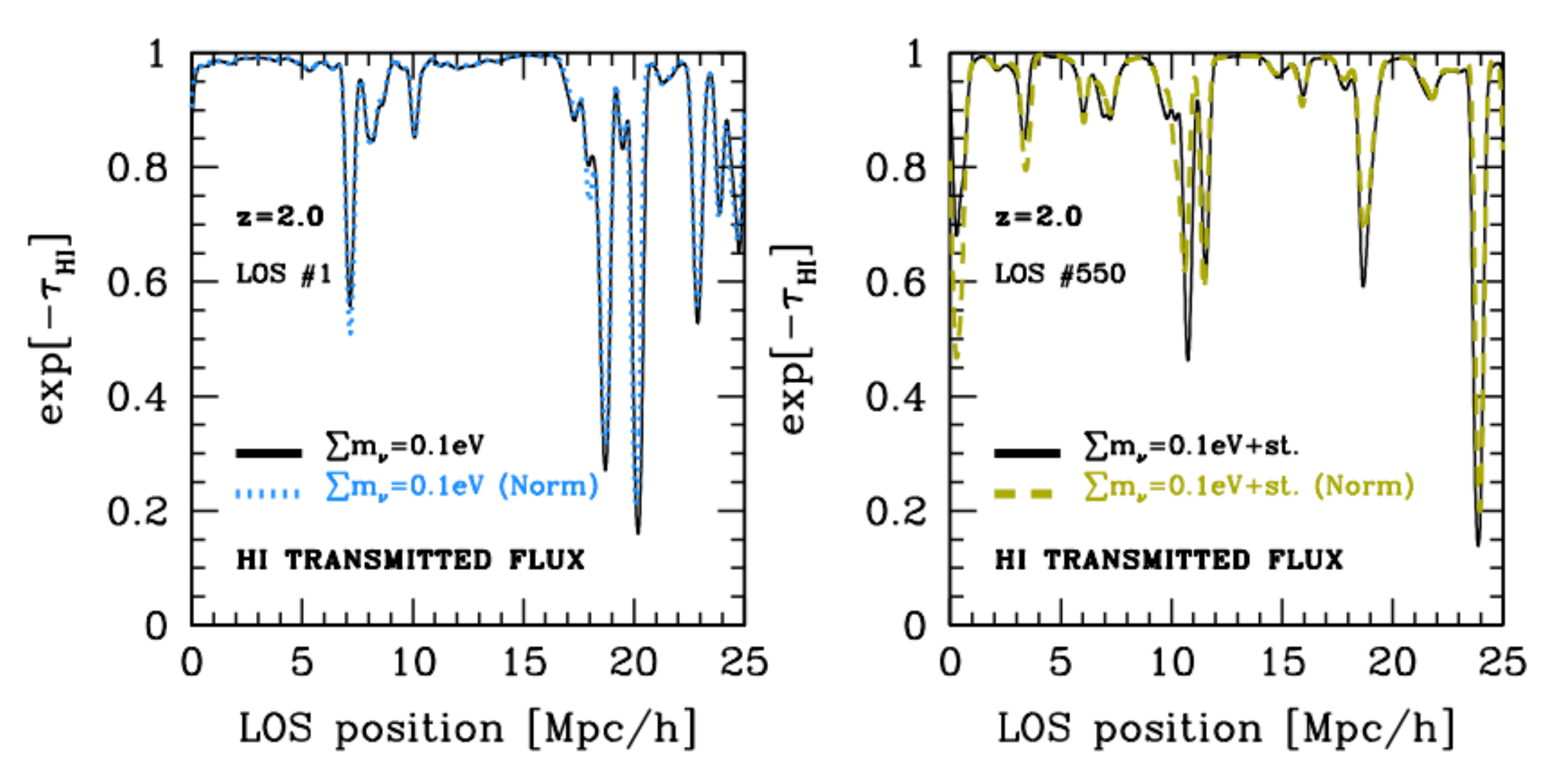} 
\caption{Normalization conventions: effects
on the HI transmitted flux. Two skewers at $z=2.0$ -- with identical random seeds -- are 
drawn from a massive neutrino cosmology where $\sum m_{\nu} =0.3$eV (left panel), and in a 
dark radiation model including three massive neutrinos
so that $N_{\rm eff} = 4.046$ (right panel). 
Solid black lines display the ``UN'' normalization convention, where $A_{\rm s}$ is fixed as in the \textit{``Best Guess"} cosmology, 
and the values of $\sigma_8$ are different at $z=0$. Dotted or dashed  colored lines are used for the `NORM' convention,
where $\sigma_8(z=0)$ is consistent with the Planck 2015 central value. 
Differences are easily perceptible, and they will eventually show up in the main Ly$\alpha$ forest observables.}
\label{fig_LyA_observables_B} 
\end{figure*}

  
The 3D total matter power spectrum ($P^{\rm 3D}_{\rm t}$) represents a key cosmological observable, among the main targets of large-volume galaxy surveys.
Its overall shape and amplitude  across a variety of $k$-ranges and redshift intervals (i.e., tomography)
have been widely used in the literature, especially for obtaining
cosmological parameter constraints.  To this end, Figure \ref{fig_3d_ps_A} shows examples of $P^{\rm 3D}_{\rm t}$ as predicted or derived from
the BG model -- the baseline cosmology of the \textit{Sejong Suite}. 
Specifically, in the top panels of the figure, solid lines display linear theory expectations,
dashed lines with error bars are measurements from simulated \textit{grid}  snapshots having $832^3$ particles/type over a 
$100h^{-1}{\rm Mpc}$ box size (denoted as 100/832), and dotted lines are nonlinear halo model predictions obtained with \textit{Halofit} (Takahashi et al. 2012). 
The tomographic evolution spanning $z=2.0$, $3.0$, $4.0$ is shown from left to right, respectively, as indicated in the plot.
The $1\sigma$ error bars displayed are related to the hydrodynamical measurements; they are 
estimated by considering the expected statistical (Gaussian) error consisting of sample variance and a Poisson shot-noise contribution (see, e.g., Schneider et al. 2016).
In addition, the highlighted green areas refer to the $k$-regime relevant for the
Ly$\alpha$ forest as mapped, for example, by eBOSS.
Departures from linear theory in that regime are clearly significant, and they cannot be neglected. Indeed, 
the DM nonlinear clustering as well as its tomographic evolution plays
a key role in shaping $P^{\rm 3D}_{\rm t}$.  
The bottom panels of the same figure show ratios between  
measurements derived from the high-resolution hydrodynamical simulations and
corresponding power spectra 
predicted by \textit{Halofit} --
indicated with solid lines, and expressed in percentages.
The extended horizontal colored areas in the plot highlight the $5\%$ scatter level.
Not surprisingly, standard halo model predictions neglecting baryonic effects are 
generally inaccurate, showing deviations  up to {\bf $\sim 10-15\%$}, especially 
in the Ly$\alpha$ region of interest, 
for all the redshift intervals considered. 
Hence, our  multicomponent high-resolution realizations  are essential -- particularly in the Ly$\alpha$ regime, where 
nonlinear effects and baryonic physics play a crucial role. 

Modeling the impact of nonlinear late-time baryonic physics is in fact essential for a percent-level
knowledge of the total matter power spectrum at scales $k \le 1 {\rm {\it h} Mpc^{-1}}$, 
which are particularly relevant for Ly$\alpha$ forest and weak lensing surveys. 
To this end, a number of works in the literature have 
addressed the topic theoretically or via hydrodynamical simulations, and    
several groups have quantified the impact of 
baryons on the matter power spectrum from their runs reporting different results (see, in particular, 
Chisari et al. 2018; Springel et al. 2018; Schneider et al. 2019). 
In general, the inclusion of baryons can suppress the matter power spectrum 
by $10\%$ or more at small scales. However,  the precise values and scales relevant for baryonic effects rely 
on the simulations adopted. And comparing among different 
simulations is nontrivial, 
because all of the results depend on the specific cosmological model, resolution, volume, sub-grid baryonic physics,
implementation of the hydrodynamics, specifics of the feedback model,  and the actual physics treated. 
In particular,  recently Springel et al. (2018) presented a detailed analysis of the impact of baryons on the 
clustering of galaxies and matter in the IllustrisTNG simulations, a set of
cosmological hydrodynamical runs spanning different volumes and physics implementations --with 
an updated  AGN feedback recipe compared to the previous 
Illustris release (Vogelsberger et al. 2014). 
Also, Chisari et al. (2018) reported accurate measurements 
of the total matter power spectrum from the Horizon cosmological hydrodynamical simulations (Dubois et al. 2014, 2016)
including the effects of AGN feedback. 
Their new AGN sub-grid model  
is similar to the one implemented in the IllustrisTNG, and differs from  
that of the Illustris simulation. 
In addition, Chisari et al. (2018) compared their results 
with the OWLS (van Daalen et al. 2011), EAGLE,
Illustris, and IllustrisTNG simulations, and  found that the impact of baryonic processes
on the total matter power spectrum
are smaller at $z=0$ in the Horizon runs.
According to their results, while the qualitative behavior of all simulations is essentially similar, with a power suppression at $k \sim 10  {\rm {\it h} Mpc^{-1}}$
due to the effect of AGN feedback on the gas,
the exact scale and strength of the suppression varies between them, even up to 30\% at $\sim 5 {\rm {\it h} Mpc^{-1}}$.
For example, the new IllustrisTNG runs show a much lower impact of baryons on the distribution of
matter, with a reduction of the overall amplitude of the effect and a restriction
to smaller scales compared to Illustris. Moreover, 
Springel et al. (2018) found that baryonic effects increase the clustering of DM
on small scales and damp the total matter power spectrum on scales up to $k \sim 10 {\rm {\it h} Mpc^{-1}}$ by $\sim 20\%$. 

Despite all of the differences related to cosmology, resolution, implementation 
of hydrodynamics, and sub-grid physics, 
in Figure \ref{fig_3d_ps_A_bis}, 
we attempt a comparison with the Horizon simulations
at $z=3$. Specifically,  the top panel shows the total matter power spectrum
as measured in the Sejong BG 100/832 run (red dashed line), 
in the Sejong BG realization characterized
by $N^3 = 832^3$ particles/type over a $25h^{-1}{\rm Mpc}$ box (denoted as ``25/832"; green dashed line),
in the Horizon run containing full baryonic physics (termed ``HR AGN", blue long dot-dashed line),
and in the Horizon run lacking AGN physics (indicated as ``HR NO AGN", pink short-dashed-dotted line).
The linear and {\it Halofit} corresponding predictions are also displayed, with solid and dotted lines, respectively. 
The bottom panels of the same figure are enlargements in the $k$-range $1 \le k {\rm {[} {\it h} \rm Mpc^{-1}]} \le 10$,
showing that the ``HR NO AGN" realization appears to be in reasonably good agreement within error bars with the
Sejong 25/832 run (as expected, since AGN feedback is switched off in the Sejong simulations), 
while the  `HR AGN' total matter power spectrum already shows deviations 
and signs of suppression at 
$z=3$ due to AGN feedback, particularly relevant at smaller scales.
Note also that
the 25/832 run is in better agreement with the  `HR NO AGN' realization
than the 100/832  simulation, since its resolution is four times higher, allowing
one to resolve small scales more accurately. 
In general, the effects of baryons on the total matter power spectrum 
are non-monotonic throughout $z=0-5$ due to an interplay between 
AGN feedback (if present), gas pressure, and the growth of structure
(see again Chisari et al. 2018).

As noted in Rossi (2017), 
while at the linear level and in
absence of massive neutrinos,  extra dark radiation, and WDM, 
the global shape of $P^{\rm 3D}_{\rm t}$ is redshift-independent for modes well inside the Hubble radius,
this is no longer the case when massive neutrinos or any {\it dark sector} particles are introduced. In fact, even at the linear level,
their addition  
induces a scale-dependent distortion of the matter power spectrum shape, along with a
combined evolution of the amplitude; the effect is then amplified in a nontrivial way
in the nonlinear regime -- see their extensive discussion about the `spoon-like' mechanism.
In particular, the amplitude and position of the maximal suppression of the `spoon-like' feature
caused by neutrinos and dark radiation on $P^{\rm 3D}_{\rm t}$ defines a characteristic nonlinear scale, 
which can potentially be used to constrain the properties of neutrinos and $N_{\rm eff}$ from LSS observables. 
Moreover, baryonic effects can mimic neutrino- or dark-radiation-induced suppressions, especially 
when $\sum m_{\nu} = 0.1$ (a limit that is approaching the normal mass hierarchy regime), and   
linear theory is unable to capture several key aspects of the small-scale evolution. 
Therefore, neglecting baryons and using linear theory extrapolations in this regime are incorrect approximations that could potentially mislead 
cosmological results.  
Finally, we recall that at $k \sim 5h {\rm Mpc}^{-1}$ the suppression on the matter
power spectrum induced by $\sum m_{\nu} = 0.1~{\rm eV}$ neutrinos can reach $\sim 4\%$ at $z\sim 3$ when compared to a massless
neutrino cosmology, and $\sim 10\%$ if a massless sterile neutrino is included (Rossi 2017). 

Next, we turn to the flux statistics and focus on the main Ly$\alpha$ forest observables.  


\subsection{Ly$\alpha$ Forest Spectra and Skewers} \label{subsec_lya_spectra_skewers}

Mapping  the properties of the gas at various redshift slices while considering different combinations
of cosmological parameters is an essential step, in order to  characterize the complex matter-to-flux relation
and eventually compute the bias of Ly$\alpha$ forest quasars. In this view, our high-resolution skewers described in Sections \ref{subsec_postprocessing} and \ref{sec_sejong_suite_components}
represent a valuable asset, as they allow one to accurately
quantify  the variations of the gas key properties within a wide range in parameter space -- particularly  when massive or sterile neutrinos, dark radiation, and WDM are included.

Figure \ref{fig_LyA_observables_A} shows a number of key
physical properties of the gas,
as inferred from such Ly$\alpha$ mocks. 
These skewers, randomly drawn from snapshots at different redshifts, are constructed from $25~h^{-1}{\rm Mpc}$ box simulations
that are part of the \textit{Supporting Grid} (see Table \ref{table_supporting_sims_dark_rad}). 
In the various panels, the normalization is adjusted such that $A_{\rm S}$ is kept fixed  as in the fiducial (BG) cosmology. Hence,
the values of $\sigma_8$  at the present epoch  differ from the Planck 2015 central value, depending on the degree of neutrino mass and dark radiation components;
we refer to this convention as `UN' in the previous tables. 
For display purposes, we consider identical skewers (i.e., LOS drawn with the same random seed) across three different cosmologies, namely,
the BG (solid orange  lines), 
a scenario with three degenerate massive neutrinos having $\sum m_{\nu} = 0.3~{\rm eV}$ (dotted green lines),
and a model where a massless sterile neutrino is added to the three active massive ones, so that $N_{\rm eff} = 4.046$ (dashed-dotted blue lines). 
The physical quantities displayed, ordered from top to bottom, are the neutral atomic Hydrogen (HI)
density fraction, the HI temperature (in kelvins),  the HI peculiar velocity (in km/s), the HI optical depth $\tau_{\rm HI}$, and the HI transmitted flux ${{\cal{F}}_{\rm HI}}$  --
where ${\cal{F}}_{\rm HI} = \exp [- \tau_{\rm HI}]$. The redshift evolution is also shown: specifically,  in the left panels $z=2.0$, in the central panels $z=3.0$, and  the right panels have $z=4.0$. 
Even visually, variations in all of the key physical properties among the three  distinctive
cosmologies are clearly noticeable. 
These differences will eventually manifest in the major Ly$\alpha$ forest observables,
and in particular in the 1D flux power spectrum -- as  
well as in the 
overall structure of the cosmic web traced by the gas component.  
In addition, note that departures from the BG are more significant when a massless sterile neutrino
is included (blue dashed-dotted lines in the panels), and thus the Ly$\alpha$ forest flux is also an excellent 
high-$z$ probe for detecting hypothetical sterile neutrinos. 

Figure \ref{fig_LyA_observables_B} highlights the effects of our normalization conventions
on the HI transmitted flux. As an illustrative example, 
we consider  
a massive neutrino cosmology with $\sum m_{\nu} =0.3~{\rm eV}$ (left panel), and a dark radiation scenario 
having  $N_{\rm eff} =4.046$ (right panel). Two different random skewers per model are displayed, at $z=2.0$.    
Solid black lines are used for the `UN' normalization convention; 
dotted or dashed colored lines are used for the `NORM' convention, where $\sigma_8(z=0)$ is enforced to be consistent with Planck 2015 expectations for all the realizations. 
The main point of the plot is to highlight the impact of distinct normalization choices on the HI transmitted flux: 
differences are visible, and they will eventually manifest in the various Ly$\alpha$ forest observables. 


\subsection{Flux Power Spectrum} \label{subsec_flux_ps}


\begin{figure}
\centering
\includegraphics[angle=0,width=0.49\textwidth]{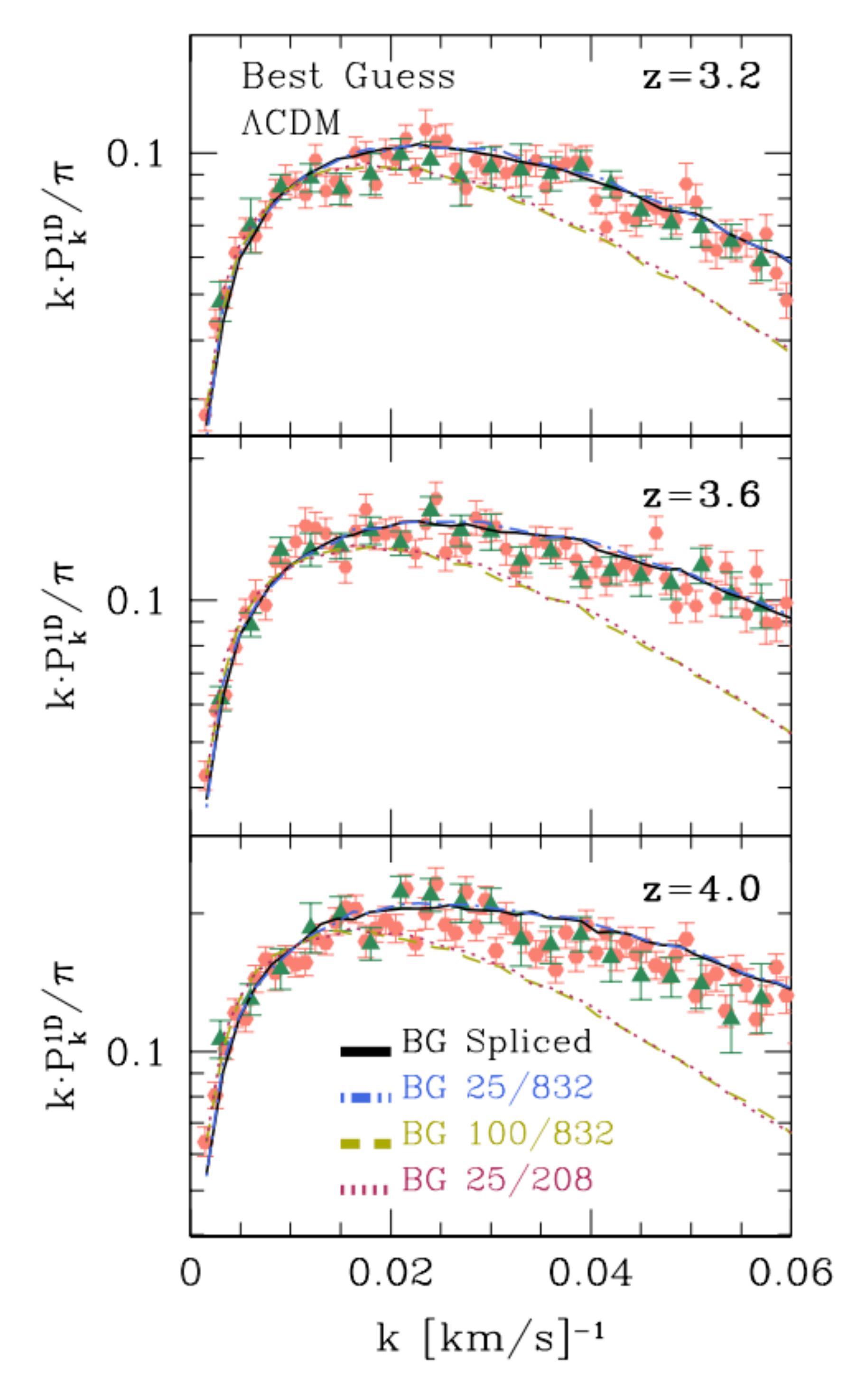}
\caption{Examples of 1D flux power spectra as inferred from the {\it Sejong Suite}, as well as from recent high-$z$ quasar catalogs. 
Green triangles are measurements reported by Irsic et al. (2017b) using the VLT/XSHOOTER legacy survey, and  
orange circles indicate those of Yeche et al. (2017), which also include data from the SDSS-III BOSS survey. 
Three redshift intervals are displayed:  $z=3.2$ (top panel), $z=3.6$ (middle panel), and $z=4.0$ (bottom panel).
In the figure, black solid lines show the {\it ``Best Guess''} {\it spliced} flux power spectrum obtained by combining a  
large-scale power run having a box size of $100~h^{-1} {\rm Mpc}$ (gold dashed lines; $N_{\rm p} = 832^3$/type)
with two small-scale power realizations characterized by a $25~h^{-1} {\rm Mpc}$ box size (blue dashed-dotted and red dotted lines; 
$N_{\rm p} = 832^3$ and $N_{\rm p} = 208^3$ per type, respectively), so that the 
equivalent resolution is $2 \times  3328^3 \sim 74$ billion particles in a $(100 h^{-1} {\rm Mpc})^3$ box size. 
All of the  synthetic power spectra  are averaged over 100,000 
mock quasar absorption spectra extracted from individual simulations at the corresponding redshifts. 
Overall, there is a remarkable consistency between data measurements and predictions from the {\it Sejong Suite} 
high-resolution hydrodynamical simulations.}
\label{fig_1d_flux_ps_data_16D_paper}
\end{figure} 


Next, we consider the 1D flux power spectrum ($P^{\rm 1D}_{{ \cal{F}}}$), a
key Ly$\alpha$ forest observable highly sensitive  to cosmological parameters, 
neutrino masses, WDM, and additional dark radiation 
components such as sterile neutrinos -- via significant attenuation effects especially at small scales.
The neutrino free streaming induces in fact a characteristic redshift- 
and mass-dependent suppression of power that affects the properties of the transmitted flux fraction;  
hence, accurately measuring the full shape, amplitude, and tomographic evolution of the small-scale
$P^{\rm 1D}_{{ \cal{F}}}$ is fundamental for inferring key properties on the formation and growth of structures at high-$z$, and for
constraining cosmological parameters and the neutrino mass.

Because of nonlinearities and baryonic physics, a careful small-scale modeling of $P^{\rm 1D}_{{ \cal{F}}}$ is only possible via sophisticated high-resolution 
hydrodynamical simulations such as those presented in the \textit{Sejong Suite}. In fact, at small scales the Ly$\alpha$ flux distribution depends
on the complex IGM spatial distribution, peculiar velocity field, and thermal properties of the gas, making the
connection with the total matter power spectrum $P^{\rm 3D}_{\rm t}$ rather complex. In this view, a detailed
knowledge of the gas-to-matter and peculiar velocity biases, of the nature of the
ionizing background radiation, and on the fluctuations in the temperature-density relation is necessary.

In previous works, using a refined technique that involves a model based on a Taylor expansion of the flux (evaluated numerically), 
we have already successfully used the information contained in $P^{\rm 1D}_{{ \cal{F}}}$ 
to obtain among the strongest individual and joint bounds on 
neutrino masses and $N_{\rm eff}$, exploiting the Ly$\alpha$ forest in synergy with the CMB and low-$z$: 
perfecting this technique using new simulations from the \textit{Sejong Suite} is the subject of ongoing work. 

To this end, Figure \ref{fig_1d_flux_ps_data_16D_paper} displays a few BG 1D flux power spectra as derived from our {\it Grid Suite}
(i.e., Table \ref{table_grid_sims_base}), confronted with recent $P^{\rm 1D}_{{ \cal{F}}}$  measurements
obtained from the SDSS-III BOSS survey and from  
the VLT/XSHOOTER legacy survey (XQ-100; L{\'o}pez et al. 2016) -- the latter is observed with the X-Shooter spectrograph on the Very Large Telescope (Vernet et al. 2011). 
In the plot, green triangles refer to the observational measurements reported by Irsic et al. (2017b), while orange circles indicate those of Yeche et al. (2017). 
Three redshift intervals are considered: namely, $z=3.2$ (top panel), $z=3.6$ (middle panel), and $z=4.0$ (bottom panel), respectively.
In all of the panels, black solid lines display the BG {\it spliced} flux power spectrum -- corresponding to the effective resolution of  
a single simulation characterized by $2 \times  3328^3 \sim 74$ billion particles in a $(100 h^{-1} {\rm Mpc})^3$ box size for the reference cosmology. 
The three individual flux power spectra  used to derive the spliced $P^{\rm 1D}_{{ \cal{F}}}$ are also shown, with different line styles. 
Specifically, gold dashed lines refer to the large-scale power run with a number of particles per component $N_{\rm p} = 832^3$ over a box size of $100~h^{-1} {\rm Mpc}$,
while blue dashed-dotted and red dotted lines are for the small-scale power realizations (i.e., $25~h^{-1} {\rm Mpc}$ box size) having $N_{\rm p} = 832^3$ and $N_{\rm p} = 208^3$ per type, respectively. 
Note that 
all of the different simulated power spectra displayed in the figure are obtained by averaging 100,000 
mock quasar absorption spectra extracted from each individual simulation at the corresponding redshift, as explained in Section \ref{subsec_postprocessing},   
and that the wavevector $k$ is now expressed in ${\rm [km/s]}^{-1}$.
Overall, there is a remarkable consistency between data measurements and predictions from our high-resolution hydrodynamical simulations. 


\begin{figure*}
\centering
\includegraphics[angle=0,width=0.75\textwidth]{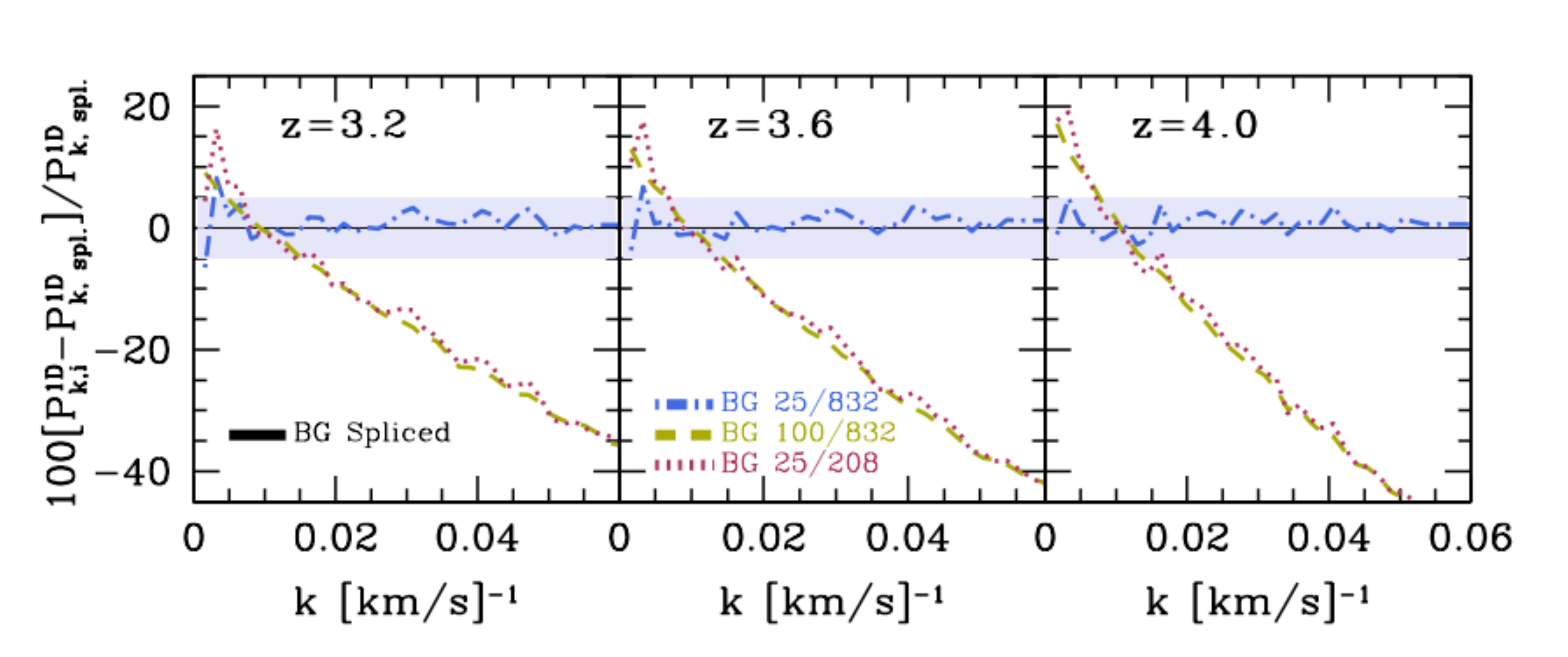}
\caption{Insights into the McDonald (2003) {\it splicing} technique.  Ratios of individual flux power spectrum components, 
normalized by the {\it spliced} $P^{\rm 1D}_{{ \cal{F}}}$ -- expressed in percentage, and assuming the reference cosmology of the {\it Sejong Suite}.   
Line styles and colors are the same as in Figure \ref{fig_1d_flux_ps_data_16D_paper} (see the indications in the various panels).   
From left to right, the redshift intervals are $z=3.2, 3.6, 4.0$, respectively. The extended horizontal cyan band highlights the $\pm 5\%$ scatter level. 
Clearly, the highest-resolution run ($25~h^{-1} {\rm Mpc}$ side, $N_{\rm p} = 832^3$/type) is quite accurate at small scales and lies always within the $5\%$ range,
while the larger-box realization ($100~h^{-1} {\rm Mpc}$) generally lacks in resolution. 
The ``transition" run ($25~h^{-1} {\rm Mpc}$ side,  $N_{\rm p} = 208^3$/type)
is used to correct the larger-box simulation for the lack of resolution, and the small one for the lack of nonlinear coupling between the highest and lowest $k$-modes -- 
allowing for an effective resolution up to  
$3 \times  3328^3 = 110$ billion particles in a $(100 h^{-1} {\rm Mpc})^3$ box size.}
\label{fig_1d_flux_ps_theory_4}
\end{figure*} 


\begin{figure*}
\centering
\includegraphics[angle=0,width=0.34\textwidth]{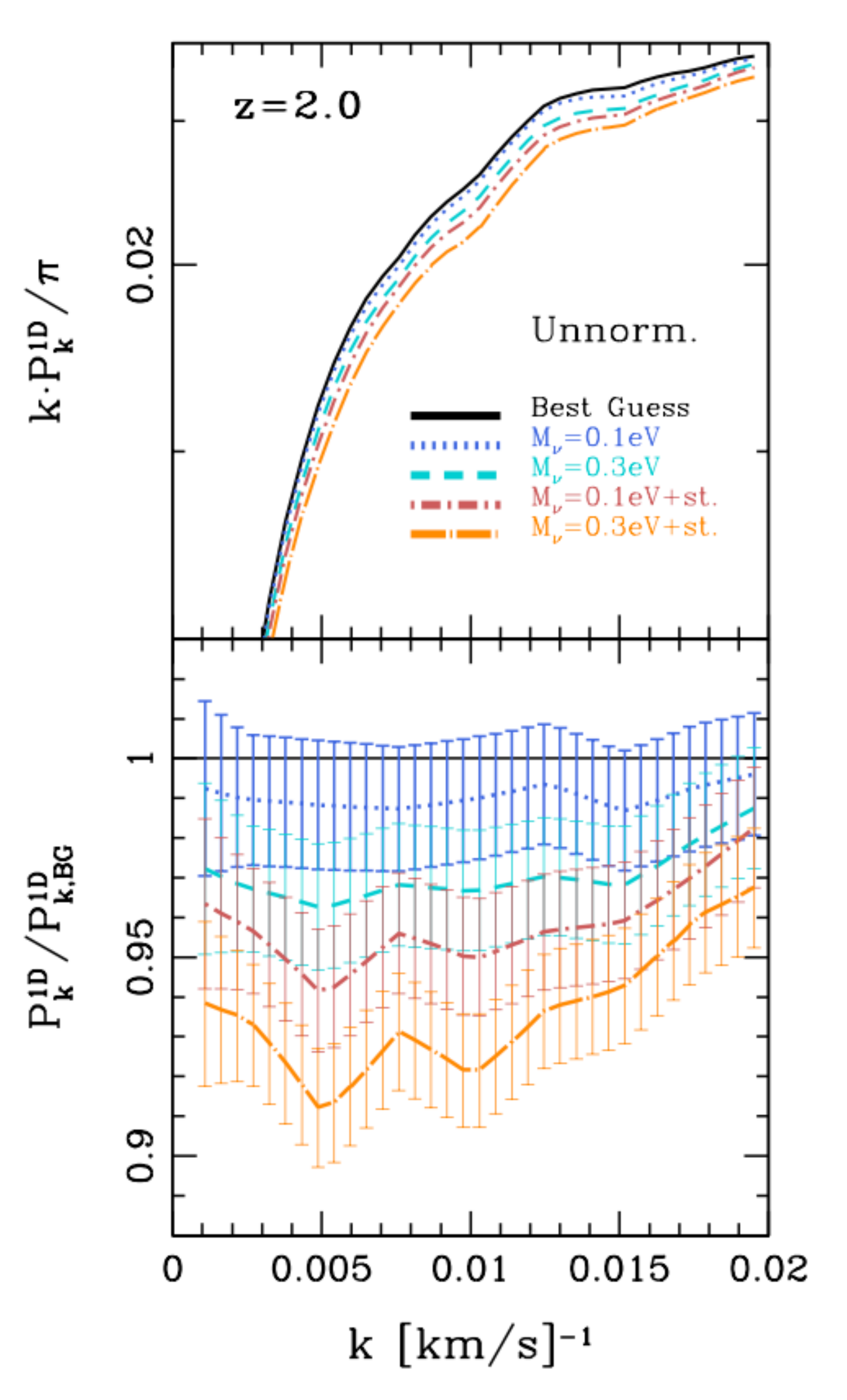}
\includegraphics[angle=0,width=0.309\textwidth]{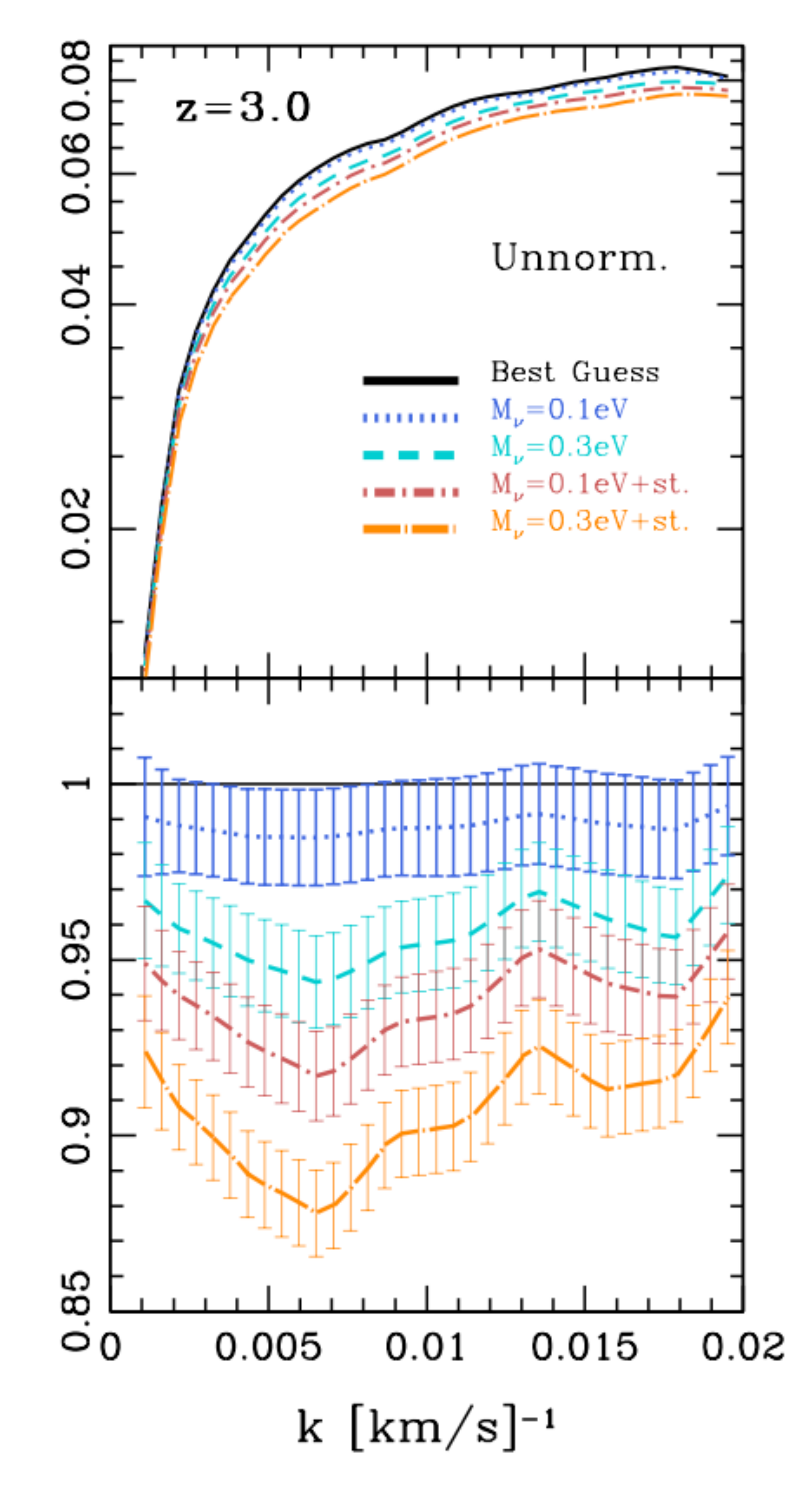}
\includegraphics[angle=0,width=0.313\textwidth]{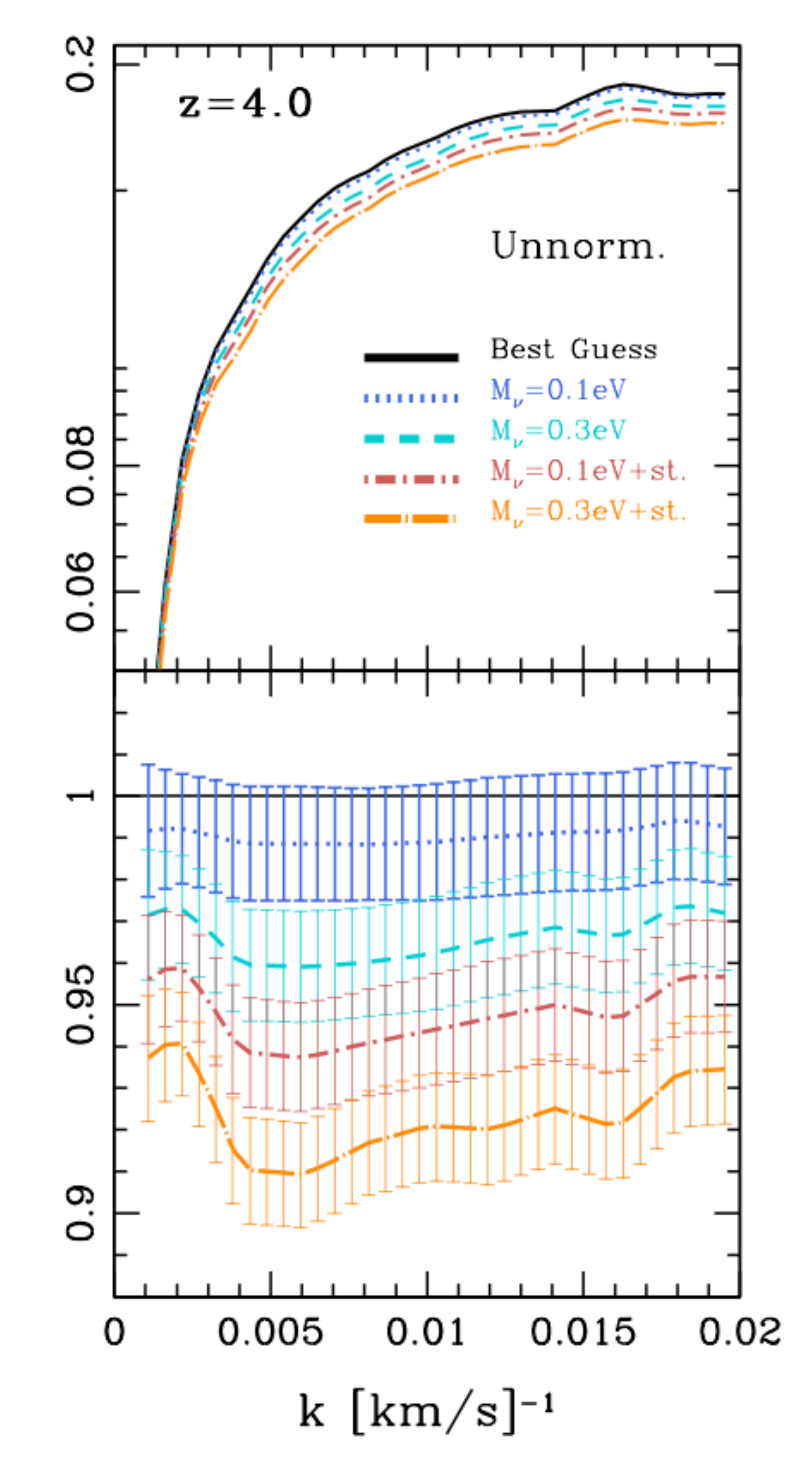}
\caption{Tomographic evolution of $P^{\rm 1D}_{{ \cal{F}}}$ in the presence of massive neutrinos and dark radiation.  
From left to right, $z=2.0, 3.0, 4.0$, respectively. 
Measurements are derived from the {\it Supporting Suite} simulations (see Table \ref{table_supporting_sims_dark_rad}; $25~h^{-1} {\rm Mpc}$ box, $N_{\rm p} = 256^3$/type resolution).  
The top panels display $P^{\rm 1D}_{{ \cal{F}}}$ for the {\it `Best Guess'} (black solid line), 
for a model with three degenerate massive neutrinos with $\sum m_{\nu} = 0.1~{\rm eV}$ (blue dotted line) or $\sum m_{\nu} = 0.3~{\rm eV}$ (cyan dashed line),
and for a corresponding dark radiation cosmology, which includes an additional  thermalized massless sterile neutrino ($N_{\rm eff} =4.046$; red dashed-dotted or orange long-dashed-dotted lines, respectively).
All of the $P^{\rm 1D}_{{ \cal{F}}}$ are averaged over 10,000 mock quasar absorption spectra. 
The bottom panels display flux power spectrum ratios for the same non-canonical models, normalized by corresponding
fiducial simulation results -- using similar line styles as in the top panels. 
Error bars are $1 \sigma$ deviations computed from the same 10,000 simulated skewers at each $z$-interval. 
The characteristic  ``spoon-like" feature is clearly seen -- more prominent  for larger neutrino masses, and more enhanced when sterile neutrinos are also present. 
See the main text for further details.}
\label{fig_1d_flux_ps_theory_10}
\end{figure*} 


As a further insight into the McDonald (2003) {\it splicing} technique to overcome resolution demands in simulations, Figure \ref{fig_1d_flux_ps_theory_4} 
quantifies the scatter (in percentage) between the BG {\it spliced} flux power spectrum
and its individual components -- assuming the reference cosmology of the {\it Sejong Suite}. 
Line styles and colors are similar to those adopted in Figure \ref{fig_1d_flux_ps_data_16D_paper}.
Namely, gold dashed lines are used for the $100~h^{-1} {\rm Mpc}$ run with $N_{\rm p} = 832^3$/type, 
blue dashed-dotted lines show the $25~h^{-1} {\rm Mpc}$ box-size realization having $N_{\rm p} = 832^3$/type, and
red dotted lines display the $25~h^{-1} {\rm Mpc}$ box-size simulation with resolution $N_{\rm p} = 208^3$/type. 
We examine the same redshift intervals as in Figure \ref{fig_1d_flux_ps_data_16D_paper} (i.e., $z=3.2, 3.6, 4.0$).
The $y$-axis reports the ratios of individual flux power spectrum components, normalized by the {\it spliced} $P^{\rm 1D}_{{ \cal{F}}}$ -- 
expressed in percentage. The extended horizontal cyan band highlights the $\pm 5\%$ scatter level. 
As evident from the figure, the highest-resolution run (blue dashed-dotted lines, $25~h^{-1} {\rm Mpc}$ side, $N_{\rm p} = 832^3$/type) lies within 
the $5\%$ range, and it is quite accurate at small scales (large $k$), while the larger-box realization ($100~h^{-1} {\rm Mpc}$) generally lacks in resolution.
To this end, the `transition' run (i.e.,  red dotted lines, $25~h^{-1} {\rm Mpc}$ side,  $N_{\rm p} = 208^3$/type)
is used to correct the larger-box simulation for the lack of resolution, and the small box for the lack of nonlinear coupling between the highest and lowest $k$-modes.

Regarding {\it dark sector} physics, Figure \ref{fig_1d_flux_ps_theory_10} shows an example of the tomographic evolution of the small-scale $P^{\rm 1D}_{{ \cal{F}}}$
in the presence of massive neutrinos and dark radiation, as inferred from simulations belonging to the {\it Supporting Suite}  (Table \ref{table_supporting_sims_dark_rad}). 
For illustrative purposes, we consider runs with $25~h^{-1} {\rm Mpc}$ box sizes and a resolution  
$N_{\rm p} = 256^3$/type. 
From left to right, the redshift intervals examined are $z=2.0, 3.0, 4.0$, respectively. Here, the normalization convention  
is the one previously indicated as ``UN", where $A_{\rm s}$ is fixed as in the reference cosmology 
-- in order to clearly isolate the effects of neutrinos and dark radiation from other possible 
degeneracies. In dimensionless units, the top panels display $P^{\rm 1D}_{{ \cal{F}}}$ for the BG (black solid line), 
for a model with three degenerate massive neutrinos having a total summed mass $\sum m_{\nu} = 0.1~{\rm eV}$ (blue dotted line) or $\sum m_{\nu} = 0.3~{\rm eV}$ (cyan dashed line),
and for a corresponding dark radiation cosmology with $N_{\rm eff} =4.046$ 
(red dotted-dashed or orange long dashed-dotted lines, respectively).  
The evolution of the flux power spectrum shape across different cosmic epochs is clearly seen, along with the characteristic global 
suppression of power induced by massive and/or sterile neutrinos -- more prominent  for larger neutrino masses, and more enhanced when sterile neutrinos are also present. 
The bottom panels display flux power spectrum ratios for the same non-canonical models, normalized by the 
reference BG simulation -- using similar line styles as in the top panels. 
In this case, all of the flux power spectra measurements are averaged over 10,000 mock quasar absorption spectra, and  in the bottom plots, 
error bars are $1 \sigma$ deviations computed from those simulated skewers at each redshift interval. 

As thoroughly examined in Rossi (2017), 
the characteristic signature of massive and/or additional sterile neutrinos on the transmitted Ly$\alpha$ forest flux 
is a global suppression of power at small scales, manifesting as  a  typical `spoon-like' feature. 
This effect is evident in the bottom panels of Figure \ref{fig_1d_flux_ps_theory_10}, where
one can clearly appreciate the small-scale $P^{\rm 1D}_{{ \cal{F}}}$ suppression along with its tomographic evolution. 
Similarly as in Rossi (2017), we confirm that the most significant deviations from the BG  reference model
occur around $z \sim 3$, with a maximal departure at scales $k \sim 0.005~[{\rm km/s}]^{-1}$ corresponding to
$k\sim0.575 ~h {\rm Mpc}^{-1}$ in a Planck 2015 cosmology. These preferred 
nonlinear scales are characterized by a maximal sensitivity to massive neutrinos and dark radiation in terms of the 1D flux statistics.
Luckily, this regime falls in the middle of the standard Ly$\alpha$ forest extension; hence,  
the Ly$\alpha$ forest has a remarkable potential in constraining neutrino masses 
and possible deviations from the canonical $N_{\rm eff} = 3.046$ value -- especially around $z \sim 3$ -- 
in synergy with complementary lower-$z$ probes.  

\begin{figure}
\begin{center}
\includegraphics[angle=0,width=0.50\textwidth]{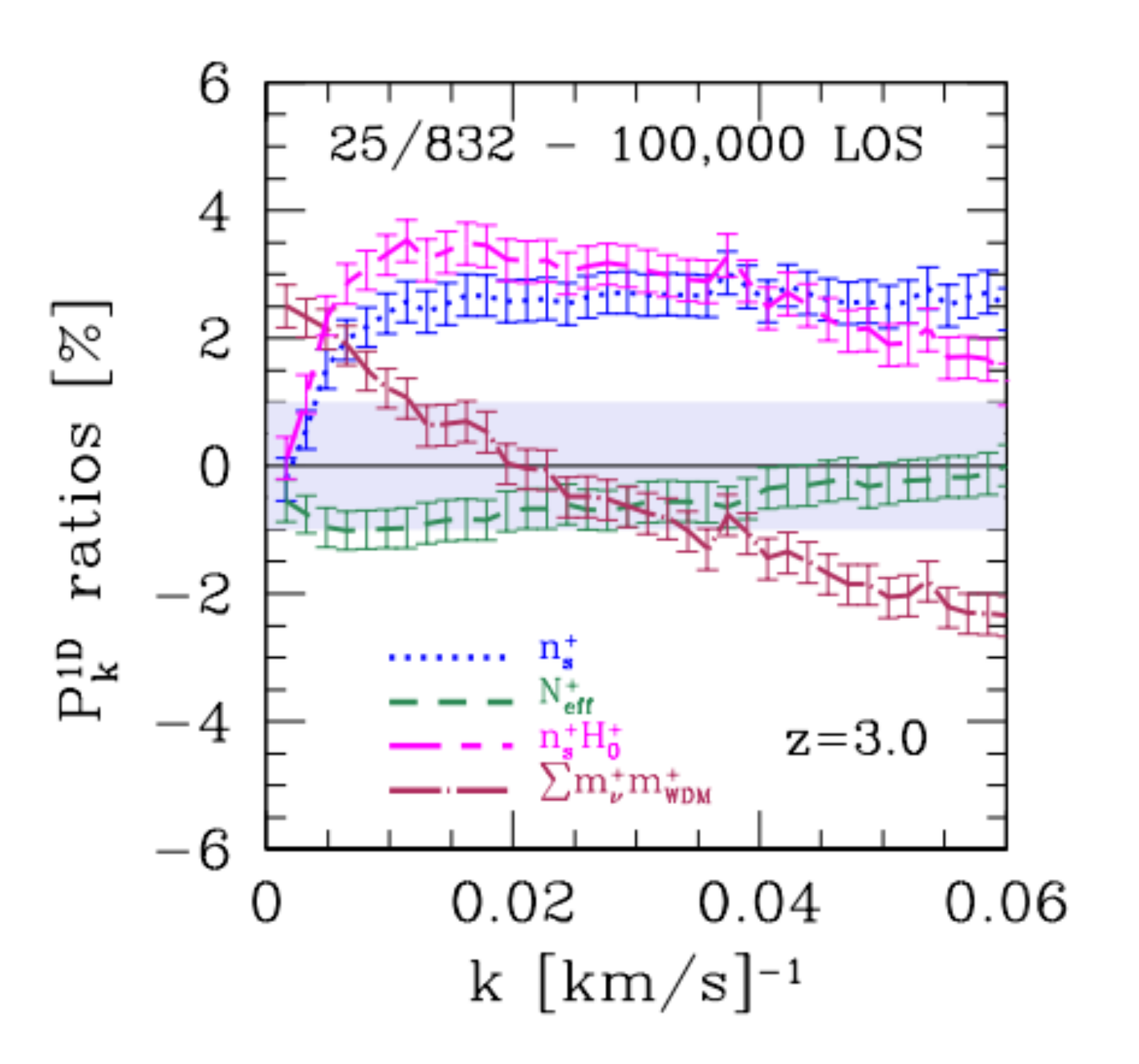}
\caption{Sensitivity of  $P^{\rm 1D}_{{ \cal{F}}}$ 
under variations of individual or combined cosmological parameters (see Table \ref{table_grid_params_variations}), as inferred from the {\it Grid Suite}  
for simulations with $832^3$ particles/type over a $25h^{-1}{\rm Mpc}$
box.  The various flux power spectra are obtained by averaging 100,000 skewers extracted 
from the corresponding simulations at $z=3.0$, with error bars showing 1$\sigma$ deviations.
The colored horizontal band in the panel represents the $1\%$ level.}
\label{fig_grid_coverage}
\end{center}
\end{figure}

Finally, Figure \ref{fig_grid_coverage} 
provides a few selected illustrations
of the parameter space coverage, as described by the flux power spectrum inferred from
simulations belonging to the {\it Grid Suite} (Tables \ref{table_grid_sims_base} and \ref{table_grid_sims_cross_terms}). 
Specifically, the plot shows the effect of changing two individual parameters ($n_{\rm s}$ and $N_{\rm eff}$)
and two joint parameter combinations ($n_{\rm s}$ and $H_0$, or  $\sum m_{\nu}$ and $m_{\rm WDM} $, respectively) 
on $P^{\rm 1D}_{{ \cal{F}}}$ (expressed in percentage),
with respect to the BG cosmology -- as 
indicated in Table \ref{table_grid_params_variations}.
The simulations considered are those with the highest resolution, having $832^3$ particles/type  over a $25h^{-1}{\rm Mpc}$
box. The various flux power spectra are obtained by averaging 100,000 LOS skewers extracted 
from the corresponding runs at $z=3.0$, and  error bars are 1$\sigma$ deviations, with
the colored horizontal band representing the $1\%$ level.
Generally, flux variations with respect to the baseline cosmology 
fluctuate from 1\% to 10\% or more,  
depending on the combinations of
cosmological and astrophysical parameters, thus covering a sensible range. 
In particular, the flux statistics is quite sensitive to changes in $n_{\rm s}$, as 
well as in modifications of the astrophysical parameters. 
Also, the individual or combined parameter 
variations are scale-dependent and 
impact the flux power spectrum differently, allowing us
to build a Taylor expansion model of the flux power spectrum.  
Moreover, note that for all of these realizations, the power spectrum normalization is such that  $\sigma_8=0.815$ at $z=0$; this is not the
case for the {\it Supporting Suite}, where we use instead a different (theoretically motivated) convention. 
The sensitivity of $P^{\rm 1D}_{{ \cal{F}}}$
under variations of different cosmological and astrophysical parameters 
has also been addressed in our previous studies. 
In particular, in Rossi (2017),  we have proven that 
 suppressions on the matter -- rather than on the flux --
power spectrum induced by a $0.1$ eV neutrino are even more significant, 
and can reach ~4\% at $z \sim 3$ when compared to a massless
neutrino cosmology, and $\sim 10\%$ if a massless sterile neutrino is included.
Furthermore, the highest response to free-streaming  effects is achieved in 
the Ly$\alpha$ forest regime, thus making it an ideal place for constraining the neutrino mass.

\subsection{IGM Properties} \label{section_igm_properties}


\begin{figure}
\centering
\includegraphics[angle=0,width=0.48\textwidth]{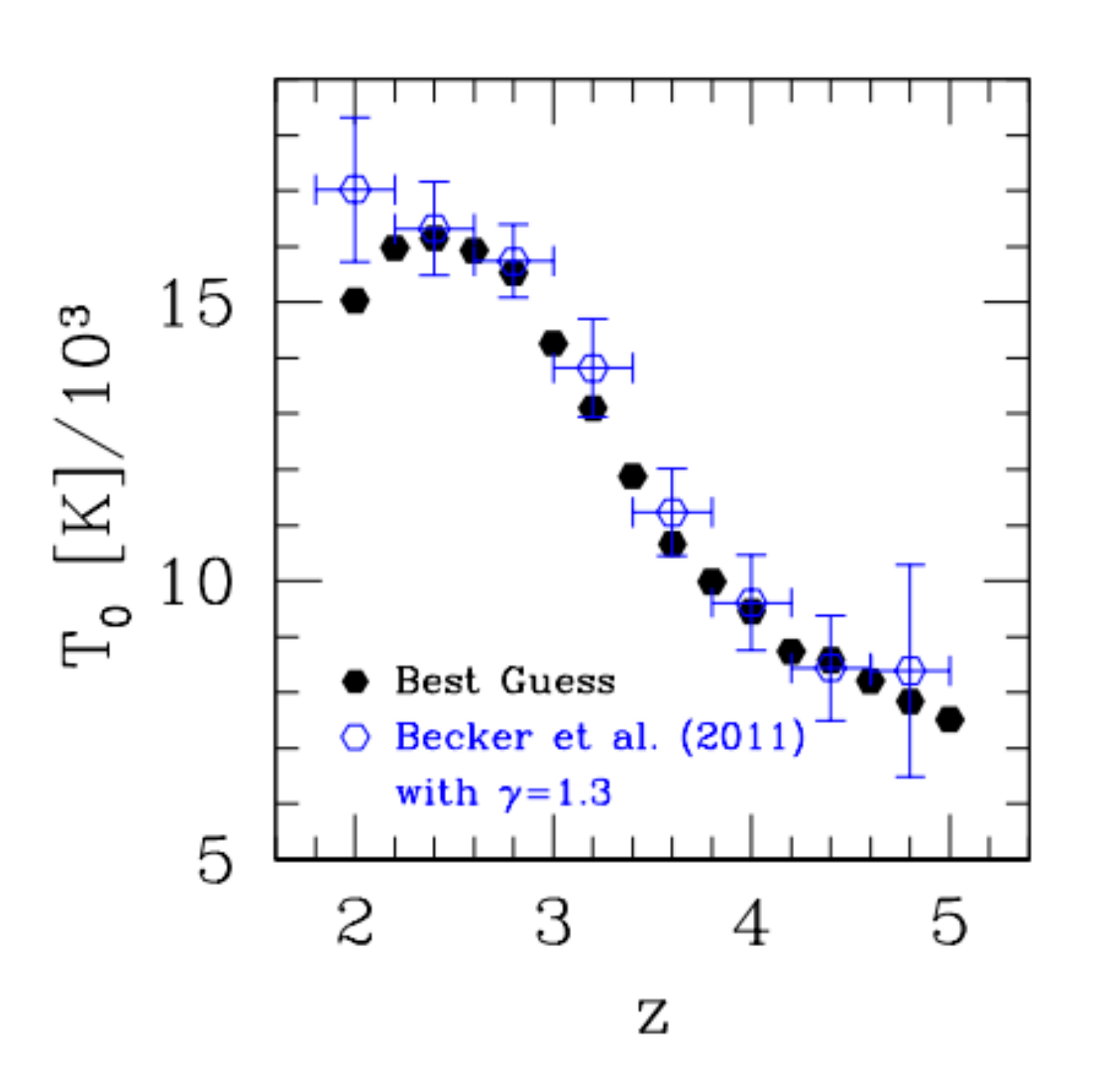}
\caption{Gas mean temperature $T_0$ as
inferred from particle subsamples extracted from the {\it `Best Guess'} at different $z$ (black filled dots), contrasted
with the well-known measurement by Becker et al. (2011) of the IGM temperature at the mean density when $\gamma = 1.3$ (empty blue dots with error bars).
Our central $T_0$ values are consistent with their reported measurements, as well as the robust detection of an increase in $T_0$ between $z=5.0$ and $z=2.0$.}
\label{fig_igm_properties_1}
\end{figure}

\begin{figure*}
\centering
\includegraphics[angle=0,width=0.31\textwidth]{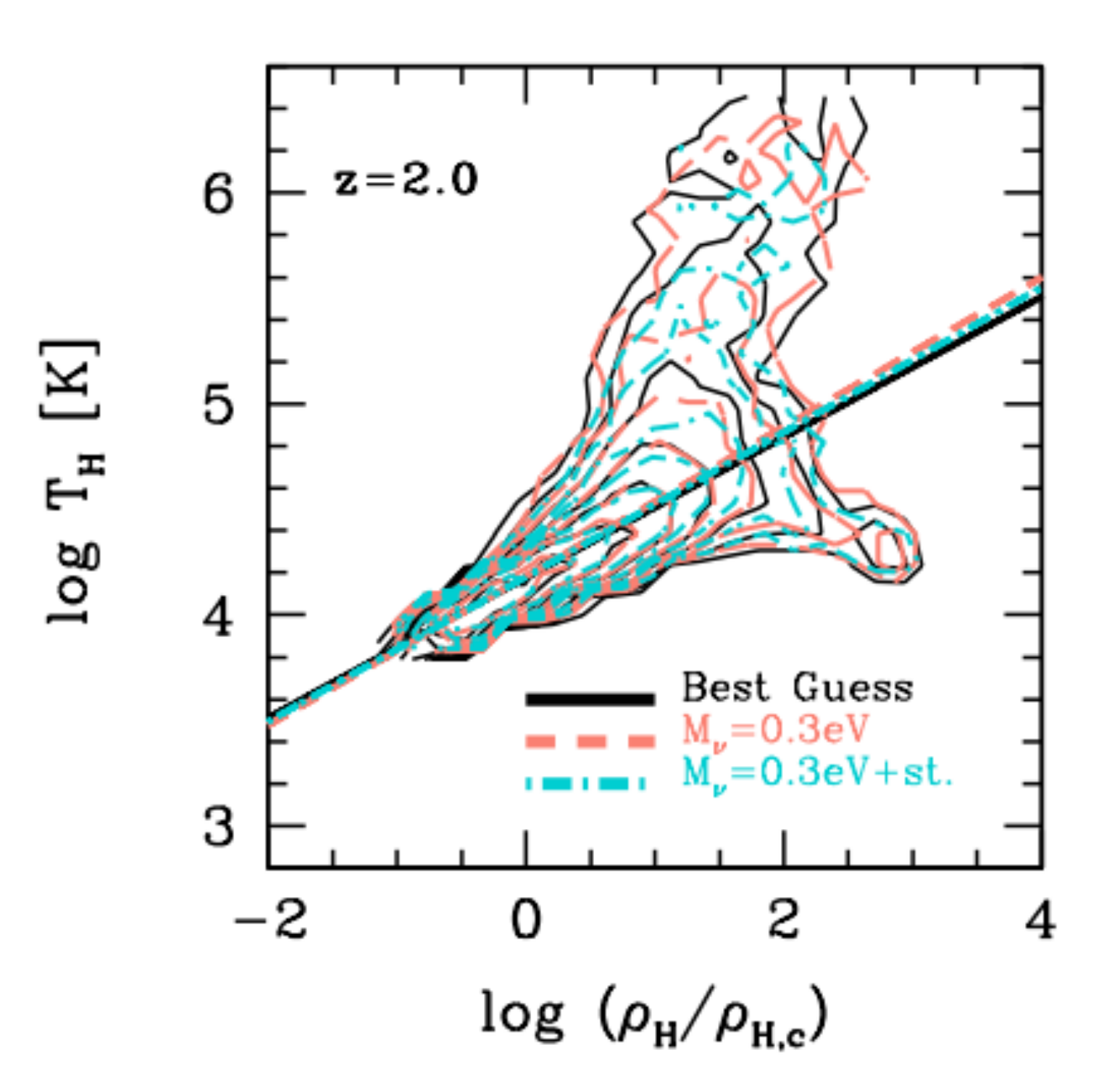}
\includegraphics[angle=0,width=0.325\textwidth]{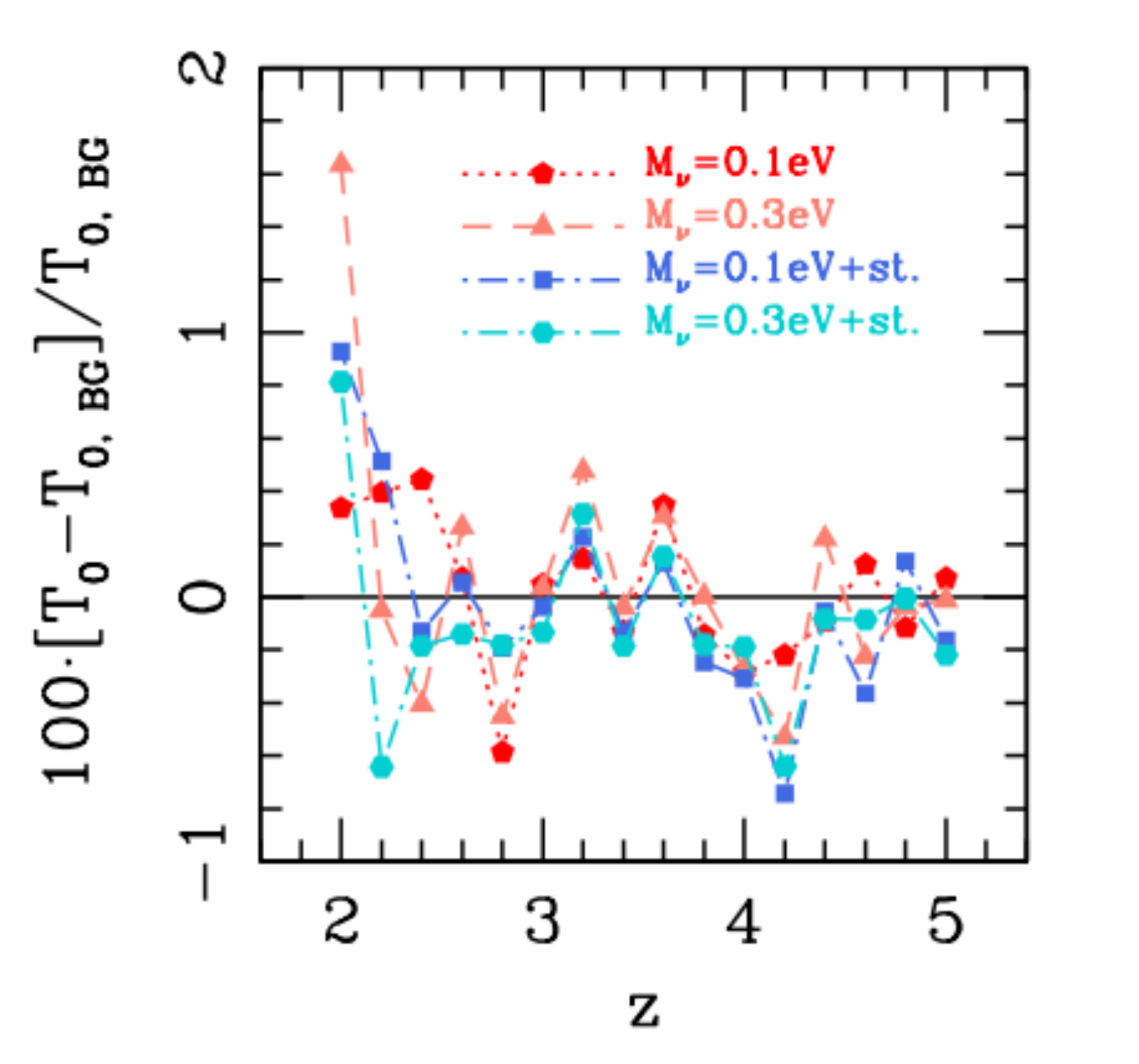}
\includegraphics[angle=0,width=0.31\textwidth]{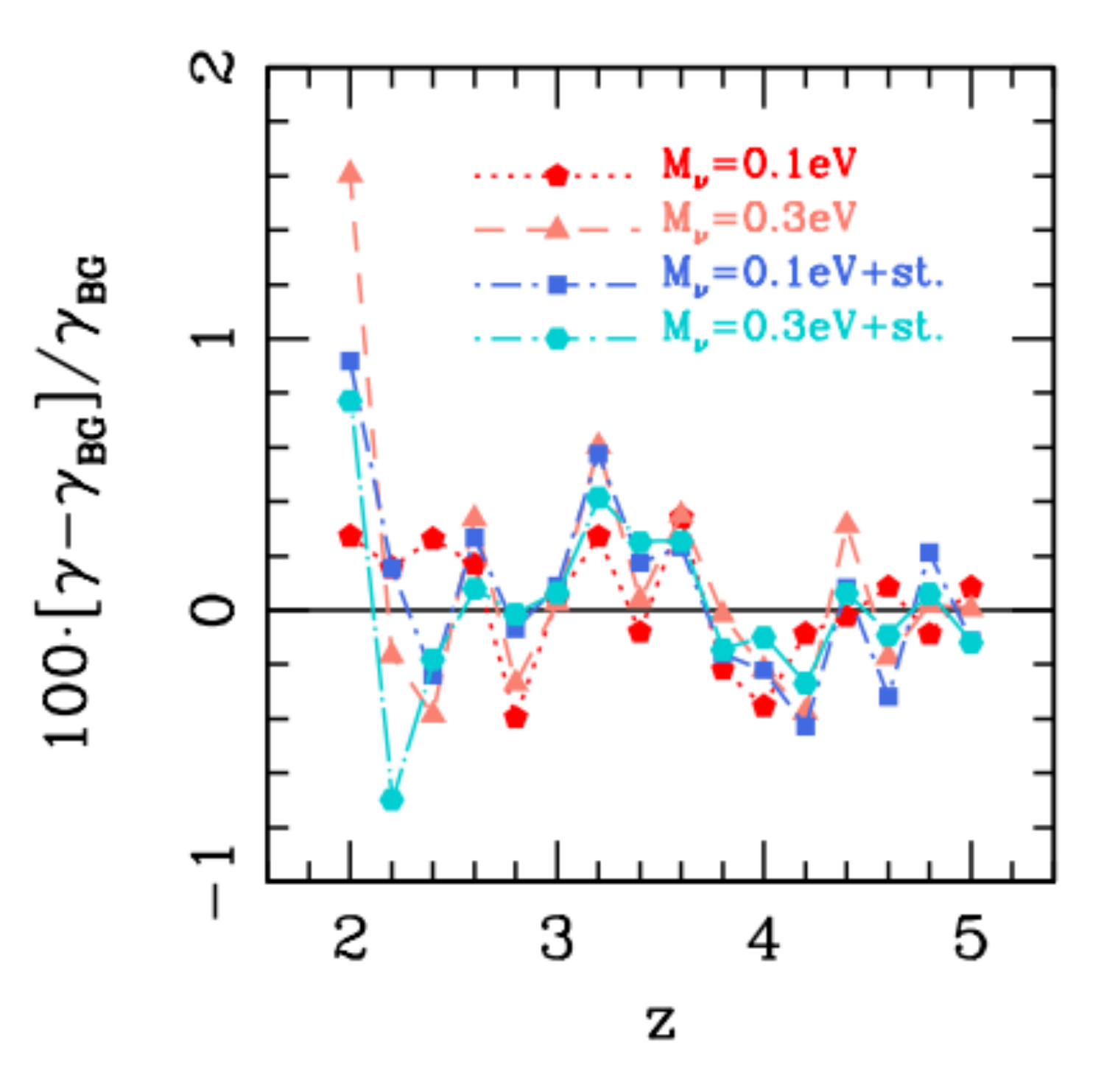}
\caption{Impact of   massive and sterile neutrinos on the   status and basic properties of the IGM at high $z$. 
[Left]    $T-\rho$ relation at $z=2.0$ for the  baseline {\it ``Best Guess"} simulation of the {\it Sejong Suite} (black solid lines),
 for a neutrino cosmology with $\sum m_{\nu} =0.3~{\rm eV}$ (orange dashed lines; $N_{\rm eff} =3.046$), and for
 a dark radiation scenario including a massless sterile neutrino  (cyan dashed-dotted lines; $N_{\rm eff} =4.046$).
Straight lines are corresponding linear fits to simulated data, in order to infer the
best-fit values of $T_0$ and $\gamma$ for the three different models. 
Contours are isodensity surfaces that are $1/2^{\rm n}$ times the height of the maximum value of the density of sources, with $n$ ranging from $1$ to $7$.  
[Middle] Tomographic evolution of $T_0$ in the redshift interval $2.0 \le z \le 5.0$, due to the presence of massive and sterile neutrinos. 
Besides the nonstandard  cosmologies  also shown in the left panel, two additional scenarios are included, but now when $\sum m_{\nu} =0.1~{\rm eV}$, as indicated by the various labels. 
Measurements are expressed in percentage, and normalized by the corresponding baseline cosmology values. 
[Right] Same as the middle panel, but for $\gamma$. Although
the tomographic variations of $T_0$ and $\gamma$ across the models are always at the sub-percent level, 
quantifying the impact of massive neutrinos
and dark radiation on the IGM is important, particularly in relation to
galaxy formation studies.}
\label{fig_igm_properties_2}
\end{figure*}


Uncertainties in the 
IGM thermal status are among the main causes of systematics in the flux power spectrum determinations, and therefore
accurately reproducing the thermal and ionization history of the IGM is 
essential for  improving the robustness of all Ly$\alpha$-based studies. 
The IGM probed by the Ly$\alpha$ forest, highly ionized at high-$z$ and progressively neutral with decreasing redshift,  
consists of mildly nonlinear gas density fluctuations; the
gas traces the DM distribution, and it is generally assumed to be in photoionization equilibrium with the UV
background produced by galaxies and quasars.
In low-density regions, the IGM gas 
density ($\rho$) and temperature ($T$) are closely connected, via a redshift-dependent polytropic power-law relation.
In simple form, the relation can be expressed as a function of redshift as:
\begin{equation}
\log T(z) = \log T_0(z) +[\gamma(z) -1] \log \delta(z), 
\label{eq_IGM_temperature_density_relation}
\end{equation}
where $T_0$ is the gas mean temperature, $\delta = \rho/ \rho_{\rm c}$,
$\rho_{\rm c}$ is the critical density,
and $\gamma$
is a redshift-dependent parameter, which is also
related to the reionization history model and spectral shape of the UV background. 

To this end, Figure  \ref{fig_igm_properties_1} compares the gas mean temperature $T_0$ as
inferred from particle subsamples extracted from the BG at different redshifts ({\it Grid Suite}, Table \ref{table_grid_sims_base})
via computation of the relevant IGM quantities that enter in
Equation (\ref{eq_IGM_temperature_density_relation})  using standard SPH techniques, with the
well-known measurement by Becker et al. (2011) of the IGM temperature at the mean density when $\gamma = 1.3$.
This corresponds to a mild flattening of the $T-\rho$ relation as expected during an extended He II reionization process.
In the figure, black filled dots are results from the BG, while empty blue dots with error bars refer to
the measurements performed by Becker et al. (2011). 
Note that, for our baseline cosmology, we set $\gamma(z=3)=1.3$ and $T_0 (z=3)=15000~{\rm K}$  
-- see Table \ref{table_baseline_params_sims}. Moreover, $T_0$ is largely insensitive to $\gamma$ at $z > 4$, as highlighted by the same authors.  
The main point of the figure is to show the consistency 
of our central $T_0$ values with their reported measurements, and to confirm the finding of a similar trend -- namely,
a robust detection of an increase in $T_0$ between $z=5.0$ and $z=2.0$. 
Measuring $T_0$ as a function of redshift may be informative
on the amount of heating or cooling that is 
occurring for a relatively consistent population of baryons in the IGM.

The additional presence of massive neutrinos and exotic particles such as sterile neutrinos (or
more generally dark radiation) does affect the status and basic properties of the IGM at high $z$.
Potentially, this may also be seen as a source of systematics, or perhaps as a way to
detect such particles through their effects on the $T-\rho$ relation, and eventually
on the formation of structures at Galactic scales. Hence, quantifying the impact of massive neutrinos
and dark radiation on the IGM is important, particularly in relation to
galaxy formation studies. 
Figure \ref{fig_igm_properties_2} illustrates the extent of such impact. 
Specifically, the left panel shows an example of the $T-\rho$ relation at $z=2.0$ as measured from selected high-resolution hydrodynamical simulations
from the {\it Supporting Suite} (Table \ref{table_supporting_sims_dark_rad}). 
Aside from the baseline BG model (black solid lines), we consider 
a neutrino cosmology characterized by  three  degenerate massive neutrinos with $\sum m_{\nu} =0.3~{\rm eV}$ (orange dashed lines),
and a dark radiation scenario with  $N_{\rm eff} =4.046$ (cyan dashed-dotted lines).
Contour levels in the figure are isodensity surfaces that are $1/2^{\rm n}$ times the height of the maximum value of the density of sources, with $n$ ranging from $1$ to $7$.  
The straight lines having analogous styles and colors are  simple linear fits to simulated data, in order to infer the
best-fit values of $T_0$ and $\gamma$ for the three different models.  As is evident from the panel, 
the impact of massive neutrinos and dark radiation on  $T-\rho$ is rather small.
To this end, the  other two panels of Figure \ref{fig_igm_properties_2}  quantify (in percentage, and normalized by the BG reference cosmology) the tomographic effect in the
redshift range $2.0 \le z \le 5.0$ on $T_0$ (middle panel) and $\gamma$ (right panel), due 
to the presence of massive and sterile neutrinos. Besides the two nonstandard models also shown in the left panel, we
display here two more scenarios:  a neutrino cosmology with three  degenerate massive neutrinos but now with a total mass of $\sum m_{\nu} =0.1~{\rm eV}$ (red dotted lines),  and 
a corresponding dark radiation model with $N_{\rm eff}=4.046$ (blue dashed-dotted lines).  
In all of the cases, the tomographic variations of $T_0$ and $\gamma$ across the models are always at the sub-percent level, and on average within $0.5\%$ with respect to the 
reference baseline cosmology; hence, they are 
hard to detect. Therefore, the temperature-density relation is only
marginally affected, showing minor modifications at small scales.


\begin{figure*}
\centering
\includegraphics[angle=0,width=0.90\textwidth]{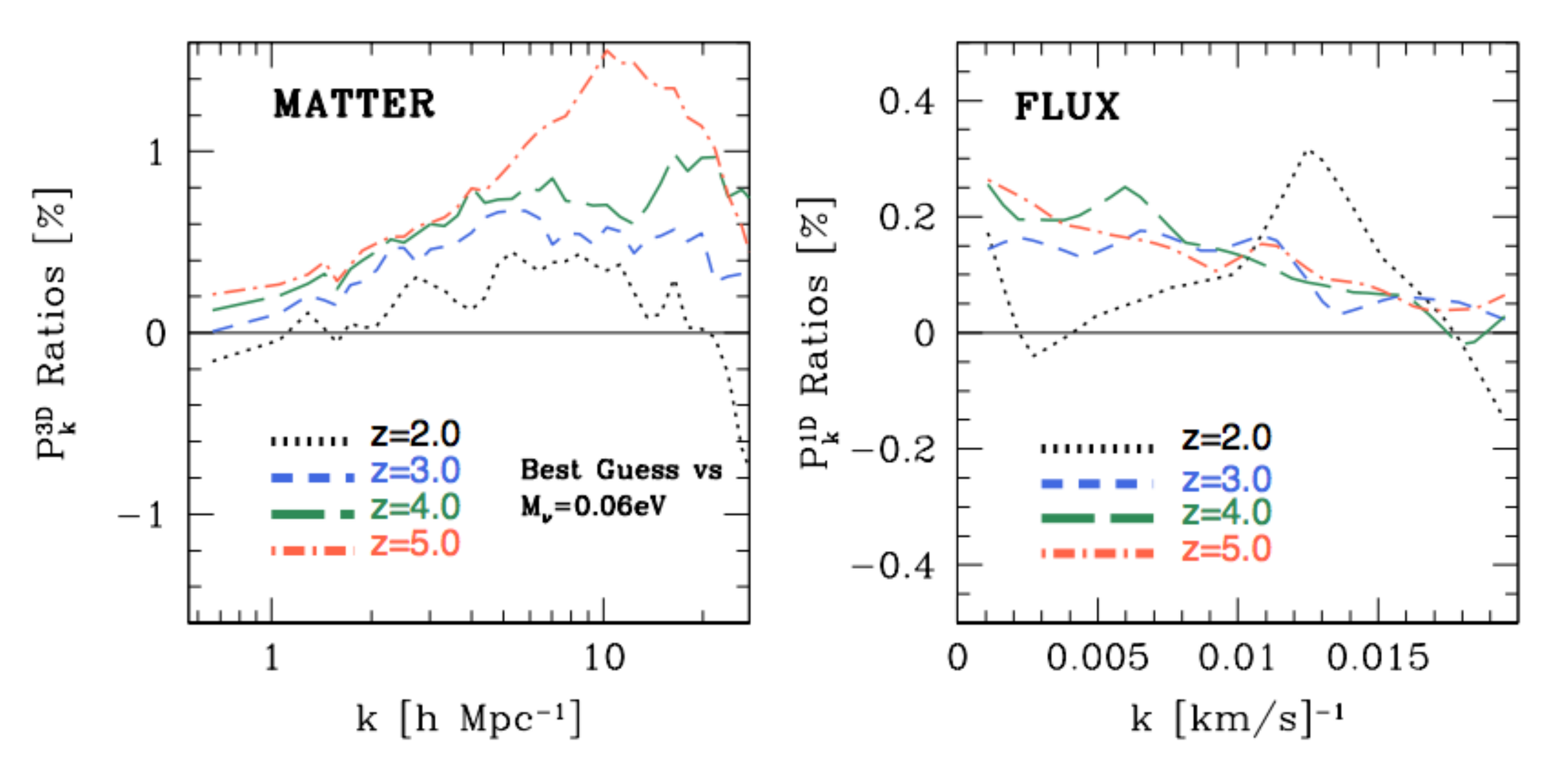}
\caption{Impact of a neutrino minimal-mass normal hierarchy on cosmological observables. [Left]
Total matter power spectrum in a baseline Planck-like model (having a neutrino minimal-mass normal hierarchy of $M_{\nu}=0.06~{\rm eV}$)
normalized by the massless neutrino baseline {\it ``Best Guess"} model of the \textit{Sejong Suite}, as a function of redshift -- as 
indicated by different colors and line styles in the panel. Deviations in terms of the total matter power spectrum are within 
$1\%$ in the  Ly$\alpha$ forest regime of interest, for all the redshift intervals considered. 
[Right] Same as the left panel, but now for the Ly$\alpha$ flux power spectrum.
Deviations in this case are always within $0.3\%$.}
\label{fig_systematics_minimal_neutrino_mass}
\end{figure*}


\begin{figure}
\centering
\includegraphics[angle=0,width=0.50\textwidth]{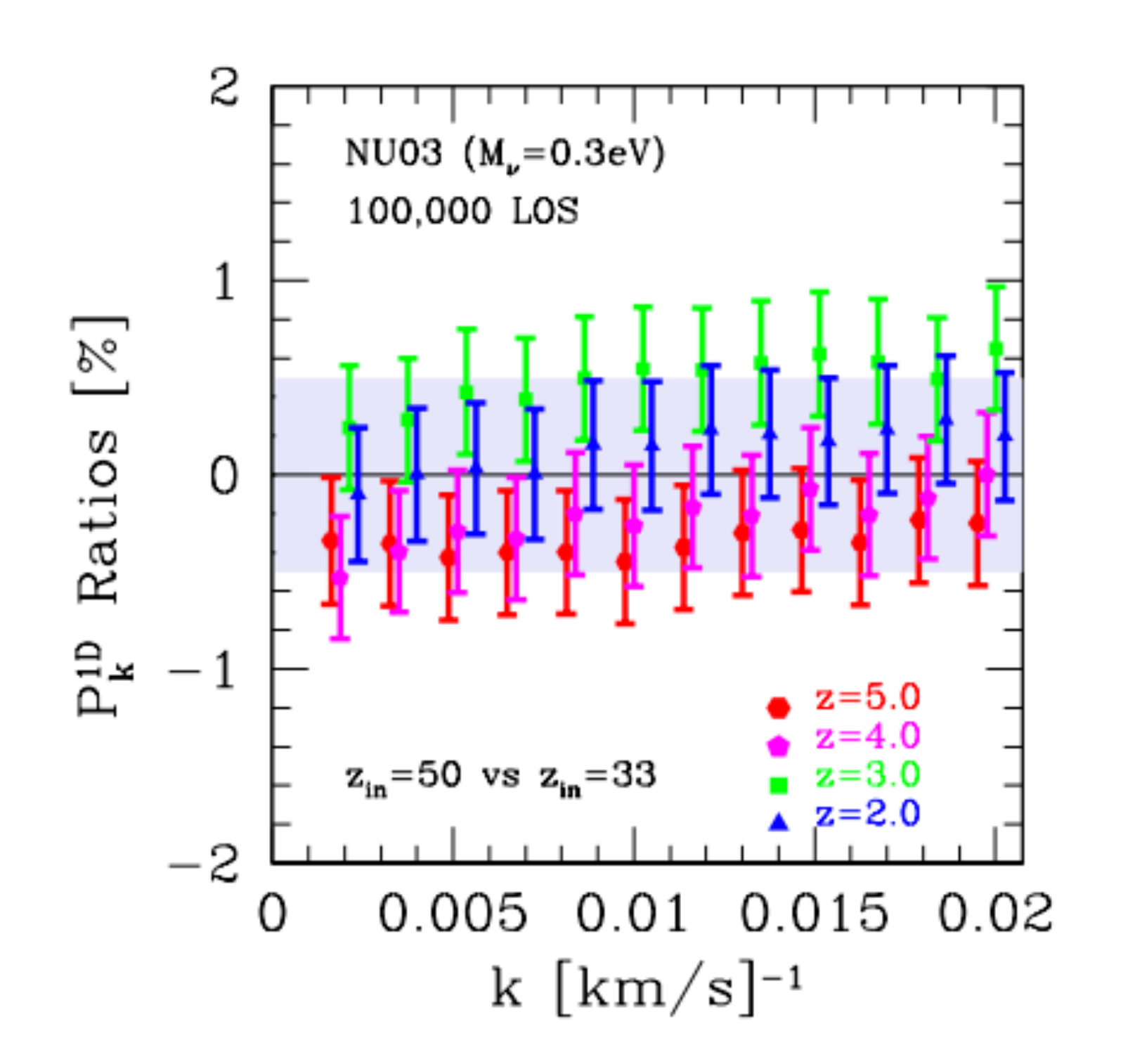}  
\caption{Systematics related to the starting redshift choice. Two
simulations, one initiated at $z_{\rm in}=50$  and the other at $z_{\rm in}=33$ (both with 2LPT), and
 characterized by 
a $100 h^{-1}{\rm Mpc}$ box and $832^3$ particles/type in a
massive neutrino cosmology with $\sum m_{\nu}=0.3~{\rm eV}$, are compared in terms of 
1D flux statistics -- with quantities expressed in power spectrum ratios. 
The colored horizontal band represents the 0.5\% level. Points in the figure are slightly displaced along the $x$-axis,  for ease
of visibility.  Error bars are $1\sigma$ deviations computed from 100,000 skewers extracted at different final $z$-intervals, as
indicated in the panel. The starting redshift effect is a sub-percent  systematics, impacting the flux statistics at the $\simeq 0.5\%$ level.}
\label{fig_starting_z}
\end{figure}  



\subsection{Systematic Tests:  Examples} \label{subsec_sys_effects}

Finally, we briefly illustrate two examples chosen from the {\it Systematics Suite}
(see Section \ref{subsec_systematics_suite}  and Table  \ref{table_systematics_sims_all}), in order to
highlight possible sources of systematics  -- among many -- that can impact parameter constraints. 

The first case, classified as a numerical artifact but also carrying some theoretical implications, 
is related to the following aspect:
the baseline BG cosmology of the {\it Grid Suite}
does not contain any massive neutrinos, but only 
three massless active neutrinos as in the current standard model of particle physics, so that  $N_{\rm eff}=3.046$.
This choice is primarily motivated by the fact that we aim at quantifying and eventually isolating the effects of massive neutrinos on
cosmic structures, with respect to a massless neutrino scenario. 
In Planck 2015 and 2018 cosmological parameter estimation, however, the baseline model adopted assumes 
a minimal-mass normal hierarchy for the neutrino masses, accurately 
approximated for current cosmological data as a single massive eigenstate with $M_{\nu} = 0.06~{\rm eV}$.
This assumption is consistent with global fits to recent oscillation and other data, but of course is not the only possibility. 
In this respect, the massive neutrino cosmologies considered in the {\it Sejong Suite} -- strictly speaking -- refer to the degenerate neutrino hierarchy, while
our baseline model neglects a minimal-mass normal hierarchy inferred from oscillation experiments. 
While this latter choice is well-motivated, at the level of parameter constraints, one may want to
take into account the implications of our baseline model assumption, and assess how it propagates into the main cosmological observables.
To this end, Figure \ref{fig_systematics_minimal_neutrino_mass} shows ratios of total matter (left panel) 
and Ly$\alpha$ flux (right panel) power spectra (expressed in percentage) as a function of redshift (i.e., $z=2.0, 3.0, 4.0, 5.0$), indicated with different colors and line styles.
Specifically, the $y$-axis displays the ratio between quantities evaluated in a baseline Planck-like model having a neutrino minimal-mass normal hierarchy of $M_{\nu} =0.06~{\rm eV}$ 
(simulation denoted as `BGM' in Table \ref{table_systematics_sims_all})
versus our BG massless neutrino reference scenario, as a function of $k$. 
As is evident from the figure, in the  Ly$\alpha$ forest regime of interest, deviations in terms of the total matter power spectrum are within 
$1\%$, while in terms of the 1D flux power spectrum, they remain within $0.3\%$. 
Therefore, one may want to account for this effect when presenting cosmological constraints.  

Another relevant source
of potential systematics, also classified as a numerical artifact, is the starting redshift ($z_{\rm in}$) of the simulations -- see e.g. Crocce et al. (2006). 
In most of our runs, we set initial conditions at $z_{\rm in}=33$ with 2LPT. 
This choice has been determined
by taking into account a number of factors,
in particular, the overall targeted accuracy, the computational cost in performing a large number of runs, 
and more importantly possible shot-noise effects introduced by the neutrino component. 
In fact, if the starting redshift 
is too high, the neutrino shot noise becomes too severe  due to high thermal velocities 
related to the particle implementation  (see Viel et al. 2010, as well as later studies from the same authors for further details).
In previous works, we have  extensively investigated the impact of this assumption.
For analogous reasons, in Rossi et al. (2014) we started our runs at an even 
slightly  lower redshift ($z_{\rm in}=30$) and 
demonstrated  that our simulations are
able to correctly recover the 1D Ly$\alpha$ flux power spectrum in the most relevant redshift range of interest for the
Ly$\alpha$ forest  ($2.2 \le z \le 4.2$, at scales up to $k=0.02$ [km/s]$^{-1}$), 
and proved those simulations to be very powerful in constraining 
massive neutrinos and dark radiation. 
Along these lines, Figure \ref{fig_starting_z} 
shows the effects of systematics associated with the choice of the initial redshift.
A run started at 
$z_{\rm in}=50$  with 2LPT and characterized by a $100 h^{-1}{\rm Mpc}$ box and $832^3$ particles per type in a
massive neutrino cosmology ($\sum m_{\nu}=0.3~{\rm eV}$, indicated as  NU03\_z50 in Table \ref{table_systematics_sims_all}), is confronted
with a similar realization, having initial conditions set at $z_{\rm in}=33$ (simulation NU03\_z33 in Table  \ref{table_systematics_sims_all}), 
in terms of 1D flux power spectrum statistics. 
The colored horizontal band represents the 
0.5\% level, with quantities expressed in power spectrum ratios; points at the same (final) redshift in the figure are slightly displaced along the $x$-axis, for ease of visibility.
Error bars are $1-\sigma$ deviations computed from 100,000 simulated skewers extracted at different final $z$-intervals ($z=2.0, 3.0, 4.0, 5.0$, respectively). 
As inferred from the plot, in the range relevant for our constraints (for example, considering eBOSS data),
the starting redshift effect is a sub-percent  systematics, impacting the flux statistics at the $\simeq 0.5\%$ level,
and thus always below the splicing uncertainty.  At the stage of cosmological constraints, however,  
this effect can also be incorporated via a nuisance parameter, and eventually
marginalized over.



\section{Summary and Conclusions: Novelties,  Applications, and Outlook} \label{sec_conclusions}


The Ly$\alpha$ forest has recently gained considerable attention as
a unique tracer of the high-redshift cosmic web, complementary to lower-$z$ probes,
with its statistical power greatly enhanced thanks to available data from the SDSS. 
In particular, the Ly$\alpha$ forest will play a leading role in unveiling the \textit{dark sector} of the universe,
as it is highly sensitive to neutrino masses, WDM,  and additional dark radiation components such as sterile neutrinos -- 
via significant attenuation effects on the matter and flux power spectra at small scales.  
Current and upcoming  surveys (i.e., eBOSS, DESI)
are or will be providing new exquisite high-quality data suitable for Ly$\alpha$ forest studies, potentially carrying novel discoveries
along with the possibility of obtaining competitive cosmological parameter and {\it dark sector} constraints.
Interpreting such statistically rich datasets demands equally high-quality numerical simulations: reliably
modeling nonlinear evolution and the complex effects of baryons, neutrinos, and 
dark radiation at small scales as required by the Ly$\alpha$ forest
is only possible via sophisticated high-resolution cosmological hydrodynamical simulations. 
Moreover, realistic numerical simulations able to 
reproduce observational surveys
are also indispensable
for controlling systematics that can spoil parameter constraints.
A complete hydrodynamical treatment is mandatory to reach the precision that data are now beginning to show.
And the most competitive bounds on neutrino and WDM  masses
will be obtained by including the key contribution of Ly$\alpha$ forest data;
being able to rule out one of the neutrino hierarchies would have a relevant impact in particle physics, as its knowledge 
will complete the understanding of the neutrino sector and shed light onto leptogenesis, baryogenesis, and the origin of mass --
with implications for neutrinoless double beta decay and double beta decay  experiments.
 
Within this context, and inspired by all of these reasons, 
we have carried out an extensive set of state-of-the-art high-resolution cosmological 
hydrodynamical simulations (over 300 runs) termed the \textit{Sejong Suite}, primarily  
developed for modeling the Ly$\alpha$ forest in the redshift interval $5.0 \le z \le 2.0$.
This paper is mainly intended as a presentation, technical description, and guide to the usage of the simulations and of 
the related post-processing products.
Motivated by practical purposes, 
the entire suite has been organized into three main categories, targeting 
different scientific and technical aspects: (1) the {\it Grid Suite}, useful for cosmological parameter constraints especially 
regarding massive and sterile neutrinos and the {\it dark sector};
(2) the {\it Supporting Suite},  aimed at studying the detailed physical effects of exotic particles and dark radiation models, as well as their impact on the high-$z$ cosmic web; 
and (3) the {\it Systematics Suite}, meant to address several systematic effects, ranging from numerical challenges till parameter degeneracies. 

While the simulations of the {\it Sejong Suite} share the same philosophy as the
runs developed in Rossi et al. (2014), there are a number of improvements and novelties at all levels in this new release related to 
{\it technical},   {\it modeling}, and {\it innovative} aspects.
In particular, on the {\it technical}  side, we devised a novel flexible pipeline able to 
produce a generic end-to-end 
simulation on a given high-performance supercomputing architecture.
We ameliorated the time step integration and its accuracy, using a finer integration step.
We customized  Gadget-3 to our supercomputing architecture, resulting in a overall faster 
and more efficient performance of the code. 
We increased the resolution with respect to our previous simulations, thanks to the 
addition of more than $123$ million particles per type for our primary realizations (i.e., \textit{Grid Suite}),
corresponding to up to 369 million particles per simulation for our heaviest runs -- 
 equivalent of adding an extra simulation characterized by  $~700^3$ particles/type
to a given simulation set. In this way, although with additional enhancing techniques, 
we are able to reach an equivalent resolution up to $3 \times  3328^3 = 110$ billion particles in a $(100 h^{-1} {\rm Mpc})^3$ box size,
corresponding to a grid resolution of $30h^{-1}{\rm kp}$ -- ideal for meeting the observational requirements of eBOSS and DESI. 
On the {\it modeling} side, we expanded the parameter space for the {\it Grid Suite}, and tightened their variation range;
we used a finer neutrino mass, introduced models with dark radiation and WDM, and simulated dark radiation scenarios without approximations. 
We also improved on the reionization history (i.e., better determination of the reionization redshift), which is well known to directly impact the parameter constraints. 
On the {\it innovative} side, the most  significant novelty is the inclusion, for the first time, of
extended mixed scenarios describing the combined effects of WDM, neutrinos, and dark radiation. 
These non-canonical models are quite interesting, particularly for constraining $N_{\rm eff}$ and WDM relic masses directly from
Ly$\alpha$ forest data. 

In future releases of the {\it Sejong Suite}, we plan to expand around this framework and provide more realizations. 
In particular,  the {\it Grid Suite} represents probably the ultimate word on splicing, as running larger-volume
high-resolution hydrodynamical simulations able to meet the requirements of upcoming surveys
is becoming progressively less prohibitive, in terms of computational costs. Therefore, abandoning splicing and interpolation
techniques and adopting emulator-based strategies
will be feasible, and we aim at extending the {\it Sejong Suite} in this direction as well. 

In addition to a thorough descriptions of the simulations,  in this 
work we also presented a first analysis of the {\it Sejong Suite}, primarily focusing on the matter and flux statistics (Section \ref{sec_sejong_suite_first_results}), 
and in particular, we showed that we are able to accurately reproduce the 
1D flux power spectrum down to scales $k = 0.06~{\rm [km/s]^{-1}}$  as mapped by recent high-resolution quasar data.  

While primarily developed for  Ly$\alpha$ forest studies, our high-resolution simulations and Ly$\alpha$ skewers may be useful for a broader 
variety of cosmological and astrophysical purposes.
In addition to cosmological parameter constraints and the characterization of systematics,
the {\it Sejong Suite} can  in fact be helpful for novel methods that aim at painting baryonic physics into
DM-only simulations, that intend to add small-scale physics to low-resolution larger-box simulations, or that
attempt to accurately reproduce observables from a limited number of realizations in parameter space
-- as suggested by several recent studies. 
In this regard, there has been interesting progress toward efficient emulators 
(Bird et al. 2019; Giblin et al. 2019; Rogers et al. 2019; Van der Velden et al. 2019; Zhai et al. 2019),  
training sets, neural networks, and machine-learning techniques (Nadler et al. 2018;  Rodr{\'i}guez et al. 2018; Mustafa et al. 2019; Ramanah et al. 2019; 
Shirasaki et al. 2019; Takhtaganov et al. 2019; 
Zamudio-Fernandez et al. 2019; Wibking et al. 2020),  
or novel approaches such as the Cosmological Evidence Modeling  (Lange et al. 2019) or the Machine-assisted Semi-Simulation Model (Jo \& Kim 2019).
The common aspect between these methods is the reliance on some (even partial) information derived from
high-resolution hydrodynamical realizations, and  
in this context, the {\it Sejong Suite} could be   useful for calibrations purposes. 
Moreover, we also foresee numerous applications toward
the galaxy-halo connection (e.g., Wechsler \& Tinker 2018; Martizzi et al. 2019, 2020), 
the  creation of mock galaxy catalogs with novel techniques (Balaguera-Antol{\'i}nez et al. 2019, 2020; Dai et al. 2019), as well
as a number of interdisciplinary and interesting 
synergies with particle physics. 
 
All of the simulations carried out in this work are presented in Tables 
\ref{table_grid_sims_base}--\ref{table_systematics_sims_all},  
and the full list of products available from the {\it Sejong Suite} is summarized in Section \ref{subsec_list_available_products}. 
We plan to make them progressively  available to the scientific community, 
and kindly ask you to refer to this publication if you use any of those products in your studies. 



\begin{acknowledgements}

This work is supported by the National Research Foundation of Korea (NRF) through Grants No. 2017R1E1A1A01077508 
and No. 2020R1A2C1005655 funded by the Korean Ministry of Education, 
Science and Technology (MoEST), and by the faculty research fund of Sejong University.
The numerical simulations presented in this work were performed using the Korea Institute of Science and Technology Information (KISTI) 
supercomputing infrastructure (Tachyon 2) under allocations KSC-2017-G2-0008 and KSC-2018-G3-0008, 
and post-processed with the KISTI KAT System (KISTI/TESLA `Skylake' and `Bigmem' architectures) under allocations KSC-2018-T1-0017, KSC-2018-T1-0033, and KSC-2018-T1-0061. 
We thank the KISTI supporting staff for technical assistance along the way, and Volker Springel for making Gadget-3 available. 
We also acknowledge extensive usage of our new computing resources (Xeon Silver 4114 master node and Xeon Gold 6126 computing node architecture) 
at Sejong University. Many thanks to the anonymous referee for insightful feedback and suggestions, and for empathizing with this challenging work.

\end{acknowledgements}

 


\end{document}